\documentstyle[preprint,aps]{revtex}
\newcommand{\vc}[1]{\mbox{$\vec{#1}$}}
\newcommand{\vcp}[1]{\mbox{$\vec{#1}^{\, \prime}$}}
\newcommand{\vcs}[1]{\mbox{$\vec{#1}^{\, 2}$}}
\newcommand{\vci}[2]{\mbox{$\vec{#1}_{#2}$}}
\newcommand{\vcpi}[2]{\mbox{$\vec{#1}_{#2}^{\, \prime}$}}
\newcommand{\vcsi}[2]{\mbox{$\vec{#1}^{\, 2}_{#2}$}}
\newcommand{\vs}[1]{\mbox{$\vec{\sigma}_{#1}$}}

\newcommand{\ps}[2]{\mbox{$( #1 \cdot #2)$}}
\newcommand{\pv}[2]{\mbox{$( #1 \times #2)$}}
\newcommand{\pth}[3]{\mbox{$ #1 \cdot ( #2 \times #3)$}}
\newcommand{\pf}[4]{\mbox{$ ( #1 \times #2 ) \cdot ( #3 \times #4 ) $}}

\newcommand{\pvso}[1]{\mbox{$ \vec{\sigma}_1 \times #1 $}}
\newcommand{\pvst}[1]{\mbox{$ \vec{\sigma}_2 \times #1 $}}
\newcommand{\pvsp}[1]{\mbox{$(\vec{\sigma}_1 + \vec{\sigma}_2)
\times #1 $}}
\newcommand{\pvsm}[1]{\mbox{$(\vec{\sigma}_1 - \vec{\sigma}_2)
\times #1 $}}

\newcommand{\psso}[2]{\mbox{$\vec{\sigma}_1 \cdot
( #1 \times #2 )$}}
\newcommand{\psst}[2]{\mbox{$\vec{\sigma}_2 \cdot
( #1 \times #2 )$}}
\newcommand{\pssp}[2]{\mbox{$(\vec{\sigma}_1 + \vec{\sigma}_2) \cdot
( #1 \times #2 )$}}
\newcommand{\pssm}[2]{\mbox{$(\vec{\sigma}_1 - \vec{\sigma}_2) \cdot
( #1 \times #2)$}}

\newcommand{\mpw}[2]{\frac{1}{#1 m^{#2}}}
\newcommand{\mpwf}[3]{\frac{#3}{#1 m^{#2}}}

\newcommand{\exnn}{\mbox{$(1 \leftrightarrow 2)$}}
\newcommand{\eh}{\mbox{$\hat{e}_1$}}
\newcommand{\kh}{\mbox{$\hat{\kappa}_1$}}
\newcommand{\chir}{\mbox{$\chi_r$}}
\newcommand{\chis}{\mbox{$\chi_{\sigma}$}}

\newcommand{\vnab}[1]{\mbox{${\vec{\nabla}}^{#1}$}}

\newcommand{\J}{\mbox{$\vec{J}(\vec{k})$}}

\newcommand{\Jz}{\mbox{$J_0 (\vec{k})$}}

\newcommand{\Jk}{\mbox{$\vec{J}(\vec{k},\vec{K})$}}
\newcommand{\Jzk}{\mbox{$ J_0(\vec{k},\vec{K})$}}
\newcommand{\Jkp}[1]{\mbox{$\vec{J} ^{\, (#1)}(\vec{k},\vec{K})$}}
\newcommand{\Jzkp}[1]{\mbox{$ J_0 ^{(#1)}(\vec{k},\vec{K})$}}

\newcommand{\jp}[1]{\mbox{$\vec{\jmath}^{\,\, (#1)}(\vec{k}\, )$}}
\newcommand{\rhp}[1]{\mbox{$\rho^{(#1)}(\vec{k}\, )$}}
\newcommand{\jpF}[1]{\mbox{$\vec{\jmath}_F^{\,\, (#1)}(\vec{k}\, )$}}
\newcommand{\rhpF}[1]{\mbox{$\rho_F^{(#1)}(\vec{k}\, )$}}
\newcommand{\rhqQ}[3]{\mbox{$
\rho_F^{(2)}(\mbox{2;#1};\vec{k},\vec{q},\vec{Q})_{#2}^{#3}$}}
\newcommand{\rhqQa}[4]{\mbox{$
\rho_{#1}^{(2)}(\mbox{2;#2};\vec{k},\vec{q},\vec{Q})_{#3}^{#4}$}}
\newcommand{\jnrq}[2]{\mbox{$
\vec{\jmath}_F^{\,\, (1)}(\mbox{2;#1};\vec{k},\vec{q}\, )_{#2}^{-}$}}
\newcommand{\jrcq}[4]{\mbox{$\vec{\jmath}_{#1}^{\,\, (3)}
(\mbox{2;#2};\vec{k},\vec{q},\vec{Q})_{#3}^{#4}$}}

\newcommand{\rhFqQ}[3]{\mbox{$
\rho_{FW}^{(2)}(\mbox{2;#1};\vec{k}\, )_{#2}^{#3}$}}
\newcommand{\jnrFq}[2]{\mbox{$
\vec{\jmath}_{FW}^{\,\, (1)}(\mbox{2;#1};\vec{k}\, )_{#2}^{-}$}}
\newcommand{\jrcFq}[3]{\mbox{$\vec{\jmath}_{FW}^{\,\, (3)}
(\mbox{2;#1};\vec{k}\, )_{#2}^{#3}$}}

\newcommand{\prop}[2]{\mbox{ $\tilde{\Delta}_{#1}(\vec{q}_{#2}^{\, 2})$}}
\newcommand{\propp}[2]{\mbox{$\tilde{
 \Delta}^{\prime}_{#1}(\vec{q}_{#2}^{\, 2})$}}
\newcommand{\wt}[2]{\mbox{ $\tilde{w}_{#1}^{(1)}(\vec{q}_{#2}\, )$}}

\newcommand{\fq}{\mbox{$f(\vec{q}_1^{\, 2},\vec{q}_2^{\, 2})$}}
\newcommand{\con}[1]{\frac{g_{#1}^2}{(2 \pi)^3}}
\newcommand{\conm}[3]{ \frac{g_{#1}^2}{(2 \pi)^3 \, #2 m^{#3} } }
\newcommand{\vtos}[1]{\mbox{$\tilde{v}_{#1}^{(1)} (\vec{q}_2 \, )$}}
\newcommand{\vto}[1]{\mbox{$\tilde{v}_{#1}^{(1)} (\vec{q} \, )$}}
\newcommand{\vtt}[1]{\mbox{$\tilde{v}_{#1}^{(3)} (\vec{q},\vec{Q})$}}

\newcommand{\Ff}[2]{\mbox{$F_{#1}^{#2}$}}
\newcommand{\Gm}[1]{\mbox{$G_{M,1}^{#1}$}}

\newcommand{\Fmn}{\mbox{$F^-$}}
\newcommand{\mut}{\mbox{$\tilde{\mu}$}}
\newcommand{\nut}{\mbox{$\tilde{\nu}$}}

\newcommand{\om}[2]{\mbox{${\omega}_{#1}^{(#2)}$}}
\newcommand{\ord}[1]{\mbox{${\cal O}(m^{#1})$}}

\newcommand{\vb}{\mbox{$\vec{\beta}$}}

\newcommand{\beq}{\begin{equation}}
\newcommand{\eeq}{\end{equation}}
\newcommand{\bea}{\begin{eqnarray}}
\newcommand{\eea}{\end{eqnarray}}

\begin{document}

\preprint{MKPH-T-96-24}

\title{{\bf Intrinsic operators for the electromagnetic nuclear current}}

\author {
    J.\ Adam, Jr.$^{1,\,2)}$
\footnote{On leave from
     Institute for Nuclear Physics, \v{R}e\v{z} n.\ Prague,
              CZ-25068, Czech Republic }
 and H.\ Arenh\"ovel$^{1)}$}
\address{
1) Institut f\"{u}r Kernphysik, Johannes Gutenberg-Universit\"{a}t,
       D-55099 Mainz, Germany\\
2) TJNAF Theory Group,
   12000 Jefferson Ave, Newport News, VA 23606, U.S.A. }
\maketitle

\begin{abstract}
The intrinsic electromagnetic nuclear meson exchange charge and 
current operators arising from a separation of the center-of-mass 
motion are derived for a one-boson-exchange model for the nuclear 
interaction with scalar, pseudoscalar and vector meson exchange 
including leading order relativistic terms. Explicit expressions 
for the meson exchange operators corresponding to the different 
meson types are given in detail for a two-nucleon system. 
These intrinsic operators are to be evaluated between intrinsic 
wave functions in their center-of-mass frame. 
\end{abstract}

\section{Introduction}

In a recent paper \cite{AdAr95}, henceforth called ``I'', we
formally have separated for an arbitrary relativistic transition
operator the center-of-mass (c.m.) motion from the intrinsic one.
This has been achieved by exploiting the general properties of the
Poincar\'{e} generators in conjunction  with a $1/m$-expansion.
As a result, the frame dependence of an arbitrary transition
operator has been derived explicitly up to the lowest-order
relativistic contributions without reference to any specific dynamic
model, leaving undetermined only the genuine intrinsic operators.
The frame dependent terms have a clear physical meaning describing
effects of a Lorentz transformation for a scalar, vector and a general
Lorentz tensor of higher rank as well as modifications due to
Lorentz contraction and Wigner rotation.

On the other hand, for the determination of the remaining intrinsic
operators one needs a specific dynamic model.
There exist several techniques for the derivation of the
electromagnetic (e.m.) currents
using a $1/m$-expansion \cite{Fr80}. In the leading relativistic
order, they all lead to unitarily equivalent descriptions, i.e.,
the various Hamiltonians and e.m.\ currents are connected by unitary
transformations. The e.m.\ charge and current densities, which we
will denote below by
$\Jz_{FW}$\ and $\J_{FW}$\  (from Foldy-Wouthuysen),
are obtained in a general
reference frame in terms of individual nucleon coordinates.
In this paper we will start from the e.m.\ currents obtained in the
framework of the extended S-matrix method \cite{ATA}. They
are listed in Appendix A for completeness. The derivation starts
from the relativistic Feynman amplitudes from which the iterated Born
contribution of the one-nucleon current is subtracted in order
to avoid double-counting.
The $1/m$-expansion is then employed at the last stage
making the technique relatively transparent and easy to use.

In order to write
down the transition amplitudes in terms of matrix elements between
states with separated c.m.\ motion, one has to include the effect of
the Lorentz boost on the rest frame wave functions, which depend on
the individual particle variables \cite{Fr75}. This is done with the help
of a unitary transformation which introduces additional so-called
boost currents to be added to the FW ones.
Then, the c.m.\ motion can be
separated also for the nuclear current operators \cite{AdAr95,FrGP},
and the transition amplitudes are expressed in terms of
matrix elements of intrinsic operators between intrinsic wave
functions and simple kinematical factors expressing the c.m.\
motion effects. The intrinsic currents
have a simpler structure than the FW ones and their explicit construction
for the one-boson-exchange (OBE) model is the main subject of this paper.

In the next section we first collect the neccessary general expressions
as obtained in I. In particular,
we give the relations of the intrinsic currents to the FW ones.
The boost contributions are written in a convenient form in
momentum representation. For simplicity, we give all explicit
expressions for the currents of a two-nucleon system, but the extension
to a larger number of nucleons is straightforward. Then, we present
in Sect.\ 3 the intrinsic currents for the one-nucleon and
the interaction-dependent meson exchange two-nucleon currents (MEC),
corresponding to the exchange of scalar, vector and pseudoscalar
mesons, both for isospin 0 and 1. Finally, we summarize our results and
give an outlook
in Sect.\ 4.

\section{General considerations}

We will start from the general expressions for a Lorentz
vector of ``type I'' having the leading order in $1/m$ in the zero 
component as derived in I and \cite{FrGP}.
Separating the c.m.\ motion of the initial and final states, we could write
the full operators in terms of pure intrinsic operators where the c.m.\
motion effects are described by a known functional dependence on
$\vc{K} = \vci{P}{f} + \vci{P}{i}$. Here, $\vec P_{i/f}$ denote the total
momentum of the initial and final hadronic system, respectively.
The intrinsic operators, introduced in I, depend only on the momentum 
transfer $\vc{k} = \vci{P}{f} - \vci{P}{i}$ and are denoted
by $\rho(\vec{k}\, ) $\ and $\vec{\jmath}\, (\vec{k}\, ) $\ with
their nonrelativistic (\rhp{0}, \jp{1}) and leading
order relativistic parts (\rhp{2}, \jp{3}). Note that the upper index in
brackets refers to the order in $1/m$.
According to Eqs.\ (91)-(92) of I,
the full operators, which have to be evaluated between intrinsic wave
functions only, are given as
\bea
\Jzkp{0} &=& \rhp{0} \, , \label{frh0I} \\
\Jkp{1} &=& \jp{1} +
 \frac{\vc{K}}{2M} \, \rhp{0} \, , \label{fj1I}\\
\Jzkp{2} &=& \rhp{2}
+ \Big( \hat{L} + \hat{W} +
 \frac{\vcs{K}}{8M^2} \Big) \rhp{0}
+ \frac{1}{2M} \vc{K} \cdot \jp{1} \, , \label{frh2I} \\
\Jkp{3} & = & \jp{3} + (\hat{L} + \hat{W}) \Jkp{1}
 + \frac{\vc{K}}{8M^2} \vc{K} \cdot \jp{1}  \nonumber\\
&&
 + \frac{\vc{K}}{2M}  \Big[ \rhp{2}  -
\frac{1}{2M} \Big( \epsilon_{f}^{(1)}+\epsilon_{i}^{(1)}
+\frac{\vcs{k}}{4M}
\Big) \rhp{0}  \Big]  \, ,
\label{fj3I}
\eea
where the operators $\hat{L}$\ and $\hat{W}$\ are defined by
\bea
\hat{L} f(\vc{k}\, ) &=&
 - \frac{2\om{fi}{1}+\om{R}{1}}{4M} \ps{\vc{K}}{\vnab{k}}  f(\vc{k}\, )
 \, , \label{Lhat} \\
\hat{W} f(\vc{k}\, ) &=& - \frac{i}{8M^2}
 \Big\{ \pth{\vc{S}}{\vc{k}}{\vc{K}} \, , f(\vc{k}\, ) \Big\} \, .
\label{What}
\eea
Here $M$ denotes the total mass of the system and
$f$\ stands for $\rho $\ or $\vc{\jmath}$.
The gradient \vnab{k}\ {\em does not} act on the nucleon form
factors \cite{AdAr95,FrGP}. \vc{S}\ is the total spin operator
of the composite system. The nonrelativistic intrinsic excitation
energy is denoted by
\beq
 \om{fi}{1}  =  \epsilon_{f}^{(1)} - \epsilon_{i}^{(1)}  \label{omfi}\,,
\eeq
with $\epsilon_{i/f}^{(1)}$ as the nonrelativistic energies for the initial
and final states, respectively.
Finally, \om{R}{1}\  is the nonrelativistic recoil energy
\beq
 \om{R}{1} = \frac{{\vci{P}{f}}^2}{2M} - \frac{{\vci{P}{i}}^2}{2M} =
 \frac{\ps{\vc{k}}{\vc{K}}}{2M}
  \, . \label{omr}
\eeq

The $\hat{L}$-term describes the Lorentz contraction of
the system, while $\hat{W}$ reflects the Wigner rotation of the
total spin associated with the transformation from the Breit
frame ($\vci{P}{f}=-\vci{P}{i}=\vc{k}/2$)
to a general one \cite{FrGP,Fr73}. Note, that
in $k$\/-congruent frames, i.e., those for which $\vci{P}{i}$
and thus $\vc{K}$ are parallel to \vc{k}, the $\hat{W}$-term vanishes.
The $\hat{L}$-term can be absorbed into the nonrelativistic operator 
by replacing $\vec k$ by an effective momentum transfer 
$\vci{k}{eff}$ having the same direction as $\vec k$ and with
$\vc{k}^{\, 2}_{eff}=\vec k^{\,2}-(k_0^{(1)})^2+
(\omega_{fi}^{(1)})^2$) \cite{AdAr95}.
The intrinsic operators, introduced in I, are obtained from the
expressions  (\ref{frh0I})--(\ref{fj3I}) if taken in the
Breit frame, i.e.,
\beq
 j_{\lambda} (\vc{k}\, ) = J_{\lambda} (\vc{k},\vc{0}) \, .
\eeq

Except for the $\hat{L}$ and $\hat{W}$ terms, all other contributions
in (\ref{frh0I})-(\ref{fj3I}) can be obtained by the Lorentz transformation
of the charge and current operators from the Breit
frame to a general one. The parameters \vb\ and $\gamma $\
of such a transformation are given by
\beq
\vb = \frac{\vc{K}}{E_f + E_i} \ \ \mbox{and} \ \
 \gamma = \frac{1}{\sqrt{1 - \beta^2}} \, .
\label{beta}
\eeq
For an arbitrary $k$\/-congruent frame, 
one can generalize the expressions in (\ref{frh0I})-(\ref{fj3I}) to
\bea
 \Jzk & = & \gamma \Big(
 \rho (\vc{k}_{eff}) + \vb \cdot \vec{\jmath}\, (\vc{k}_{eff}) \Big)  \, ,
\label{lorch}\\
 \Jk & = & \Big( \vec{\jmath}\, (\vc{k}_{eff}) +
 \frac{\gamma-1}{\beta^2} (\vb \cdot \vec{\jmath}\, (\vc{k}_{eff}) ) \vb
 + \gamma \vb \,  \rho (\vc{k}_{eff}) \Big)
 \, . \label{lorcur}
\eea
Then all kinematical effects related to the Lorentz vector
structure of the current are taken into account exactly.
Only in the intrinsic charge and current densities remain approximations
with respect to the $1/m$-expansion in the derivation
and with the introduction of the effective momentum transfer 
$\vci{k}{eff}$.

As described in detail in I, the current operator
$J_{\lambda}(\vc{k},\vc{K})$\ acting
in the space of intrinsic wave functions is defined by
the matrix element of the e.m.\ operators between plane waves
describing the c.m.\ motion of the system
\beq
\langle \vci{P}{f} |  J_{\lambda}(\vc{k}\,) | \vci{P}{i} \rangle =
 \Big( \frac{M_f M_i}{E_f E_i} \Big) ^{1/2}
  J_{\lambda}(\vc{k},\vc{K}) \, \delta ( \vci{P}{f} - \vci{P}{i} -\vc{k} )
\, , \label{totj}
\eeq
with nonrelativistic normalization of the plane waves
\beq
\langle \vci{P}{f} | \vci{P}{i} \rangle =
 \delta ( \vci{P}{f} - \vci{P}{i} ) \, .
\label{cmnorm}
\eeq
Therefore, the factor in front of $J_{\lambda}(\vc{k},\vc{K})$\ in
(\ref{totj}) ensures that \Jzk\ and \Jk\
represent {\em covariantly} normalized operators in the space of intrinsic
wave functions.
We have chosen the normalization convention (\ref{cmnorm})
since it is usually adopted implicitly in the derivations of the FW operators
on which the construction of the intrinsic operators is based. For this
reason we introduce in addition {\em noncovariantly} normalized operators of
the full current by
\beq
\tilde  J_{\lambda}(\vc{k},\vc{K})=
\Big( \frac{M_f M_i}{E_f E_i} \Big) ^{1/2}
J_{\lambda}(\vc{k},\vc{K}) \, .
\label{totjtilde}
\eeq

Up to the order considered here,
the full charge and current operators on the l.h.s.\ of (\ref{totj})
are  given by
\beq
 J_{\lambda}(\vc{k}\,) = J_{\lambda}(\vc{k}\,)_{FW} +
 i \, \Big[ \chi \, , J_{\lambda}(\vc{k}\,)_{FW} \, \Big] \, ,
\label{boostop}
\eeq
where $\chi $\ describes the wave function boost \cite{Fr75}.
The commutator term is
usually called the boost contribution to the charge and current.

The derivation of the intrinsic currents as well as their final form simplify
considerably, if one splits their relativistic parts in the following way,
employed implicitly by Friar, Gibson and Payne \cite{FrGP},
\bea
 \rhp{2} &  = &   \rhpF{2}   -
\frac{\vcs{k}}{8M^2} \ps{\vc{k}}{\vnab{k}} \rhp{0}  +
 \frac{\vcs{k}}{8M^2}  \rhp{0}  \, , \label{FGPch}\\
 \jp{3}  &  = & \jpF{3} -
 \frac{\vcs{k}}{8M^2} \ps{\vc{k}}{\vnab{k}} \jp{1} \, .
\label{FGPcr}
\eea
The gradient terms can then be absorbed in the operator $\hat{L}$ 
leading in turn to a new effective momentum transfer $\vc{k}_{F}$ 
given by
\beq
\vc{k}_{F}^{\, 2}  =  \vc{k}^{\, 2} - (k_0^{(1)})^2 + (\om{fi}{1})^2
 - \frac{\vc{k}^{\, 4}}{4M^2}  \, ,
\eeq
where the direction of $\vc{k}_{F}$\ is again parallel to \vc{k}.
Note, that $\vc{k}_{F}^{\, 2}$\ is still effectively a Lorentz scalar,
up to the order considered. The introduction of this new effective 
momentum transfer $\vec k_F$ leads to a rearrangement of the intrinsic 
operators in the following way
\bea
 \rhp{0} & \rightarrow &   \rho^{(0)}(\vec k_F)\,,
\label{ch0eff}\\
 \jp{1}  & \rightarrow & \vec\jmath^{\,\,(1)}(\vec k_F) \,,
\label{j0eff}\\
 \rhp{2} &  = &   \rhpF{2}  
+ \frac{\vcs{k}}{8M^2}  \rhp{0}  \, , \label{ch2eff}\\
 \jp{3}  &  = & \jpF{3} \,.
\label{j2eff}
\eea
Note, that now part of the relativistic effects are contained in 
$\rho^{(0)}$ and $\vec \jmath^{\,\,(1)}$. 

Therefore, it is sufficient to determine the operators $\rho_F$\ and
$\vc{\jmath}_F$. In terms of noncovariantly normalized operators,
they are in general given by
\bea
 \rhpF{0} & = &
\tilde J_0^{(0)}(\vec k,0)
\, , \label{intrh0}\\
 \jpF{1} & = & \vec {\tilde J}^{(1)}(\vec k,0)
\, , \label{intj1}\\
 \rhpF{2} & = &
\tilde J_0^{(2)}(\vec k,0)
 + \rho_{sep}^{(2)}(\vc{k}\,)
             \, , \label{intrh2}\\
 \jpF{3} & = &
\vec {\tilde J}^{(3)}(\vec k,0)
 + \vec{\jmath}_{sep}^{\,\,(3)}(\vc{k}\,)
\, , \label{intj3}
\eea
where we have introduced the separation charge and current operators
\bea
 \rho_{sep}^{(2)}(\vc{k}\,) & = &
 \frac{\vcs{k}}{8M^2} \ps{\vc{k}}{\vnab{k}} \, \rhpF{0}
             \, , \label{seprh2}\\
 \vec{\jmath}_{sep}^{\,\,(3)}(\vc{k}\,) & = &
\frac{\vcs{k}}{8M^2}\Big(1+ \ps{\vc{k}}{\vnab{k}}\Big) \, \jpF{1}
\, , \label{sepj3}
\eea
and $\rhpF{0}=\rhp{0}$\ and
$\jpF{1}=\jp{1}$ in order to unify our notation.

The e.m.\ current should satisfy the continuity equation, which means in
our notation
\beq
 \vc{k} \cdot \vc{J} (\vc{k},\vc{K})  =  k_0 J_0 (\vc{k},\vc{K}) \,.
\eeq
According to I, it implies for the intrinsic operators the following
relations
\bea
 \vc{k} \cdot \jpF{1}  &=&  \bigg[h^{(1)},\,\rhpF{0}\bigg]\,, \label{con1}\\
 \vc{k} \cdot \jpF{3}  &=&  \bigg[h^{(1)},\,\rhpF{2}\bigg] +
\bigg[h^{(3)} - \frac{\vcs{k}}{8M^2}h^{(1)},\, \rhpF{0}\bigg] \,.
\label{con3}
\eea
where $h$ denotes the intrinsic Hamiltonian.
It is useful to separate the contributions of the one-body  and the meson
exchange currents, denoted by
$a(1;k)$ and $a(2;k)$, respectively. Splitting $h$ into kinetic and
potential energy $h=t+v$, one would expect for the one-body charge and 
current operators of the two-nucleon system ($M=2m$ denoting by $m$ 
the nucleon mass) to satisfy the relations 
\bea
 \vc{k} \cdot \vc{\jmath}_{F}^{\, \, (1)}(\mbox{1};\vc{k})&=&
\bigg[t^{(1)},\,\rho_F^{(0)}(\mbox{1};\vc{k})\bigg]
\,,\label{con11body}\\
 \vc{k} \cdot \vc{\jmath}_{F}^{\, \, (3)}(\mbox{1};\vc{k})&=&
\bigg[t^{(1)},\,\rho_F^{(2)}(\mbox{1};\vc{k})\bigg]+
\bigg[t^{(3)}-\frac{\vcs{k}}{32m^2}t^{(1)},\,
\rho_F^{(0)}(\mbox{1};\vc{k})\bigg]\,.
\label{con31body}
\eea
Consequently, the two-body MEC operators would have to fulfil
\bea
 \vc{k} \cdot \vc{\jmath}_{F}^{\, \, (1)}(\mbox{2};\vc{k})&=&
\bigg[v^{(1)},\,\rho_F^{(0)}(\mbox{1};\vc{k})\bigg]
\,,\label{con12body}\\
 \vc{k} \cdot \vc{\jmath}_{F}^{\, \, (3)}(\mbox{2};\vc{k})&=&
\bigg[h^{(1)},\,\rho_F^{(2)}(\mbox{2};\vc{k})\bigg]
+\bigg[v^{(1)},\,\rho_F^{(2)}(\mbox{1};\vc{k})\bigg]
+ \bigg[v^{(3)}-\frac{\vcs{k}}{32m^2}v^{(1)}
,\,\rho_F^{(0)}(\mbox{1};\vc{k})\bigg]\,,
\label{con32body}
\eea
where we already have used the fact that $\rho_F^{(0)}(\mbox{2};\vc{k})$
vanishes. However, the relations (\ref{con31body}) and (\ref{con32body})
will be slightly modified in Sect.\ III.A (see (\ref{cont21}) and
(\ref{cont23})), because first of all
we will incorporate into the one-body current
$\vc{\jmath}_{F}^{\, \, (3)}(\mbox{1};\vc{k})$ a two-body part effectively,
and secondly, the first commutator on the r.h.s.\ of (\ref{con32body}) will
contain only $t^{(1)}$ because the commutator with $v^{(1)}$ will be of
higher order in the meson nucleon coupling constants not considered here.

In order to get the intrinsic charge and current densities from
(\ref{intrh0}) through (\ref{intj3}) one needs the FW and boost
contributions
expressed in the Breit frame. The FW currents are listed in the appendices
and their evaluation in the Breit frame is straightforward.
The wave function boost $\chi$ contains in general
a kinetic $\chi_0$ and an interaction dependent part $\chi_V$. 
In the OBE model, the interaction boost is non-zero only for a 
pseudoscalar exchange interaction. It is dealt with
explicitly in Appendix C. The leading order kinetic contribution 
($ \sim 1/m^2$) reads in terms of c.m.\ and relative variables
\beq
 \chi_0 = - \frac{1}{2} \sum_{a=1}^A
 \frac{ \ps{\vci{\rho}{a}}{\vc{P}} \ps{\vci{\pi}{a}}{\vc{P}} + h.c.}{2M^2} -
  \frac{1}{2} \sum_{a=1}^A
 \frac{\ps{\vci{\rho}{a}}{\vc{P}} (\vci{\pi}{a}^2) + h.c.}{2M m_a}
 + \sum_{a=1}^A
 \frac{\vci{s}{a} \times \vci{\pi}{a} \cdot \vc{P}}{2M m_a} \, ,
\label{chi0}
\eeq
where $\vci{\rho}{a} = \vci{r}{a} - \vc{R}, \, \vci{\pi}{a} = \vci{p}{a} -
\frac{m_a}{M} \vc{P} $, \vc{P}\ is the total momentum operator
and \vc{R}\ the c.m.\ coordinate given by the usual expression
\beq
\vc{R} = \sum_{a=1}^A m_a \vci{r}{a}/ M \, .
\label{coor}
\eeq
For two particles with equal mass ($m$), the second term
of (\ref{chi0}) vanishes and one gets
\beq
 \chi_0 = - \mpw{16}{2}
 \Big( \ps{\vc{r}}{\vc{P}} \ps{\vc{p}}{\vc{P}} +
        \ps{\vc{p}}{\vc{P}} \ps{\vc{r}}{\vc{P}} \Big) +
 \mpw{8}{2} (\vs{1} - \vs{2}) \times \vc{p} \cdot \vc{P} \, ,
\label{chi0d}
\eeq
where $\vc{r} = \vci{r}{1} - \vci{r}{2}$ and $\vc{p} =
\frac{1}{2} (\vci{p}{1} - \vci{p}{2})$.
Let us now evaluate the boost commutator in the Breit frame
using the momentum representation. To this end, we denote an
intrinsic operator $a(\vc{k})$\ in momentum representation by
\beq
  a(\vc{k},\vc{q},\vc{Q} ) =
 \langle \vcp{p} | a(\vc{k}) | \vc{p} \, \rangle \, ,
\label{aq}
\eeq
where $\vc{q} = \vcp{p} - \vc{p}$\ and $\vc{Q} = \vcp{p} + \vc{p}$.
Noting, that the boost operator $\chi $\ is diagonal with respect to
the c.m.\ plane waves, i.e.,
\beq
 \langle \vcp{P} \, |\,  \chi \, | \vc{P} \, \rangle =
  \chi (\vc{P},\vc{r},\vc{p}\, )\,
 \delta ( \vcp{P} - \vc{P} )  \,,
\eeq
one finds, that the intrinsic commutator of the kinetic boost with an operator
$a(\vc{k})$\ reads
\beq
 \langle \, \vcp{p}|\, i \Big( \chi_0 (\frac{\vc{k}}{2},\vc{r},\vc{p}\, )\,
 a(\vc{k}) \, - a(\vc{k}) \,
 \chi_0 (- \frac{\vc{k}}{2},\vc{r},\vc{p}\, ) \, \Big)
 | \vc{p} \, \rangle = a_{\chi_r}(\vec k,\vec q, \vec Q) +
a_{\chi_\sigma}(\vec k,\vec q, \vec Q)\,,
\label{boostq}
\eeq
where
\bea
a_{\chi_r}(\vec k,\vec q, \vec Q)& = &
 \mpw{32}{2} \Big( \vcs{k} +
 \ps{\vc{k}}{\vc{q}\,}\, \ps{\vc{k}}{\vnab{q}} +
 \ps{\vc{k}}{\vc{Q}} \,\ps{\vc{k}}{\vnab{Q}}
 \Big) \,  a(\vc{k},\vc{q},\vc{Q})\,, \label{chir}\\
a_{\chi_\sigma}(\vec k,\vec q, \vec Q)& = &
\mpwf{32}{2}{i} \Big(\Big\{ a(\vc{k},\vc{q},\vc{Q}), \,
\pssm{\vc{Q}}{\vc{k}} \, \Big\}\nonumber\\
&& - \Big[ a(\vc{k},\vc{q},\vc{Q}), \, \pssm{\vc{q}}{\vc{k}} \Big]
 \Big) \, .
\label{chis}
\eea

Since the FW currents are usually given in momentum representation,
the boost contributions follow directly from (\ref{boostq}).
The extension to systems with more than two nucleons is straightforward. 
The plane wave basis in the intrinsic space can be constructed with 
respect to the set of Jacobi coordinates and,
employing their commutator relations, one can derive an analogue of
(\ref{boostq}). In this case, the contribution of the second term of
(\ref{chi0}) has to be included.

\section{Intrinsic currents}

In this section, explicit expressions are given for the intrinsic currents
$\rho_F$\ and $\vec{\jmath}_F$\
of the two-nucleon system as they follow from
(\ref{intrh0})-(\ref{intj3}).
For the two-body currents, we distinguish different terms 
$ a_{F}^{(n)}(\mbox{2};\beta;\vc{k},\vc{q},\vc{Q})_B^{iso}$, 
labelled by $\beta$ ($= \mbox{pro, mes, and ret}$), according
to their meson propagator structure. Furthermore,
we denote the isospin structure of the MECs by the superscript ``$iso$''
$(=+,\,-)$, and by the subscript ``$B$'', the exchanged meson type.
Each of the currents may be decomposed in general according to the different
contributions arising from the FW-part, the boost and the so-called
separation part. Labelling them by $\alpha$ ($=FW,\,\chir, \,\chis$, and
$sep$), this reads in momentum space
\bea
 a_F^{(n)}(\mbox{1}; \vc{k},\vc{q},\vc{Q}) &=&
 \sum_{\alpha}
 a_{\alpha}^{(n)}(\mbox{1}; \vc{k},\vc{q},\vc{Q}) \, , \\
 a_F^{(n)}(\mbox{2};\beta; \vc{k},\vc{q},\vc{Q})_B^{iso} &=&
 \sum_{\alpha}
 a_{\alpha}^{(n)}(\mbox{2};\beta;\vc{k},\vc{q},\vc{Q})_B^{iso} \, .
\label{form2}
\eea
Further details of our notation are explained in Appendix A.

For each case (one-nucleon or MEC operator) we will
begin with the FW terms in the Breit frame,
then consider the boost commutators (\ref{boostq}) and finally the
additional terms from the r.h.s.\ of (\ref{intrh2})-(\ref{intj3}).
The latter ones are referred to as ``separation'' operators 
and labelled by ``$sep$'' as already introduced above.
Finally, the total intrinsic currents are listed in a way
which makes clear which part of the intrinsic continuity equations
(\ref{con11body})-(\ref{con32body}) or their modified forms 
(\ref{cont11s})-(\ref{cont23}) they saturate.

However, if the corresponding nonrelativistic operator does
not exist, as in the case of the interaction dependent charge
densities or ``$+$'' parts of MECs, the intrinsic operators
$\rho_F$\ and $\vec{\jmath}_F$\ are simply
equal to the FW ones taken in the Breit frame. In such cases,
we do not write them down repeatedly, but list them together
with other intrinsic operators. Also, some nonrelativistic
operators do not depend  on the momentum $\vc{Q}$ which further
simplifies our notation.

\subsection{One-nucleon currents}

We will consider explicitly the currents of the  nucleon labelled ``1'',
while those of the second one follow by a replacement $(1 \leftrightarrow 2
)$.
This replacement, of course, affects also the relative  variables introduced
above, changing the sign of
$ \vc{r}, \vc{p}, \vcp{p}, \vc{q}, \mbox{ and}\ \vc{Q}$.

For the nonrelativistic operators, one finds immediately from (\ref{r10}) and
(\ref{j11}) in the Breit frame
\bea
\rho_F^{(0)}(\mbox{1};\vc{k},\vc{q}\, )_1
 &=& \eh \, \delta(\frac{\vc{k}}{2} - \vc{q}\, ) \, ,
\label{irh10}\\
 \vc{\jmath}_{F}^{\, \, (1)}(\mbox{1};\vc{k},\vc{q},\vc{Q} )_1
 &=& \frac{1}{2m} \Big(\eh \, \vc{Q} +
  i (\eh + \kh ) \, \pvso{\vc{k}} \Big)
\, \delta(\frac{\vc{k}}{2} - \vc{q}\, ) \nonumber\\
 &=& \vc{\jmath}_c^{\, \, (1)}(\mbox{1};\vc{k},\vc{q},\vc{Q})_1 +
  \vc{\jmath}_s^{\, \, (1)}(\mbox{1};\vc{k},\vc{q}\,)_1
\, , \label{ij11}
\eea
where $\vc{\jmath}_{c,s}^{\,\,(1)}$\  stand for the usual nonrelativistic
convection and spin currents.

Now we turn to the relativistic contributions. We first note that,
since the currents of the first nucleon contain
$\delta(\vc{k}/2 - \vc{q}\, )$, the contributions to the boost
commutator (\ref{boostq}) simplify to
\bea
a_{\chi_r}(1;\vec k,\vec q, \vec Q)_1& = &\mpw{32}{2}  \Big(
\frac{\vcs{k}}{2} \ps{\vc{k}}{\vnab{q}}
+  \ps{\vc{k}}{\vc{Q}}\, \ps{\vc{k}}{\vnab{Q}}
 \Big) \, a(\rm{1};\vc{k},\vc{q},\vc{Q})_1\,,\\
a_{\chi_\sigma}(1;\vec k,\vec q, \vec Q)_1& = & \frac{i}{32m^2}
 \Big\{ a(\rm{1};\vc{k},\vc{q},\vc{Q})_1\, ,
  \, \pssm{\vc{Q}}{\vc{k}} \, \Big\}
  \, .
\label{boostqia}
\eea
For the FW part of the charge density, one obtains from (\ref{r12})
in the Breit frame
\beq
\rho_{FW}^{(2)}(\mbox{1};\vc{k},\vc{q},\vc{Q})_1
 = - \frac{\eh + 2 \kh}{8m^2} \,
 \Big( \vcs{k} + i \psso{\vc{Q}}{\vc{k}} \Big)
 \delta(\frac{\vc{k}}{2} - \vc{q}\, ) \, . \label{rh12fwb}
\eeq
The boost and separation contributions to the charge density are
\bea
\rho_{\chir}^{(2)}(\mbox{1};\vc{k},\vc{q},\vc{Q})_1
 &=& - \frac{\eh \, \vcs{k}}{32m^2} \,\ps{\vc{k}}{\vnab{k}}
 \delta(\frac{\vc{k}}{2} - \vc{q}\, ) \,  , \label{rh1chr}\\
\rho_{\chis}^{(2)}(\mbox{1};\vc{k},\vc{q},\vc{Q})_1
 &=& i\frac{\eh}{16m^2} \,  \pssm{\vc{Q}}{\vc{k}}
 \delta(\frac{\vc{k}}{2} - \vc{q}\, ) \, , \label{rh1chs}\\
\rho_{sep}^{(2)}(\mbox{1};\vc{k},\vc{q},\vc{Q})_1
 & = & \frac{\eh \, \vcs{k}}{32m^2}\, \ps{\vc{k}}{\vnab{k}}
 \delta(\frac{\vc{k}}{2} - \vc{q}\, ) \,  . \label{rh1sep}
\eea
The expressions (\ref{rh1chr}) and (\ref{rh1sep}) cancel completely.
This is, of course, the reason why $\rho_F(\vc{k}\, )$\ and
$\vc{\jmath}_F(\vc{k}\, )$\  are introduced in
(\ref{FGPch})-(\ref{FGPcr}). The relativistic part of
the intrinsic one-nucleon charge density is then
\bea
\rho_F^{(2)}(\mbox{1};\vc{k},\vc{q},\vc{Q})_1
 &=&- \frac{\eh}{16m^2}\,
 \Big( 2 \vcs{k} + i \pssp{\vc{Q}}{\vc{k}} \Big)
 \delta(\frac{\vc{k}}{2} - \vc{q}\, )  \nonumber\\
&& - \frac{\kh}{4m^2}\,
 \Big( \vcs{k} + i \psso{\vc{Q}}{\vc{k}} \Big)
 \delta(\frac{\vc{k}}{2} - \vc{q}\, )  \nonumber\\
 &=& \rho_{F,e}^{(2)}(\mbox{1};\vc{k},\vc{q},\vc{Q})_1 +
     \rho_{F,\kappa}^{(2)}(\mbox{1};\vc{k},\vc{q},\vc{Q})_1
 \, . \label{irh12}
\eea
In a similar way, one gets for the spatial current
\bea
\vc{\jmath}_{FW}^{\, \, (3)}(\mbox{1};\vc{k},\vc{q},\vc{Q})_1
 &=& - \mpw{16}{3} \Big[ (\vcs{Q} + \vcs{k})
 \Big(\eh \, \vc{Q} + i(\eh + \kh ) \, \pvso{\vc{k}} \Big)  \nonumber\\
&&
 + \Big(\eh \, \ps{\vc{k}}{\vc{Q}}+ 4\kh m \om{fi}{1} \Big)\,
 (\vc{k} + i\pvso{\vc{Q}} ) \nonumber\\
&&
  + \kh \, \vc{k} \times [ \vc{Q} \times ( \vc{k} + i\pvso{\vc{Q}})]
\Big] \delta(\frac{\vc{k}}{2} - \vc{q}\, ) \, , \label{j13fwb}\\
 \vc{\jmath}_{\chir}^{\, \, (3)}(\mbox{1};\vc{k},\vc{q},\vc{Q})_1
 &=& - \frac{\vcs{k}}{32m^2} \, \ps{\vc{k}}{\vnab{k}}
 \vc{\jmath}_{F}^{\, \, (1)}(\mbox{1};\vc{k},\vc{q},\vc{Q} )_1
  \nonumber\\
&&  + \mpw{32}{2} \Big( \vc{k}\, ( \vc{k} \cdot
 \vc{\jmath}_c^{\, \, (1)}(\mbox{1};\vc{k},\vc{q},\vc{Q})_1) +
  \vcs{k} \, \vc{\jmath}_s^{\, \, (1)}(\mbox{1};\vc{k},\vc{q}\, )_1 \Big)
\,  , \label{j1chr}\\
 \vc{\jmath}_{\chis}^{\, \, (3)}(\mbox{1};\vc{k},\vc{q},\vc{Q})_1
 &=& \mpw{32}{3} \Big[
 i\eh \, \pssm{\vc{Q}}{\vc{k}} \, \vc{Q}
  \nonumber\\
&& - (\eh + \kh)\, [\vc{k} \times (\vc{k} \times \vc{Q} )
 + \pth{\vs{2}}{\vc{Q}}{\vc{k}}\pv{\vc{k}}{\vs{1}} ]
 \Big]
 \delta(\frac{\vc{k}}{2} - \vc{q}\, ) \, , \label{j1chs}\\
 \vc{\jmath}_{sep}^{\, \, (3)}(\mbox{1};\vc{k},\vc{q},\vc{Q})_1
 & = &  \frac{\vcs{k}}{32m^2} \Big(1+ \ps{\vc{k}}{\vnab{k}}\Big)
 \vc{\jmath}_{F}^{\, \, (1)}(\mbox{1};\vc{k},\vc{q},\vc{Q} )_1
 \,  . \label{j1sep}
\eea
There is again a cancellation between the separation and boost
terms. The final expression for the relativistic part of the
one-nucleon intrinsic current operator then is
\bea
 \vc{\jmath}_{F}^{\, \, (3)}(\mbox{1};\vc{k},\vc{q},\vc{Q})_1
 &=&  - \mpw{16}{3} \Big[\eh (\vcs{Q} + \frac{\vcs{k}}{4})
\vc{Q} + i(\eh + \kh )(\vcs{Q} + \frac{\vcs{k}}{2})\,
  \pvso{\vc{k}}  \nonumber\\
&&
 + (\eh\, \ps{\vc{k}}{\vc{Q}} + 4\kh m \om{fi}{1} )
 (\vc{k} + i \pvso{\vc{Q}} ) \nonumber\\
&&
 - \frac{i}{2} \eh \,  \pssm{\vc{Q}}{\vc{k}}\,\vc{Q}  +
 \frac{1}{2} (\eh + \kh)  \pth{\vs{2}}{\vc{Q}}{\vc{k}} \pv{\vc{k}}{\vs{1}}
 \nonumber\\
&&
+\kh \, \vc{k} \times [ \vc{Q} \times (\frac{1}{2} \vc{k} + i\pvso{\vc{Q}})]
\Big]
 \delta(\frac{\vc{k}}{2} - \vc{q}\, ) \nonumber\\
&& - \mpw{32}{2}   \vc{k} \, (\vc{k} \cdot
 \vc{\jmath}_{F}^{\, \, (1)}(\mbox{1};\vc{k},\vc{q},\vc{Q} )_1 )
 \, . \label{ij13}
\eea

Now we will consider the divergence of these currents in view of the
general continuity equations for the intrinsic operators in (\ref{con1}) and
(\ref{con3}). Using  $\vc{k}=2\vc{q}$, one finds for the divergence of
the nonrelativistic intrinsic current
\beq
  \vc{k} \cdot  \vc{\jmath}_{F}^{\, \, (1)}(\mbox{1};\vc{k},\vc{q},\vc{Q})_1
 =  \frac{\ps{\vc{q}}{\vc{Q}}}{m} \,
\rho_F^{(0)}(\mbox{1};\vc{k},\vc{q}\, )_1 \, .
\label{cont11}
\eeq
and for the relativistic one
\bea
\vc{k} \cdot \vc{\jmath}_{F}^{\, \, (3)}(\mbox{1};\vc{k},\vc{q},\vc{Q})_1
&=& -\frac{1}{16m^3}\,\bigg[2\eh (\vcs{Q} + \frac{\vcs{k}}{4})
\ps{\vc{q}}{\vc{Q}}+  \eh( 2\vcs{k} + i\pssp{\vc{Q}}{\vc{k}} )
\ps{\vc{q}}{\vc{Q}} \nonumber\\
&&+ 4\kh m \om{fi}{1} ( \vcs{k} + i \psso{\vc{Q}}{\vc{k}})\bigg]
\delta(\frac{\vc{k}}{2} - \vc{q}\, )
 - \mpwf{32}{2}{\vcs{k}}\, (\vc{k} \cdot
 \vc{\jmath}_{F}^{\, \, (1)}(\mbox{1};\vc{k},\vc{q},\vc{Q} )_1 )\nonumber\\
&=&  - \mpw{8}{3}(\vcs{Q}+\vcs{q})\ps{\vc{q}}{\vc{Q}}\rho_F^{(0)}(\mbox{1}
;\vc{k},\vc{q}\, )_1 + \frac{1}{m}
\ps{\vc{q}}{\vc{Q}}\rho_{F,e}^{(2)}(\mbox{1};\vc{k},\vc{q},\vc{Q})_1
\nonumber\\
&&+\om{fi}{1}\rho_{F,\kappa}^{(2)}(\mbox{1};\vc{k},\vc{q},\vc{Q})_1
 - \mpwf{32}{3}{\vcs{k}}\, \frac{\ps{\vc{q}}{\vc{Q}}}{m} \,
\rho_F^{(0)}(\mbox{1};\vc{k},\vc{q}\, )_1\,,
\label{cont13}
\eea
where $\rho^{(2)}_{F,e} (\rm{1};\vc{k}\, )$ and $\rho^{(2)}_{F,\kappa}
(\rm{1};\vc{k}\, )$ are defined in (\ref{irh12}).

Since for an intrinsic operator in momentum representation
$a(\vc{q},\vc{Q})$ the following relations hold
\bea
\langle \vc{p}^{\,\prime}|\Big[t^{(1)},a\Big]|\vc{p}\,\rangle& = &
\frac{\ps{\vc{q}}{\vc{Q}}}{m}a(\vc{q},\vc{Q})\label{comt1}
\,,\\
\langle \vc{p}^{\,\prime}|\Big[t^{(3)},a\Big]|\vc{p}\,\rangle& = &
-\frac{\ps{\vc{q}}{\vc{Q}}}{8m^3}(\vcs{Q}+\vcs{q})a(\vc{q},\vc{Q})
\,,\label{comt3}
\eea
one finds that the r.h.s.\ of
(\ref{cont11}) is just the commutator of the nonrelativistic intrinsic kinetic
energy $\vc{p}^{\, 2}/m$ with the charge density
\beq
  \vc{k} \cdot  \vc{\jmath}_{F}^{\, \, (1)}(\mbox{1};\vc{k},\vc{q},\vc{Q})_1
=\langle \vc{p}^{\,\prime}|\Big[t^{(1)},\rho_F^{(0)}(\mbox{1};\vc{k})_1
\Big]|\vc{p}\,\rangle\,,\label{cont11s}
\eeq
i.e.\ the relation (\ref{con11body}), and
for the divergence of the relativistic contribution in (\ref{cont13})
\bea
\vc{k} \cdot \vc{\jmath}_{F}^{\, \, (3)}(\mbox{1};\vc{k},\vc{q},\vc{Q})_1
&=& \langle \vc{p}^{\,\prime}|\Big[t^{(1)},
\rho_{F,e}^{(2)}(\mbox{1};\vc{k})_1\Big]|\vc{p}\,\rangle
+\langle \vc{p}^{\,\prime}|\Big[h^{(1)},\rho_{F,\kappa}^{(2)}(\mbox{1};\vc{k})_1
\Big]|\vc{p}\,\rangle
\nonumber\\
&&+
\langle \vc{p}^{\,\prime}|\Big[t^{(3)} - \mpwf{32}{2}{\vcs{k}}\,
t^{(1)},\rho_F^{(0)}(\mbox{1};\vc{k})_1 \Big]|\vc{p}\,\rangle\,,
\label{cont13s}
\eea
which almost equals (\ref{con31body}) with the sole exception, that in the
commutator of $\rho_{F,\kappa}^{(2)}(\mbox{1};\vc{k})$ the full nonrelativistic
intrinsic Hamiltonian $h^{(1)}$ appears instead of the kinetic energy
$t^{(1)}$. This modification of (\ref{con31body}), we had already alluded to,
is a consequence of the fact that we have kept in the $\hat \kappa$-part of
$\vc{\jmath}_{F}^{\, \, (3)}(\mbox{1};\vc{k})$ in (\ref{ij13})
the total intrinsic energy transfer, 
thus including implicitly a two-body contribution.

That means, that in order to satisfy the full
continuity equation for the intrinsic currents, one should
find for the intrinsic
model-dependent interaction currents the following relations involving
commutators with the $NN$ potential $v$
\bea
 \vc{k} \cdot \vc{\jmath}_F^{\, \, (1)} (\rm{2};\vc{k},\vc{q},\vc{Q})
 & = & \langle \vcp{p} \, |
 \Big[ v^{(1)}\, , \, \rho_F^{(0)}(\rm{1};\vc{k}\, ) \Big]
 | \vc{p} \, \rangle \, ,
\label{cont21}\\
 \vc{k} \cdot \vc{\jmath}_F^{\, \, (3)} (\rm{2};\vc{k},\vc{q},\vc{Q})
 & = &  \langle \vcp{p} \, |\Big[ t^{(1)},\rho_F^{(2)}(2;\vc{k})
 \Big] +  \Big[ v^{(1)} \, , \, \rho^{(2)}_{F,e} (\rm{1};\vc{k}\, ) \Big]
 | \vc{p} \, \rangle\nonumber\\
&& + \langle \vcp{p} \, |
 \Big[ v^{(3)} - \frac{\vcs{k}}{32m^2}v^{(1)} , \,
 \rho_F^{(0)}(\rm{1};\vc{k}\, ) \Big]  | \vc{p} \, \rangle
 \, , \label{cont23}
\eea
where the last term is the interaction-dependent
part of the recoil commutator in (\ref{con3}). Note, that only $\rho_{F,e}^{(2)
}(\rm{1};\vc{k}\, )$ appears in the commutator with $v^{(1)}$ 
in (\ref{cont23}) because
$\rho_{F,\kappa}^{(2)}(\rm{1};\vc{k}\, )$ is already contained in
(\ref{cont13s}). It is clear that
(\ref{cont21}) and (\ref{cont23}) should hold also for each meson
contribution separately.

For the explicit evaluation of the commutators of the potential with the
one-body charge density, we first note for a one-body operator
\beq
a(1;\vec k, \vec q, \vec Q)= a(\vec k, \vec Q)_1\delta(\frac{\vec k}{2}
-\vec q) +\exnn 
\eeq
the general relation
\bea
 \langle \vcp{p} \, | \Big[ v_B \, ,
 a(\rm{1};\vc{k} \, ) \Big] | \vc{p} \, \rangle &=&
{v}_B (- \vci{q}{2}, \vc{Q} + \frac{1}{2} \vc{k} \, ) \,
a(\vc{k}, \vc{Q} + \vci{q}{2} \, )_1
 - a(\vc{k}, \vc{Q} - \vci{q}{2} \, )_1
 {v}_B (- \vci{q}{2}, \vc{Q} - \frac{1}{2} \vc{k} \, )
\nonumber\\
&&+\exnn \,.
\eea
Specializing now to the charge density operator as given in (\ref{irh10})
and (\ref{irh12}) and using the isospin dependence as introduced in 
(\ref{fisoe})-(\ref{gisom}), we find the following relations
\bea
 \langle \vcp{p} \, | \left[ v_B^{(1)}\, ,
 \rho^{(0)}_F(\rm{1};\vc{k} \, ) \right] | \vc{p} \, \rangle &=&
-\Ff{e,1}{-}\tilde v_B^{(1)}(\vc{q_2}\,) +\exnn \,,
\label{comv1rh0}\\
 \langle \vcp{p} \, | \left[ v_B^{(1)}\, ,
 \rho^{(2)}_{F,e}(\rm{1};\vc{k} \, ) \right] | \vc{p} \, \rangle &=&
- i\frac{\Ff{e,1}{+}}{32m^2}\Big(\Big[\tilde v_B^{(1)}(\vc{q_2}\,),\,
\pssp{\vc{Q}}{\vc{k}}\Big]
\nonumber\\
&&+\Big\{\tilde v_B^{(1)}(\vc{q_2}\,),\,\pssp{\vc{q}_2}{\vc{k}}\Big\}\Big)
\nonumber\\
&&+ \frac{\Ff{e,1}{-}}{32m^2}
\Big(4\vcs{k}\tilde v_B^{(1)}(\vc{q_2}\,)+i\Big\{\tilde v_B^{(1)}(\vc{q_2}\,),\,
\pssp{\vc{Q}}{\vc{k}}\Big\}
\nonumber\\
&&+i\Big[\tilde v_B^{(1)}(\vc{q_2}\,),\,\pssp{\vc{q}_2}{\vc{k}}\Big]\Big)
+\exnn \,.
\label{comv1rh2}\\
 \langle \vcp{p} \, | \left[ v_B^{(3)}\, ,
 \rho^{(0)}_F(\rm{1};\vc{k} \, ) \right] | \vc{p} \, \rangle &=&
\frac{\Ff{e,1}{+}}{2}
\Big(\tilde v_B^{(3)}(-\vci{q}{2},\vc{Q} + \frac{1}{2}\vc{k}\,)
-\tilde v_B^{(3)}(-\vci{q}{2},\vc{Q} - \frac{1}{2}\vc{k}\,)\Big)
\nonumber\\
&&
-\frac{\Ff{e,1}{-}}{2}
\Big(\tilde v_B^{(3)}(-\vci{q}{2},\vc{Q} + \frac{1}{2}\vc{k}\,)
+\tilde v_B^{(3)}(-\vci{q}{2},\vc{Q} - \frac{1}{2}\vc{k}\,)\Big)
\nonumber\\
&& +\exnn \,,
\label{comv3rh0}
\eea
These relations will be useful for checking the continuity condition
(\ref{cont23}) for the various meson contributions. Since the interaction
currents and the
commutators in (\ref{cont23}) separate into ``+''  and ``$-$'' parts with
respect to $F^\pm$ and $G^\pm$, we can check the continuity condition
separately for these parts.

There are two interesting
properties of the relativistic part of the continuity equation
that are demonstrated explicitly below for particular meson exchanges.
First, since we use the
representation in which there is no intrinsic retardation potential
($\nu = 1/2$), the retardation part of the current has to satisfy
\bea
 \vc{k} \cdot \vc{\jmath}_F^{\, \, (3)} (\mbox{2;ret};\vc{k},\vc{q},\vc{Q})
  &=&  \frac{\ps{\vc{q}}{\vc{Q}}}{m}\,
 \rho_F^{(2)}(\mbox{2;ret};\vc{k},\vc{q},\vc{Q})
\nonumber\\
&=& \langle \vcp{p} \, | \left[ t^{(1)}\, ,
 \rho^{(2)}_F(\mbox{2;ret};\vc{k} \, ) \right] | \vc{p} \, \rangle
\, .
\label{contret}
\eea
This is possible only when the boost contributions from $\chi_r$\
are included. Consequently, if retardation is considered, it makes little
sense to take the corresponding FW operators neglecting at the same time
the effects of the wave function boost. Another consequence is that in
(\ref{cont23}) we need to consider the ``pro'' and ``mes'' MEC 
contributions as defined in Appendix A only.

Second, between the com\-mu\-ta\-tors
$\Big[ v^{(1)}\, , \, \rho^{(2)}_{F,e} (\rm{1};\vc{k}\, ) \Big]$
and $\Big[ v_{LS}^{(3)}\, , \, \rho_F^{(0)} (\rm{1};\vc{k}\, ) \Big]$
there are cancellations, where $v_{LS}^{(3)}$\ is the spin-orbit part 
of the potential. Recall, that also in the relativistic charge density
$\rho^{(2)}_{F,e}(\rm{1};\vc{k}\, )$\
a can\-cel\-la\-tion occurs between the spin-orbit FW term and 
the $\chi_{\sigma}$
term. This suggests that for a proper description of
the relativistic intrinsic MECs the inclusion of the spin
part of the boost is also important.

Now we will derive the detailed expressions for the intrinsic meson
exchange current operators. The FW-operators are listed in Appendix A. They
have to be evaluated in the Breit frame using (\ref{q12b})-(\ref{QBreit}).
In order to keep track of the various pieces contributing to the intrinsic
operators, we give in Tables \ref{tab1} and \ref{tab2} a survey on the
nonvanishing terms with reference to the corresponding equations where the
explicit expressions are given. In the case of nonrelativistic
contributions, the intrinsic operators are given by the FW-ones.

\subsection{Scalar meson exchange}

The intrinsic potential contributions including leading relativistic 
order following from the exchange of a single scalar meson are 
\bea
 \vto{S} &=& - \con{S}\, \prop{S}{} \, , \label{vs1}\\
 \vtt{S} &=& - \mpw{8}{2}\, \vto{S} \Big(
 2 \vcs{Q} + i \pssp{\vc{q}}{\vc{Q}} \Big) \nonumber\\
&=& \vtt{S,Q} + \vtt{S,LS} \, . \label{vs3}
\eea
Here, $\prop{S}{}$ denotes the meson propagator including a hadronic form
factor (see Appendix A). The
last term in (\ref{vs3}) is the spin-orbit potential.
In some OBE potentials the following replacement is done in the first term
of (\ref{vs3})
\beq
 \vcs{Q} = \vcs{Q} + \vcs{q} - \vcs{q} \rightarrow
 \vcs{Q} + \vcs{q} + m_S^2 \, ,
\label{mrep}
\eeq
since $\vto{S}(\vcs{Q} + \vcs{q})/2$\, corresponds to the intrinsic
anticommutator $\{ \hat{\vc{p}}^{\, \, 2},\,v_S^{(1)} \}$.

Turning now to the corresponding exchange currents, we will
start from its ``$+$'' part (see Appendix A).
Since in this case there are no nonrelativistic contributions
neither to charge nor to current densities, the intrinsic
operators are simply given by the FW ones taken in the Breit frame. In all
following expressions we have kept the notation
$\vec q_2=\frac{1}{2} \vec k -\vec q$. Hence, we get
from (\ref{j2spro+})-(\ref{j2sret+})
\bea
 \jrcq{F}{pro}{S}{+} &=&
  - \Ff{e,1}{+} \mpw{4}{2} \vtos{S} ( \vc{Q} + i \pvso{\vc{k}} ) +\exnn \, ,
\label{intjspro+}\\
 \rhqQ{ret}{S}{+}  &=& \Ff{e,1}{+}\, \conm{S}{8}{ } \,
 \ps{\vc{k}}{\vci{q}{2}}\, \propp{S}{2} +\exnn \, , \label{intrhsret+}\\
 \jrcq{F}{ret}{S}{+} &=&  \conm{S}{16}{2} \Big\{
 \Ff{e,1}{+}\, \Big[ \ps{\vc{k}}{\vci{q}{2}}\, \vc{Q} -
 2 \ps{\vc{Q}}{\vci{q}{2}}\, \vci{q}{2} \Big]
  \nonumber\\
&&   + i\Gm{+}\, \ps{\vc{k}}{\vci{q}{2}}\, \pvso{\vc{k}}
  \Big\} \, \propp{S}{2} +\exnn \, , \label{intjsret+}
\eea
where $\propp{S}{2}$ is defined in (\ref{propp}). Note that for the exchange
\exnn\ one has $\vec q_1=\frac{1}{2} \vec k +\vec q$.

As next we will look at the consequences of the continuity equation.
The retardation charge and current obviously satisfy (\ref{contret}).
Thus remains the divergence of the ``pro'' current which is
\beq
 \vc{k} \cdot \jrcq{F}{pro}{S}{+} =
  - \frac{\Ff{e,1}{+}}{4m^2} \, \ps{\vc{k}}{\vc{Q}} \, \vtos{S} \, +\exnn \, .
\label{contspro+}
\eeq
This indeed is the sum of the commutators on the r.h.s.\ of (\ref{cont23})
because one finds directly from (\ref{comv1rh0}) through (\ref{comv3rh0})
with (\ref{vs1}) and (\ref{vs3})
\bea
 \langle \vcp{p} \, | \Big[ v_{S}^{(1)}\, ,
 \, \rho^{(0)}_{F} (\rm{1};\vc{k}\, ) \Big]^+
 | \vc{p} \, \rangle &=& 0\,,\\
 \langle \vcp{p} \, | \Big[ v_{S}^{(1)}\, ,
 \, \rho^{(2)}_{F,e} (\rm{1};\vc{k}\, ) \Big]^+
 | \vc{p} \, \rangle \, &=&-i \frac{\Ff{e,1}{+}}{16m^2}\,
 \vtos{S}\pssp{\vci{q}{2}}{\vc{k}} +\exnn
\,,\\
 \langle \vcp{p} \, | \Big[ v_{S}^{(3)}\, ,
 \, \rho^{(0)}_{F} (\rm{1};\vc{k}\, ) \Big]^+
 | \vc{p} \, \rangle \, &=&- \frac{\Ff{e,1}{+}}{16m^2}\,
 \vtos{S}\Big(4\ps{\vc{k}}{\vc{Q}}-i
\pssp{\vci{q}{2}}{\vc{k}}\Big)
\nonumber\\
&&+\exnn\,,
\eea
where the superscript ``+'', referring to the isospin dependence, 
indicates that only the ``$+$'' part of the commutator is retained.
Clearly, the resulting current and the continuity equation it satisfies
are completely different from those one would obtain, e.g., 
by a minimal replacement in the LS potential neglecting further 
relativistic effects.

Let us now consider the ``$-$'' part of the operators.
There is one nonrelativistic term in (\ref{j2smes1-}) which we will not
list here again.
The relativistic contributions to the charge and current densities
follow from (\ref{j2spro-})-(\ref{j2sret-}) and the boost and separation
ones are listed in Appendix B.

After summing up these various terms one obtains finally for the
relativistic intrinsic operators
\bea
 \rhqQ{mes}{S}{-} &=&
  \Ff{e,1}{-} \conm{S}{2}{} \, \ps{\vc{k}}{\vc{Q}}
 \, \fq \, , \label{intrhsmes-}\\
 \rhqQ{ret}{S}{-} &=&
  \Ff{e,1}{-} \conm{S}{4}{ } \, \ps{\vci{q}{2}}{\vc{Q}}
 \, \propp{S}{2} +\exnn \, . \label{intrhsret-}\\
 \jrcq{F}{pro}{S}{-} &=&
  - \Ff{e,1}{-} \conm{S}{4}{2} \,
 (\frac{7}{8} \vc{k} + i \pvso{\vc{Q}} ) \, \prop{S}{2}+\exnn  \nonumber\\
&=&
  - \Ff{e,1}{-} \conm{S}{4}{2} \,
 (\frac{3}{4} \vc{k} + i \pvso{\vc{Q}} ) \, \prop{S}{2}+\exnn  \nonumber\\
&& - \frac{\vc{k}}{32m^2} \, \vc{k} \cdot  \jnrq{mes}{S}
    \, , \label{intjspro-}\\
 \jrcq{F}{mes}{S}{-} &=&
  - \Ff{e,1}{-}\, \conm{S}{4}{2} \fq \, \vc{q}\,
 \Big[ 2 \vcs{Q} -
 i \pssp{\vc{Q}}{\vc{q}\, } \nonumber\\
&& - i\pssm{\vc{Q}}{\vc{k}}
 - 2 \frac{\ps{\vc{k}}{\vc{Q}}}{\ps{\vc{k}}{\vc{q}\, }}\,
 \ps{\vc{q}}{\vc{Q}} \, \Big] \, , \label{intjsmes3-}\\
 \jrcq{F}{ret}{S}{-} &=& - \conm{S}{8}{2}
\Big[ \frac{\Ff{e,1}{-}}{\ps{\vc{k}}{\vc{q}\, }} \,
 \Big( \frac{1}{4} \ps{\vc{k}}{\vci{q}{2}}\,
 \vc{k} \times (\vc{k} \times \vc{q}\, )  \nonumber\\
&&
 + \ps{\vc{Q}}{\vci{q}{2}}\, \Big[\vc{k} \times (\vc{q} \times \vc{Q}\, )
 -  2 \vc{q}\, \ps{\vc{q}}{\vc{Q}} \, \Big]
 \Big)  \nonumber\\
&&  - i\Gm{-}\, \ps{\vc{Q}}{\vci{q}{2}} \, \pvso{\vc{k}} \,
  \Big] \, \propp{S}{2} +\exnn \, . \label{intjsret-}
\eea

The nonrelativistic current satisfies
\beq
 \vc{k} \cdot  \jnrq{mes}{S} = - \Ff{e,1}{-}\, \vtos{S} + \exnn \, ,
\label{contsmes1-}
\eeq
which in conjunction with (\ref{comv1rh0}) is in agreement with (\ref{cont21}).
Obviously, the retardation charge and
current densities satisfy (\ref{contret}).
For the remaining divergence of the relativistic ``pro'' and ``mes''
currents, one finds from the above expressions
\bea
\vc{k} \cdot\Big( \jrcq{F}{pro}{S}{-}+ \jrcq{F}{mes}{S}{-}\Big)&=&
\nonumber\\
&&\hspace*{-6cm}\frac{\Ff{e,1}{-}}{8m^2} \vtos{S}
 \Big[ 2 \vcs{Q} + \frac{3}{2} \vcs{k}
+ i \, (\vs{1} + \vs{2}) \times \vc{Q} \cdot (\vc{k} - \vc{q}\, )
 \Big]
+ \exnn
\nonumber\\
&&\hspace*{-6cm}
+  \Ff{e,1}{-} \conm{S}{2}{2} \,\ps{\vc{q}}{\vc{Q}} \ps{\vc{k}}{\vc{Q}}
 \, \fq
-\frac{\vcs{k}}{32m^2} \, \vc{k} \cdot  \jnrq{mes}{S}
\, .
\label{cons-}
\eea
The first term on the r.h.s.\ is equal to the sum of the following two
commutators
\bea
 \langle \vcp{p} \, |
\Big[ v_{S}^{(1)}\, ,
 \, \rho^{(2)}_{F,e} (\rm{1};\vc{k}\, ) \Big]^- | \vc{p} \, \rangle&=&
\frac{\Ff{e,1}{-}}{16m^2} \vtos{S}\,
 \Big[  2\vcs{k} +
 i \, (\vs{1} + \vs{2}) \cdot (\vc{Q} \times \vc{k} \, )
 \Big] + \exnn \, ,
\\
 \langle \vcp{p} \, |\Big[ v_{S}^{(3)}\, , \,
\rho_F^{(0)} (\rm{1};\vc{k}\, ) \Big]^-| \vc{p} \, \rangle &=&
\frac{\Ff{e,1}{-}}{8m^2} \vtos{S}\,
 \Big[ 2\vcs{Q} + \frac{1}{2} \vcs{k}
+ i \, (\vs{1} + \vs{2}) \cdot ( \vc{Q} \times \vc{q_2}\, )
\Big]\nonumber\\
&& + \exnn \,,
\eea
while the next term is just the commutator of $t^{(1)}$ with
the mesonic density (\ref{intrhsmes-}) according to (\ref{comt1}).
The last term in (\ref{cons-}) is the recoil current contribution
in (\ref{cont23}).
This completes the verification of the ``$-$'' part of
the continuity equation.

Finally, let us shortly describe how one can obtain the conserved current
if the replacement (\ref{mrep}) is made in the relativistic part of the
NN potential. First of all, it follows from (\ref{comv3rh0})
that the ``$+$'' part of the continuity
equation is not affected. For the ``$-$'' part, an additional term arises
in the commutator of the nonrelativistic charge density with the potential,
namely
\beq
 - \Ff{e,1}{-} \conm{S}{4}{2} (m_S^2 + \vcsi{q}{2}) \prop{S}{2}
 + \exnn \, .
\label{mrepcons}
\eeq
Notice, that due to the nucleon exchange term this additional contribution
vanishes in the absence of strong form factors. This suggests the
following prescription to handle the more general case with form factors:
(i) Rearrange the different terms of the current without form factors,
i.e., shift some part from the ``mes'' to the ``pro'' component,
so that its divergence explicitly contains the additional piece
(\ref{mrepcons}), (ii) Introduce then the
strong form factors into the modified potential and the rearranged current.
Obviously, this procedure is not unique, but there is no other more consistent
way of solving this problem.

The resulting modified current is a little more complicated for scalar meson
exchange, but for vector meson exchange it simplifies significantly.
Let us consider the vector
\beq
 - \frac{1}{2} (\vci{q}{1} - \vci{q}{2}) \Big[ \prop{S}{2} + \exnn -
 (2 m_S^2 + \vcsi{q}{1} + \vcsi{q}{2} ) \prop{S}{1} \prop{S}{2}
 \Big] \, ,
\label{mrepcur}
\eeq
which vanishes if the strong form factors are disregarded and whose
divergence is just
\beq
 ( m_S^2 + \vcsi{q}{2} ) \prop{S}{2} - \exnn \, .
\eeq
Multiplying (\ref{mrepcur}) with $ - \Ff{e,1}{-} \conm{S}{4}{2} $\
and adding it to the intrinsic current
above, one gets a new conserved current for the modified
potential. For the currents of this section
this means the following replacements 
in (\ref{intjspro-}) $\frac{3}{4} \vc{k} \rightarrow (\frac{3}{4} \vc{k} -
\vc{q}\,)$ and in (\ref{intjsmes3-}) $\vcs{Q} \rightarrow ( \vcs{Q} + m_S^2 +
 \vcs{q} + \frac{1}{4} \vcs{k} )$.

\subsection{Vector meson exchange}

The exchange of a vector meson contributes to the potential
by
\bea
 \vto{V} &=&  \con{V}\, \prop{V}{} \, , \label{vv1}\\
 \vtt{V} &=& - \mpw{4}{2}\, \vto{V} \Big[
 (1+ 2 \kappa_V) \vcs{q} + (1+\kappa_V)^2
 \pf{\vs{1}}{\vc{q}\, }{\vs{2}}{\vc{q}\, }  \nonumber\\
 &&  - \vcs{Q} - i(\frac{3}{2} + 2 \kappa_V)
  \pssp{\vc{q}}{\vc{Q}} \Big] \nonumber\\
&=& \vtt{V,q} + \vtt{V,\sigma q} + \vtt{V,Q} + \vtt{V,LS} \, . \label{vv3}
\eea
The term $\vtt{V,\sigma q}$\ contains a central spin-spin and a tensor
part. Again, \vcs{q}\ is often replaced by $ - m_V^2$\
and then $\vtt{V}$\ may be redefined to be
\bea
\hat v_V^{(3)}(\vc{q},\vc{Q}) &=&  \mpw{4}{2}\, \vto{V} \Big\{
 (1+ 2 \kappa_V) m_V^2 + (1+\kappa_V)^2
 \Big[ m_V^2 \ps{\vs{1}}{\vs{2}} +
 \ps{\vs{1}}{\vc{q}\, } \ps{\vs{2}}{\vc{q}\, } \Big]
  \nonumber\\
 &&  + \vcs{Q} + i(\frac{3}{2} + 2 \kappa_V)
  \pssp{\vc{q}}{\vc{Q}} \Big\} \, . \label{vv3r}
\eea
Like the nonrelativistic potential \vto{V}, many currents are obtained
from those of scalar exchange given in the previous section by the 
replacements $ m_S \rightarrow m_V$ and $g_S^2 \rightarrow - g_V^2$.

For the ``$+$'' part, one gets in this way
the retardation current from (\ref{intrhsret+})
and (\ref{intjsret+}). In addition, there is the ``pro'' part of the current
which follows from (\ref{j2vpro+})
\beq
 \jrcq{F}{pro}{V}{+} =
   \mpwf{4}{2}{\Ff{e,1}{+}} \vtos{V}
 \Big[ \vc{Q} - i(1+\kappa_V) \pvsp{\vci{q}{2}} \Big] +\exnn \, .
\label{intjvpro+}
\eeq
Its divergence should equal the sum of the following commutators
\bea
 \langle \vcp{p} \, | \Big[ v_{V}^{(1)}\, ,
 \, \rho^{(2)}_{F,e} (\rm{1}) \Big]^+ | \vc{p} \, \rangle \, &=&
 - i\frac{\Ff{e,1}{+}}{16m^2} \vtos{V} \pssp{\vci{q}{2}}{\vc{k}}
 + \exnn \, ,\\
 \langle \vcp{p} \, | \Big[ v_{V}^{(3)}\, ,
 \, \rho_F^{(0)} (\rm{1};\vc{k}\, ) \Big]^+ | \vc{p} \, \rangle &=&
 \frac{\Ff{e,1}{+}}{8m^2} \vtos{V}\Big[ 2\ps{\vc{k}}{\vc{Q}}
-i (\frac{3}{2} + 2 \kappa_V)  \pssp{\vci{q}{2}}{\vc{k}}\Big]
\nonumber\\
&&+\exnn \, ,
\eea
which is easy to verify.

For the ``$-$'' part the replacements
$ m_S \rightarrow m_V$ and $g_S^2 \rightarrow - g_V^2$\ yield
the nonrelativistic current \jnrq{mes}{V}  from (\ref{j2smes1-}),
the retardation charge \rhqQ{ret}{V}{-}  and current
\jrcq{F}{ret}{V}{-}  from (\ref{intrhsret-}) and (\ref{intjsret-}),
respectively, and the mesonic charge \rhqQ{mes}{V}{-}
from (\ref{intrhsmes-}). In order to get the remaining currents,
one should take (\ref{j2vpro-})-(\ref{j2vmestr-}) and add
the nonretardation parts of the boost and separation 
contributions, i.e., (\ref{jschrpro}) and (\ref{jschs})
with the appropriate replacement of the
meson parameters. Keeping the ``mes-tr'' mesonic current separately
(see appendix A), we finally obtain
\bea
 \jrcq{F}{pro}{V}{-} &=&
   \Ff{e,1}{-} \conm{V}{4}{2} \,
 \Big[ (1+\kappa_V) \vci{q}{2} - \kappa_V \vci{q}{1} -
\frac{1}{4} \vc{k}
 - i (1+ 2 \kappa_V) \pvso{\vc{Q}}  \nonumber\\
&& - (1+\kappa_V)^2 \vs{1} \times ( \vs{2} \times \vci{q}{2})\Big]
 \, \prop{V}{2}+\exnn  \nonumber\\
&& - \frac{\vc{k}}{32m^2} \, \vc{k} \cdot  \jnrq{mes}{V}
    \, , \label{intjvpro-}\\
 \jrcq{F}{mes}{V}{-} &=&
  \Ff{e,1}{-}\, \conm{V}{2}{2} \fq \, \vc{q}
 \Big[ (1+2 \kappa_V)(\vcs{q} + \frac{\vcs{k}}{4}) - \vcs{Q}
  \nonumber\\
&&
 - (1+\kappa_V)^2
 \pf{\vs{1}}{\vci{q}{1}}{\vs{2}}{\vci{q}{2}}
 + i(\frac{3}{2} + 2 \kappa_V) \pssp{\vc{Q}}{\vc{q}\, }\nonumber\\
&& + i(\frac{1}{2} + \kappa_V) \pssm{\vc{Q}}{\vc{k}}
 -  \frac{\ps{\vc{k}}{\vc{Q}}}{\ps{\vc{k}}{\vc{q}\, }}\,
 \ps{\vc{q}}{\vc{Q}}
 \Big] \, , \label{intjvmes-}\\
 \rhqQ{mes-tr}{V}{-} &=&
  \Ff{e,1}{-}\, \conm{V}{2}{} \fq \nonumber\\
&& \Big[ 2 \ps{\vc{k}}{\vc{Q}} +
i (1+\kappa_V) \pssp{\vc{q}}{\vc{k}} \Big] \, ,
\label{intrhvmestr-} \\
 \jrcq{F}{mes-tr}{V}{-} &=&
  \Ff{e,1}{-}\, \conm{V}{4}{2} \fq \nonumber\\
&& \Big\{ 2 \ps{\vc{q}}{\vc{Q}}
\Big[ 2 \vc{Q} +
 i(1+\kappa_V) \Big(\pvsp{\vc{q}} + \frac{1}{2} \pvsm{\vc{k}}\Big) \Big]
 \nonumber\\
&&
+ \vc{k} \times \Big[ (1+\kappa_V)^2 \, (\vs{1} \times \vci{q}{1})
 \times  (\vs{2} \times \vci{q}{2})  \nonumber\\
&&
 + i (1+\kappa_V) \Big( \pvsm{\vc{q}} + \frac{1}{2} \pvsp{\vc{k}}
\Big)\times \vec Q \Big] \Big\}
\, .
\label{intjvmestr-}
\eea

Let us now examine the continuity equation for the ``$-$'' part
of the current. Its nonrelativistic and retardation parts
are satisfied exactly as in the case of scalar meson exchange.
Since also the transverse mesonic current satisfies
\beq
 \vc{k} \cdot  \jrcq{F}{mes-tr}{V}{-} =
 \frac{\ps{\vc{q}}{\vc{Q}}}{m}\,  \rhqQ{mes-tr}{V}{-}\,,
\eeq
we are left with the divergence of the relativistic ``pro'' and ``mes''
currents for which one finds
\bea
\vc{k} \cdot\Big( \jrcq{F}{pro}{V}{-}+ \jrcq{F}{mes}{V}{-}\Big)&=&
\nonumber\\
&&\hspace*{-6.5cm}\frac{\Ff{e,1}{-}}{4m^2} \vtos{V}
 \bigg( \frac{1}{4} \vcs{k}-\vcs{Q} +(1 + 2 \kappa_V) \vcsi{q}{2}
+(1 + \kappa_V)^2 \pf{\vs{1}}{\vci{q}{2}}{\vs{2}}{\vci{q}{2}}
\nonumber\\
&&\hspace*{-6.5cm}+i \Big( \frac{3}{2}+
 2 \kappa_V \Big) \pssp{\vc{Q}}{\vc{q}\, }
-i \Big( \frac{1}{2}+ \kappa_V \Big) \pssp{\vc{Q}}{\vc{k}}\bigg)
+ \exnn
\nonumber\\
&&\hspace*{-6.5cm}
-  \Ff{e,1}{-} \conm{V}{2}{2} \,\ps{\vc{q}}{\vc{Q}} \ps{\vc{k}}{\vc{Q}}
 \, \fq
-\frac{\vcs{k}}{32m^2} \, \vc{k} \cdot  \jnrq{mes}{V}
\, .
\label{conv-}
\eea
Again, the first term on the r.h.s.\ is equal to the sum of the following
two commutators
\bea
 \langle \vcp{p} \, |
\Big[ v_{V}^{(1)}\, ,
 \, \rho^{(2)}_{F,e} (\rm{1};\vc{k}\, ) \Big]^- | \vc{p} \, \rangle &=&
\frac{\Ff{e,1}{-}}{16m^2} \,\vtos{V}
 \Big( 2 \vcs{k} +  i \pssp{\vc{Q}}{\vc{k}} \Big) +\exnn \,, \\
 \langle \vcp{p} \, |\Big[ v_{V}^{(3)}\, , \,
\rho_F^{(0)} (\rm{1};\vc{k}\, ) \Big]^-| \vc{p} \, \rangle &=&
\frac{\Ff{e,1}{-}}{4m^2} \,\vtos{V}
\Big((1 + 2 \kappa_V)\, \vcsi{q}{2} +  (1 + \kappa_V)^2\,
  \pf{\vs{1}}{\vci{q}{2}}{\vs{2}}{\vci{q}{2}} \nonumber\\
&& \hspace*{-0.5truecm}
  - \vcs{Q} - \frac{1}{4} \vcs{k}
- i ( \frac{3}{2}+ 2 \kappa_V ) \pssp{\vc{Q}}{\vci{q}{2}} \Big)
 +\exnn \, ,
\eea
while the next term is just the commutator of $t^{(1)}$ with
the mesonic density and
the last term in (\ref{conv-}) is the recoil current contribution
in (\ref{cont23}). Thus the continuity condition is satisfied also for
vector meson exchange.

Finally, in order to obtain the conserved currents for the
modified potential (\ref{vv3r}), one can use again the same procedure
as in the previous section. Namely, switching off the strong form
factors for a moment and using in the mesonic current the following
identities
\bea
 \ps{\vci{q}{1}}{\vci{q}{2}} & = & \frac{\vcs{k}}{2} -
 ( \vcsi{q}{1} + \vcsi{q}{2} )   \, , \\
 \vcsi{q}{1} + \vcsi{q}{2} &=&
 - 2 m_V^2 + (m_V^2 + \vci{q}{1}) + (m_V^2 + \vci{q}{2}) \, ,
\eea
one gets the modified intrinsic currents
\bea
 \jrcq{F}{pro}{V}{-} &=&
   \Ff{e,1}{-} \conm{V}{4}{2} \,
 \Big( \frac{1}{4} \vc{k}
 + (1+\kappa_V)^2 \Big( \frac{\vc{k}}{2} \ps{\vs{1}}{\vs{2}}
 - \vs{2}  \ps{\vs{1}}{\vci{q}{2}} \Big)
  \nonumber\\
&&
 - i (1+ 2 \kappa_V) \pvso{\vc{Q}} \Big)
 \, \prop{V}{2}+\exnn  \nonumber\\
&& - \frac{\vc{k}}{32m^2} \, \vc{k} \cdot  \jnrq{mes}{V}
    \, , \label{intjvpror-}\\
 \jrcq{F}{mes}{V}{-} &=&
  \Ff{e,1}{-}\, \conm{V}{2}{2} \fq \, \vc{q}
 \,\Big[ - (1+2 \kappa_V) m_V^2  - \vcs{Q}
  \nonumber\\
&&
 + i(\frac{3}{2} + 2 \kappa_V) \pssp{\vc{Q}}{\vc{q}\, }
 + i(\frac{1}{2} + \kappa_V) \pssm{\vc{Q}}{\vc{k}}
\nonumber\\
&&
 + (1+\kappa_V)^2 \Big(
  \ps{\vs{1}}{\vci{q}{2}} \ps{\vs{2}}{\vci{q}{1}}
 - (m_V^2 + \frac{\vcs{k}}{2}) ( \vs{1} \cdot \vs{2} )
 \Big)
\nonumber\\
&& -  \frac{\ps{\vc{k}}{\vc{Q}}}{\ps{\vc{k}}{\vc{q}}}\,
 \ps{\vc{q}}{\vc{Q}}
 \Big] . \label{intjvmes3r-}
\eea
Unlike for the scalar exchange, the modified current is in this case
somewhat simpler than the original one.

\subsection{Pseudoscalar meson exchange}

For pseudoscalar exchange one faces the problem of different, unitarily
equivalent representations which can be characterized by a parameter \mut\
(see Appendix A).
The simplest form of the pseudoscalar exchange potential
corresponds to the representation with $\mut = 0$.
In this case, only one simple relativistic contribution appears besides the
nonrelativistic potential
\bea
 \vto{PS} &=&  - \conm{PS}{4}{2}\,
 \ps{\vs{1}}{\vc{q}\, } \ps{\vs{2}}{\vc{q}\, } \prop{PS}{} \, ,\label{vps1}\\
 \vtt{PS} &=&  \conm{PS}{16}{4}\, (\vcs{Q} + \vcs{q} )\,
 \ps{\vs{1}}{\vc{q}\, } \ps{\vs{2}}{\vc{q}\, } \prop{PS}{} \, . \label{vps3}
\eea
Since the operator form of the last term is  $ -\{ \vcs{p}, \, v_{PS}^{(1)}\}
$, its inclusion into the Lippmann-Schwinger equation causes numerical
instabilities \cite{GAr}. Therefore, most realistic $NN$ potentials 
disregard any relativistic correction to the OPEP.

Let us first consider the ``$+$'' part of the e.m.\ operators. The intrinsic
currents follow from (\ref{rprounit}) and (\ref{jprounit}) in conjunction
with the expressions of (\ref{rh2pspro+})-(\ref{j2psret+}),
(\ref{rh2psvos})-(\ref{j2psvos}) and (\ref{j2pschiv})
\bea
 \rhqQ{pro}{PS}{+}  &=& \Ff{e,1}{+}\, \conm{PS}{32}{3}
 \Big[ 3  \ps{\vs{1}}{\vc{k}\, } \ps{\vs{2}}{\vci{q}{2}} - \frac{1}{2} \,
 \pf{\vc{k}}{\vci{q}{2}}{\vs{1}}{\vs{2}} \, \Big]\!\!\prop{PS}{2} \nonumber\\
&& +\exnn \, , \label{intrhpspro+}\\
 \jrcq{F}{pro}{PS}{+} &=&  \conm{PS}{16}{4} \Big\{ \Ff{e,1}{+} \Big[
 \vc{Q} \Big( \vs{1} \cdot(\vc{k}-\vc{q}\,) \ps{\vs{2}}{\vci{q}{2}}
 +\frac{1}{8}\, \Sigma^{(+)}(\vec q_2,\,\vec k)
\Big)
\nonumber\\
&&
 -\vs{1} \ps{\vci{q}{2}}{\vc{Q}}\ps{\vs{2}}{\vci{q}{2}}
 - i \vc{k} \times \vci{q}{2} \ps{\vs{2}}{\vci{q}{2}}
 - \frac{1}{4} \vci{q}{2}\, \Sigma^{(+)}(\vec q_2,\,\vec Q)
\Big]
 \nonumber\\
&&
 + \frac{1}{4} \Gm{+} \,
 \vc{k} \times \Big[ 3\vs{1}\times\vec Q \,\ps{\vs{2}}{\vci{q}{2}}
 -\vs{1}\times \vci{q}{2} \,\ps{\vs{2}}{\vc{Q}}
\nonumber\\
&&
 - \frac{i}{2} \vci{q}{2} \ps{\vs{2}}{\vc{k}\, } \Big] \prop{PS}{2}
 +\exnn \, , \label{intjpspro+}\\
 \rhqQ{ret}{PS}{+}  &=& \Ff{e,1}{+}\, \conm{PS}{32}{3} \,
 \ps{\vc{k}}{\vci{q}{2}} \ps{\vs{1}}{\vci{q}{2}} \ps{\vs{2}}{\vci{q}{2}}
  \, \propp{PS}{2} +\exnn \, , \label{intrhpsret+}\\
 \jrcq{F}{ret}{PS}{+} &=&  \conm{PS}{64}{4} \Big\{
 \Ff{e,1}{+}\, \Big[ 2 \vci{q}{2}\, \ps{\vc{Q}}{\vc{q}\, } + \vc{k} \times
 \pv{\vc{Q}}{\vci{q}{2}} \Big] \ps{\vs{1}}{\vci{q}{2}}
  \nonumber\\
&&
 - \Gm{+}\, \vc{k} \times \Big[
 2 \vs{1} \times \vci{q}{2}  \ps{\vc{Q}}{\vci{q}{2}}
 +i\vci{q}{2} \ps{\vc{k}}{\vci{q}{2}} \Big] \Big\} \ps{\vs{2}}{\vci{q}{2}}\,
 \propp{PS}{2} \nonumber\\
&& +\exnn \, , \label{intjpsret+}
\eea
where $\Sigma^{(\pm)}(\vec a, \,\vec b)$ is defined in (\ref{Bab}). 
With respect to the continuity condition (\ref{cont23}), one notes that
again the retardation operators satisfy (\ref{contret}),
while for the divergence of the ``pro'' current one gets
\bea
 \vc{k} \cdot  \jrcq{F}{pro}{PS}{+} &=&
 \Ff{e,1}{+} \conm{PS}{16}{4}\,
 \Big\{ \Big[\ps{\vc{k}}{\vc{Q}} \ps{\vs{1}}{\vci{q}{2}}
+ \ps{\vc{q}}{\vc{Q}} \ps{\vs{1}}{\vc{k}} \Big]\ps{\vs{2}}{\vci{q}{2}}
  \nonumber\\
&&
 - \frac{1}{4} \ps{\vc{k}}{\vci{q}{2}} \, \Sigma^{(+)}(\vec q_2,\,\vec Q)
 + \frac{1}{8} \ps{\vc{k}}{\vc{Q}}\, \Sigma^{(+)}(\vec q_2,\,\vec k)
\Big\}
 \prop{PS}{2} \nonumber\\
&& +\exnn \, . \label{conps1}
\eea
Explicit evaluation of the commutators on the r.h.s.\ of (\ref{cont23})
leads to
\bea
 \langle \vcp{p} \, | \Big[ v^{(1)}_{PS} \, , \,
\rho^{(0)} (1;\vc{k}\, ) \Big]^+
 | \vc{p} \, \rangle &=& 0\,,\\
 \langle \vcp{p} \, |
\Big[ t^{(1)}\, ,
 \, \rho^{(2)}_{F} (\rm{2};\mbox{pro};\vc{k}\, )^+_{PS} \Big]
| \vc{p} \, \rangle &=&
  \Ff{e,1}{+} \conm{PS}{64}{4}\, \ps{\vc{q}}{\vc{Q}}
 \Big[ 5  \ps{\vs{1}}{\vc{k}} \ps{\vs{2}}{\vci{q}{2}} \nonumber\\
&& +
 \ps{\vs{1}}{\vci{q}{2}} \ps{\vs{2}}{\vc{k}} \Big]
\prop{PS}{2} +\exnn \, ,
\label{conps2}\\
 \langle \vcp{p} \, | \Big[ v^{(1)}_{PS} \, , \,
 \rho^{(2)}_{F,e} (1;\vc{k}\, )\Big]^+
 | \vc{p} \, \rangle &=& \Ff{e,1}{+} \conm{PS}{64}{4}
\Big[\ps{\vc{Q}}{\vci{q}{2}}\Sigma^{(+)}(\vec q_2,\,\vec k)
\nonumber\\
&&
- \ps{\vc{k}}{\vci{q}{2}}\Sigma^{(+)}(\vec q_2,\,\vec Q)
\Big]\!
 \prop{PS}{2} +\exnn \, ,
\label{conps4}\\
 \langle \vcp{p} \, | \Big[ v^{(3)}_{PS}
 , \, \rho^{(0)} (1;\vc{k}\, ) \Big]^+ | \vc{p} \, \rangle &=&
 \Ff{e,1}{+} \conm{PS}{16}{4}\,  \ps{\vc{k}}{\vc{Q}}
 \ps{\vs{1}}{\vci{q}{2}} \ps{\vs{2}}{\vci{q}{2}}
 \prop{PS}{2}
\nonumber\\
&&+\exnn \, , \label{conps3}
\eea
and their appropriate sum equals (\ref{conps1}).

With respect to the ``$-$'' currents, we do not repeat the expressions
for the nonrelativistic currents (\ref{j2pspro1-})-(\ref{j2psmes1-}), unchanged
they define the corresponding nonrelativistic intrinsic ones.
The FW-currents for $\mut = 0$ follow from
(\ref{rprounit})-(\ref{jprounit}). Let us first collect the
charge densities. There are no additional contributions to them
from the \chis\ and \chir\ boost commutators since there
is no nonrelativistic exchange charge operator, and the ``$-$'' part of the
interaction dependent boost $\chi_V$ of (\ref{rh2pschiv}) disappears in the
Breit frame. Thus one finds
\bea
 \rhqQ{pro}{PS}{-} &=&  \Ff{e,1}{-}\, \conm{PS}{32}{3}
\Sigma^{(+)}(\vec q_2, \,\vec Q)
 \prop{PS}{2} + \exnn \, ,\label{intrhpspro-}\\
 \rhqQ{mes}{PS}{-} &=&
 - \Ff{e,1}{-} \conm{PS}{8}{3} \,  \ps{\vc{k}}{\vc{Q}}  \,
  \ps{\vs{1}}{\vci{q}{1}}  \ps{\vs{2}}{\vci{q}{2}} \fq \, ,
\label{intrhpsmes-}\\
 \rhqQ{ret}{PS}{-} &=&
 \Ff{e,1}{-} \conm{PS}{16}{3}\, \ps{\vc{Q}}{\vci{q}{2}}\,
   \ps{\vs{1}}{\vci{q}{2}} \ps{\vs{2}}{\vci{q}{2}}
 \ \propp{PS}{2}  + \exnn \, .
\label{intrhpsret-}
\eea

The spatial currents are more complicated than those for the scalar and
vector meson exchanges, since first of all the nonrelativistic meson 
nucleon vertex depends on
the nucleon spin and, furthermore, there is a nonrelativistic ``pro'' current
generating additional boost and separation contributions, which are listed in
Appendix B.
The \mut -independent FW-current densities are listed in
(\ref{j2pspro3-}), (\ref{j2psmes3-}) and (\ref{j2psret-}).
Combining all terms, we find for the relativistic intrinsic current operators 
\bea
 \jrcq{F}{pro}{PS}{-} &=&   \conm{PS}{32}{4}\, \Big\{ \Ff{e,1}{-}\,
 \Big[ \vs{1} \Big\{ \Big(3 \ps{\vc{k}}{\vc{q}\,}-4\vcs{q} - 2\vcs{Q}\Big)
 \ps{\vs{2}}{\vci{q}{2}}
  \nonumber\\
&&
 +\ps{\vc{q}}{\vci{q}{2}}\ps{\vs{2}}{\vc{q}\,}
 -\ps{\vc{Q}}{\vci{q}{2}}\ps{\vs{2}}{\vc{Q}}
 +\frac{i}{2}(\vec k \times \vec q\,)\cdot \vec Q \Big\}
  \nonumber\\
&&  
 + \frac{1}{4}\vc{k}\, \Big\{ \Big(7\ps{\vs{1}}{\vc{q}\,} - 
 4\ps{\vs{1}}{\vc{k}}\Big) \ps{\vs{2}}{\vci{q}{2}}-\frac{1}{2}
 \ps{\vs{1}}{\vci{q}{2}} \ps{\vs{2}}{\vc{k}\, }\Big\}
  \nonumber\\
&&
 -\vc{q}\,\Big\{\vs{1}\cdot (\vc{k} + \vci{q}{1})
 \ps{\vs{2}}{\vci{q}{2}}
 +\frac{1}{2}\ps{\vs{1}}{\vc{q}\,} \ps{\vs{2}}{\vc{k}\, }\Big\}
  \nonumber\\
&&
 -i(2\vec k -\vec q\,)\times \vec Q\, \ps{\vs{2}}{\vci{q}{2}}
 +\frac{1}{2} \vec Q\,\Sigma^{(-)}(\vec q_2, \,\vec Q)
\Big]
  \nonumber\\
&&
 + \frac{1}{4} \Gm{-}  \vc{k} \times \Big[
 \Big( 6i\, \vc{Q} + 5\, \vc{k} \times \vs{1} \Big) \ps{\vs{2}}{\vci{q}{2}} +
 \vci{q}{2} \times \vs{1}\, \ps{\vs{2}}{\vc{k}\, } 
\nonumber\\
&&
 - 2i \vci{q}{2}\,  \ps{\vs{2}}{\vc{Q}} \, \Big] 
 \Big\}\prop{PS}{2} + \exnn \,,
\label{intjpspro-}\\
\jrcq{F}{mes}{PS}{-} &=&  \Ff{e,1}{-}\, \conm{PS}{32}{4}\, \vc{q}\, \fq\ 
 \Big\{\vcs{q}\Big[\ps{\vs{1}}{\vc{k}} \ps{\vs{2}}{\vci{q}{2}}
 + \ps{\vs{1}}{\vci{q}{1}} \ps{\vs{2}}{\vc{k}}\Big]
 \nonumber\\
&&
 + 4 \Big(\vcs{q} +  \vcs{Q} - 
 \frac{\ps{\vc{k}}{\vc{Q}}\ps{\vc{q}}{\vc{Q}}}{\ps{\vc{k}}{\vc{q}\, }}
 \Big) \ps{\vs{1}}{\vci{q}{1}} \ps{\vs{2}}{\vci{q}{2}}  
\nonumber\\ 
&&
 + 2\ps{\vci{q}{2}}{\vc{Q}} \ps{\vs{1}}{\vci{q}{1}}  \ps{\vs{2}}{\vc{Q}} 
 + 2\ps{\vci{q}{1}}{\vc{Q}} \ps{\vs{1}}{\vc{Q}}  \ps{\vs{2}}{\vci{q}{2}} 
 \nonumber\\
&&
 -i\,(\vec k \times \vec q\,)\cdot \vec Q\,\Big(\ps{\vs{1}}{\vci{q}{1}}+
 \ps{\vs{2}}{\vci{q}{2}}\Big) \Big\} \, , 
\label{intjpsmes-}\\
 \jrcq{F}{ret}{PS}{-} &=& - \conm{PS}{64}{4} \Big\{ \Ff{e,1}{-}\, 
 \Big[ 
 \Big( 4 \, \ps{\vci{q}{2}}{\vc{Q}}^2 -  \ps{\vc{k}}{\vci{q}{2}}^2 \, \Big)
 \Big( \vs{1} -  \vc{q}\,\frac{\ps{\vs{1}}{\vci{q}{1}}}{\ps{\vc{k}}{\vc{q}\,}}
 \Big)
     \nonumber\\
&&   + \ps{\vs{1}}{\vci{q}{2}} 
 \Big(  \vci{q}{2}\, \ps{\vc{k}}{\vci{q}{2}} - 2
   \vc{Q}\,  \ps{\vci{q}{2}}{\vc{Q}} \, \Big) \,  \Big] \nonumber\\
&& 
   + \Gm{-} \vc{k} \times \Big[ 2i \vci{q}{2}\, 
 \ps{\vci{q}{2}}{\vc{Q}} 
 + \vs{1} \times \vci{q}{2}\,  \ps{\vc{k}}{\vci{q}{2}}\, 
\Big] \Big\}\ps{\vs{2}}{\vci{q}{2}} \propp{PS}{2} 
\nonumber\\
&& + \exnn \, . 
\label{intjpsret-} 
\eea

Finally, with respect to the continuity equation (\ref{cont23}), it is easy 
to see that the ``ret''-part satisfies (\ref{contret}). The 
verification for the remaining current is straightforward but lengthy. Since
it is desirable to see in detail the correspondance between the various 
commutators and the associated currents, we have outlined it in detail 
in Appendix E. 

\section{Summary and Outlook}

In this work we have
consistently derived a complete set of intrinsic operators of the
e.m.\ charge and current density operators for one- and two-body
interaction currents including leading order relativistic contributions for
a one-boson-exchange potential consisting of scalar, pseudoscalar and
vector meson exchanges. These operators have to be evaluated between
intrinsic rest frame wave functions only, since they include already the
effects of the wave function boosts. Furthermore, these operators respect
gauge and Lorentz invariance up to the terms of leading relativistic order.

The definition of the intrinsic operators is based
upon the general separation of the center-of-mass motion from the intrinsic
one as described in \cite{AdAr95}. It relies on the general properties of
the Poincar\'{e} generators in conjunction  with a $1/m$-expansion.
Although the explicit construction of the operators has been done here 
in the framework of the extended S-matrix method of \cite{ATA}, other 
methods are known to give unitarily equivalent results.

These operators will now allow to study relativistic effects in e.m.\ processes
in nuclei in a more systematic and consistent way, at least at not too high
energy and momentum transfer, since they rely on the $1/m$-expansion.
They will also allow a systematic comparison with covariant approaches
\cite{Tj} currently being applied only for the simplest two-nucleon
system. Confrontation of their results \cite{HuTj} with more conventional
ones using the $1/m$-expansion seems to indicate that it might be more
important to incorporate relativistic effects into the
nuclear currents than into the nuclear dynamics.

At present, there are only a few consistent treatments within the
$1/m$-expansion, namely for deuteron photodisintegration
using a simplified dynamical model of a pure one-pion-exchange interaction
\cite{GAr} and
for electrodisintegration using a realistic interaction model \cite{TrA,Tam}.
However, additional simplifications have been introduced in \cite{BWAr} by
considering only quasifree kinematics where the interaction currents in 
general are negligible, and in \cite{TrA} by keeping only local
velocity-independent terms. In those approximate treatments,
effects due to genuine relativistic components of the intrinsic wave
functions have been neglected, too.
However, a thorough treatment has also to include the relativistic effects
in the internal dynamics. This would in particular
imply a readjustment of realistic
potential models, because at least some relativistic effects
are, implicitly or explicitly, accounted for by fitting some
phenomenological parameters of
a realistic nucleon-nucleon potential to experimental scattering data,
although it is then often inserted into the nonrelativistic Schr\"{o}dinger
equation.

Therefore, the next task will be to include all
leading order terms in a realistic calculation for the two-body system,
which is not very difficult for the charge operator.
However, when one turns to the spatial current,
the number of terms increases enormously. But with present day methods, it
is not impossible. Particularly interesting will be the study of
polarization observables because some of them appear to be very
sensitive to relativistic e.m.\ currents already at rather
small momentum transfer \cite{BWAr,ArL95}. In this region it seems reasonable
to include them in the framework of the $1/m$-expansion.

\section*{Acknowledgements}

This work has been supported by the Deutsche Forschungsgemeinschaft
(SFB 201), by grants GA AV CR 148410 and GA CR 202/94/037 and by the
US Department of Energy under the contract \#DE-AC05-84ER40150 .
J.~A.\ thanks the Alexander von Humboldt-Foundation for
supporting his stay in Mainz.

\renewcommand{\theequation}{A.\arabic{equation}}
\setcounter{equation}{0}
\section*{Appendix A: Foldy-Wouthuysen currents}

In this appendix we list the FW one-nucleon and exchange currents
as obtained by the extended S-matrix technique of \cite{ATA}.
The equations from \cite{ATA} are
referred to as ATA-(number of eq.). The one-nucleon currents and the
one-pion-exchange ones were carefully compared to the results of \cite{GAr}.
A couple of misprints in \cite{ATA} were found and we correct
them below.

We follow the notation of \cite{ATA} for particle momenta
($i= 1,\,2$)
\bea
\vci{q}{i} &=& \vcpi{p}{i} - \vci{p}{i} \, , \\
\vci{Q}{i} &=& \vcpi{p}{i} + \vci{p}{i} \, ,
\eea
and for the e.m.\ form factors
\bea
\hat{e}_i       &=& \frac{1}{2}
 \Big( F_1^{is} (k^2) + F_1^{iv} (k^2) \, \tau_i^3 \, \Big) \, , \\
\hat{\kappa}_i       &=& \frac{1}{2}
 \Big( F_2^{is} (k^2) + F_2^{iv} (k^2) \, \tau_i^3 \, \Big) \, ,
\eea
where $\tau_i^3$\ is the third isospin component of the ith nucleon.
$F_{1,2}^{is/iv}$\ denote the isoscalar and isovector Dirac-Pauli nucleon form
factors, respectively.
In passing to the Breit frame, one finds for the two nucleon system
\bea
\vci{q}{1,2} & = & \frac{1}{2} \vc{k} \pm \vc{q} \, ,
\label{q12b} \\
\vci{Q}{1,2} & = &\pm \vc{Q} \, ,
\eea
with $\vc{q} = \vcp{p} - \vc{p}$, $\vc{Q} = \vcp{p} + \vc{p}$ and
$\vec p^{\,(\prime)}=(\vec p^{\,(\prime)}_1-\vec p^{\,(\prime)}_2)/2$.
However, we keep the notation $ \vci{q}{1,2}$\ in the expressions for
the intrinsic currents instead of (\ref{q12b}), whenever it
simplifies the formulas. The time components of $q_i$, that
describe the energy transferred in the corresponding vertex, are
expressed in terms of the on-shell nucleon kinetic energies,
i.e., up to the order considered
\beq
 q_{i0} = \frac{1}{2m} \ps{\vci{q}{i}}{\vci{Q}{i}} +\ord{-3}\, ,
\label{QBreit}
\eeq
where $m$ denotes the nucleon mass.

With respect to the meson exchange currents associated with the various
mesons of a given OBE potential, it is useful to separate the isospin
dependence in the potential contribution.
$V_B = T\, \tilde{v}_B$,
where $T= \ps{\vci{\tau}{1}}{\vci{\tau}{2}}$ for an
isovector meson, and $T= 1$\ for an isoscalar one.
Because the isospin dependence of the MECs originate from the e.m.\ form
factors and from the isospin operators of the potential, 
it is convenient to separate the MECs into pieces proportional
to $F_{e/\kappa,i}^{\pm}$\ and $G_{M,i}^{\pm}$ as defined by
\bea
 \Ff{e,i}{\pm} &=& \hat{e}_i \, T \, \pm \, T \, \hat{e}_i \, , 
\label{fisoe}\\
 \Ff{\kappa,i}{\pm} &=& \hat{\kappa}_i \, T \, \pm \, T \, \hat{\kappa_i} \, ,
\label{fisok}\\
    G_{M,i}^{\pm} &=& \Ff{e,i}{\pm} \,  + \, \Ff{\kappa,i}{\pm} \, ,
\label{gisom}
\eea
We would like to emphasize that this is {\em not} the separation into
isoscalar and isovector parts.
Notice also that $F_{e/\kappa,i}^{-}$\ and $G_{M,i}^{-}$ are 
odd under nucleon interchange and thus vanish for isoscalar exchange.

\subsection*{One-nucleon currents}

For completeness we will begin with the well-known
expressions for the one-nucleon current in two-nucleon space
in an arbitrary frame of reference. To this end we remind the reader
at the general definition for the matrix element of an operator $a(\vc{k})$
in two-nucleon momentum space transferring a momentum $\vec k$ 
\beq
\langle \vcp{P}\,\vcp{p}|a(\vc{k})|\vc{P}\,\vc{p}\,\rangle
= a(\vc{k},\vc{K},\vc{q},\vc{Q})\,\delta(\vcp{P}-\vc{P}-\vc{k}) \, ,
\eeq
with $\vc{K} = \vcp{P} + \vc{P}$.
For a one-body operator $a(1;\vc{k})$ one finds
\bea
\langle \vcp{P}\,\vcp{p}|a(1;\vc{k})|\vc{P}\,\vc{p}\,\rangle &=&
\langle \vcp{p_1}\,\vcp{p_2}|a(1;\vc{k})|\vc{p_1}\,\vc{p_2}\rangle
\nonumber\\
&=& a(1;\vc{k},\vc{Q_1})_1 \delta(\vc{q_2}) + \exnn
\nonumber\\
&=& \Big(a(1;\vc{k},\vc{Q_1})_1 + \exnn\Big)
\delta(\vcp{P}-\vc{P}-\vc{k})\,.
\eea
The last line follows from the fact that $a(1;\vc{k},\vc{Q_1})_1$ contains
$\delta(\vc{q_1}-\vc{k})$.
Therefore one gets
\beq
 a(1;\vc{k},\vc{K},\vc{q},\vc{Q}) = a(1;\vc{k},\vc{Q_1})_1
+ \exnn\,,\label{akQ}
\eeq
where $\vci{q}{1,2}$ is given in (\ref{q12b}) and
$\vci{Q}{1,2}=\vc{K}/2\pm \vc{Q}$.
Thus it is sufficient to list the contribution of nucleon ``1'', i.e.,
$a(1;\vc{k},\vc{Q_1})_1$, since the contribution of the other one is
obtained by adding the exchange \exnn.

For the one-body current including the leading
relativistic order, one has
\bea
\rho_{FW}^{(0)} (1;\vc{k},\vci{Q}{1})_1 &=& \eh \delta(\vc{q_1}-\vc{k})\, ,
\label{r10}\\
 \rho_{FW}^{(2)} (1;\vc{k},\vci{Q}{1})_1 &=&  - \frac{\eh + 2 \kh}{8m^2}
   \Big( \vcs{k} + i \psso{\vc{Q}_1}{\vc{k}} \Big)\, \delta(\vc{q_1}-\vc{k})
\, ,\label{r12} \\
 \vc{\jmath}_{FW}^{\, \, (1)} (1;\vc{k},\vci{Q}{1})_1 &=&  \frac{1}{2m}
  \Big( \eh \vci{Q}{1}  + i (\eh + \kh )\, \pvso{\vc{k}} \, \Big)\,
\delta(\vc{q_1}-\vc{k})\, ,\label{j11}\\
 \vc{\jmath}_{FW}^{\, \, (3)} (1;\vc{k},\vci{Q}{1})_1 &=&
- \frac{\vcsi{Q}{1} + \vcs{k}}{8m^2}
\vc{\jmath}_{FW}^{\, \, (1)} (1;\vc{k},\vci{Q}{1})_1
\nonumber\\
&&  - \frac{1}{16m^3}\Big[
\Big( \eh \,\ps{\vc{k}}{\vci{Q}{1}} + 4 \kh m  k_0 \Big)
 (\vc{k} + i \pvso{\vci{Q}{1}} )  \nonumber\\
&&
  +\kh \, \vc{k} \times\Big[\vci{Q}{1} \times (\vc{k}+i\pvso{\vci{Q}{1}} )
\Big] \Big]
\delta(\vc{q_1}-\vc{k})  \, .\label{j13}
\eea
Note that ATA-(4.1d), corresponding to (\ref{j13}), 
has a wrong power of $m$ and the vector product is missing in the
last term. In the second line of (\ref{j13}), $k_0$ stands for the total
energy transfer. This means that the corresponding one-nucleon current
contains implicitly some interaction effects.
The divergence of this part of the current equals
the commutator of the full nonrelativistic Hamiltonian with
that part of the Darwin-Foldy and spin-orbit charge density of (\ref{r12})
which is proportional to \kh.
For the free nucleon one has the relation $k_0 = \ps{\vc{k}}{\vci{Q}{1}}/
2m$. It is possible to redefine the
one-nucleon current by replacing
$k_0$\ by its free-nucleon value or, alternatively, introducing the
full momentum transfer also in the preceeding term proportional
to \eh. In this case, the MECs have to be redefined consistenly \cite{ATA}.
In this paper we adopt the particular form (\ref{j13}), since then the
corresponding MECs have the simplest form.

In order to get the currents in the Breit frame, one simply replaces
$\vci{Q}{1}$\ by $\vc{Q}$, and the $\delta$-function in (\ref{akQ})
then becomes $\delta (\frac{\vc{k}}{2} - \vc{q}\, )$. Furthermore, for the
$\exnn$-exchange one has to make the replacements $\vec Q \rightarrow -\vec
Q$ and $\delta (\frac{\vc{k}}{2} - \vc{q}\, )\rightarrow
\delta (\frac{\vc{k}}{2} + \vc{q}\, )$.

\subsection*{Meson exchange currents}

The various contributions from the exchange of a given meson type to
the OBE potential are derived in \cite{ATA} for an arbitrary reference frame.
Here we need only the intrinsic (\vc{P}-independent) parts
of the potential. Their momentum representation
is denoted by $\tilde{v}_B (\vc{q}, \vc{Q})$ where the isospin dependence
has been separated. Thus, up to the order considered one has
\beq
 \tilde{v}_B (\vc{q}, \vc{Q}) = \vto{B} + \vtt{B} + \ord{-5} \, ,
\label{potform}
\eeq
where \vto{B}\ is an even function of \vc{q}.

Unlike in \cite{ATA}, we include here explicitly
the hadronic form factors into the potentials and exchange currents. This is
done by modifying expressions containing a single propagator of a meson
of mass $m_B$ by
\bea
{\Delta}_B (z)= \frac{1}{m_B^2 + z } \rightarrow
 \tilde{\Delta}_B (z) & = &
 \frac{f_B^2 (z)}{m_B^2 + z} \, , \label{prop} \\
{\Delta}_B^{\, \prime} (z)= -  \frac{1}{(m_B^2 + z)^2} \rightarrow
 \tilde{\Delta}_B^{\, \prime} (z) & = &
  \frac{d}{d \, z} \ \tilde{\Delta}_B (z) \, ,
\label{propp}
\eea
where $z$ stands for $\vcs{q}$ and $f_B$ is the
strong form factor at the meson nucleon vertex.

For the meson-in-flight or mesonic currents the introduction
of a hadronic form factor
consistent with gauge invariance is achieved by the replacement
\beq
{\Delta}_B (z_1){\Delta}_B (z_2) \rightarrow
f_B(z_1,\,z_2) =
 - \,  \frac{1}{z_1-z_2} \Big(\tilde{\Delta}_B (z_1) -
\tilde{\Delta}_B (z_2) \Big) \, ,
\label{fqq}
\eeq
where $z_i$ stands for $\vcs{q_i}$ ($i=1,\,2$). This is the minimal form
which fulfills the continuity equation \cite{Ris84}.
In the case that we assume the hadronic form factor to be of the form
\beq
f_{B} (z) = \Big( \frac{\Lambda^2 - m_B^2}{\Lambda^2 + z}
\Big)^{n/2} ,\,\, n = 1, 2, \ldots
\eeq
where $\Lambda$ is a range parameter, it can be shown that
then (\ref{fqq}) corresponds to the minimal substitution in the meson
propagator yielding
\bea
f_B(z_1,\,z_2) &=& (-)^{n-1} \frac{(\Lambda^2 - m_B^2)^{n}}{(n-1)!}
\nonumber\\
&&
 \frac{d^{n-1}}{d (\Lambda^2)^{n-1}}
 \Big[ \frac{1}{\Lambda^2- m_B^2}
 \Big( \frac{1}{z_1 + m_B^2}  \frac{1}{z_2 + m_B^2}
- \frac{1}{z_1 +\Lambda^2} \frac{1}{z_2 + \Lambda^2} \Big)\Big] \,.
\eea
All mesonic currents are of nonrelativistic order.
In order to get higher order terms, one has to consider the
higher order contributions to the meson-nucleon vertices and the
effects of retardation of the exchanged mesons, i.e., dependence
of the propagator function on the energies transferred. For the general
expression (\ref{fqq}) this means using the Taylor decomposition
\bea
 f_B(z_1-\delta_1, z_2-\delta_2)& =& f_B(z_1, z_2)
 + \, \frac{1}{(z_1- z_2)} \Big(
 (\delta_1 - \delta_2) f_B(z_1, z_2) \nonumber\\
& &+ \delta_1 \tilde{\Delta}_B^{\, \prime} (z_1)
-\delta_2 \tilde{\Delta}_B^{\, \prime} (z_2) \Big)+{\cal O}(\delta^2) \,
,\label{ffret}
\eea
where $\delta_i=q_{i0}^2$. This relation generalizes ATA-(4.10e).
The first term on the r.h.s.\ of (\ref{ffret}) contributes to
the nonrelativistic mesonic current, the second one also has a
propagator structure of the mesonic current and thus will be added to the
relativistic ``mes'' part. It can be shown that
the divergence of the corresponding current saturates the commutator
of the kinetic energy with the mesonic charge density. The last
terms have to be combined with the retardation currents.

The MECs as derived in \cite{ATA} are associated with particular
relativistic Feynman diagrams: nucleon Born, contact, and mesonic
contribution. In other techniques, the same currents formally 
arise from a different set of
(time-ordered) diagrams. Therefore we rather group the currents
according to their propagator structure, this also allows us to
combine several contributions and to present the results in a 
more compact form. The MECs with a single meson propagator 
$\prop{B}{2}$ attached are labelled
``pro'', those corresponding to meson-in-flight currents
containing $f_B(\vec q_1^{\,\,2},\, \vec q_2^{\,\,2})$ 
``mes'', and finally those containing
the derivative of a propagator ``ret'', since they arise
from the retardation or boost contributions.
 The ``pro'' and ``ret'' operators can be separated
according to which of the two nucleons interacts with the e.m.\ field. We always
give only the terms corresponding to the e.m.\ interaction of
the first nucleon.

The retardation of the exchanged mesons gives rise to contributions
both to the potential and to the MECs. The different prescriptions 
of how to treat the retardation yield different, but unitarily equivalent
results \cite{Fr80}. The unitary transformation is acting in the
intrinsic space only.
This unitary equivalence can be parametrized by
a parameter $\nu $. The value $\nu = 1/2$\ corresponds to the nonretardation
potential in the c.m.\ frame, but it is impossible to find a suitable
value for $\nu$\
which would simplify {\em all} MECs. Nevertheless, the representation
corresponding to  $\nu = 1/2$\ is most frequently used and we give
all intrinsic operators for this choice. For completeness, we list the
additional operators for a general  $\nu $\ in Appendix D
where we also have corrected several misprints of \cite{ATA}. For this
reason, we will directly refer to this appendix with respect to the
``ret'' contributions.

For pseudoscalar meson exchange, 
there is an additional unitary freedom related to
various off-energy shell continuations of the energy dependence
of the $\pi NN$ vertex. The corresponding unitary transformation 
is parametrized by a parameter \mut\ .
The simplest form for potential and MEC results
from the choice   $\mut = 0$, and we have adopted this choice
 in the main body of this paper. However, the widely used
OBEPQ Bonn potentials correspond to the value  $\mut = -1$
\cite{AGA}.
Therefore, we present the additional  \mut-dependent
operators in Appendix A.

We have already mentioned in Sect.\ 3,
that the MECs simplify for all exchanges if
a part of the relativistic one-nucleon current density is considered
to contain interaction effects through its dependence on
the total energy transfer $k_0$ (see (\ref{j13})). Then, the currents, denoted
as ``external'' in \cite{ATA}, are implicitly included and should
not be added to the MECs (see section 5 of \cite{ATA}).

For simplicity, all MECs are multiplied by the Dirac e.m.\
form factor  $F_1$. For this reason, one has to replace in the explicit
expressions of \cite{ATA} the term 
$-iF_B^{mes/con}(\vec \tau_1 \times \vec \tau_2\,)^3$ 
by $F_{e,1}^-$.
It is shown in the appendix B of \cite{ATA} how one can modify the MECs
in the $1/m$ expansion framework in order to incorporate
independent e.m.\ form factors as suggested in
\cite{GrR}. The resulting additional currents of ATA-(B.17)
are proportional to the differences $F_{\gamma BB}- F^V_1$ or
$F_{\gamma NNB}- F^V_1$.
 These differences are of the order
of $\vcs{k}/\Lambda^2$, where $\Lambda$ is the corresponding
cut-off parameter. The currents ATA-(B.17) are therefore of
leading relativistic order. Moreover, they are frame-independent
and therefore can be added to the intrinsic currents considered
in this paper without any further modification.

With respect to our notation, we would like to remark that the currents in the
momentum space representation depend in general on \vc{k}, \vc{K}, \vc{q},
and \vc{Q}, but for the sake of briefness only their \vc{k}-dependence
is retained in our notation of the operators while in the explicit
expressions on the r.h.s.\ of the equations we use \vc{k},
$\vci{q}{1/2}=\frac{1}{2}\vc{k}\pm \vc{q}$, and $\vci{Q}{1/2}=\frac{1}{2}
\vc{K}\pm \vc{Q}$ for convenience.

\subsubsection*{Scalar meson exchange}

With respect to the ``$+$'' part of the MECs for scalar exchange,
there is one single nonretardation current from ATA-(4.4b)
\beq
 \jrcFq{pro}{S}{+} =
- \Ff{e,1}{+} \mpw{4}{2} \vtos{S} ( \vci{Q}{1} + i \pvso{\vc{k}} )
 + \exnn \, ,
\label{j2spro+}
\eeq
where $\vtos{S}$ is given in (\ref{vs1}).
Furthermore, one gets the retardation charge and current densities from
(\ref{rhret})-(\ref{jretSV}) for $\tilde \nu = 0$
\bea
 \rhFqQ{ret}{S}{+} &=& \Ff{e,1}{+}\, \conm{S}{8}{ } \,
 \ps{\vc{k}}{\vci{q}{2}}\, \propp{S}{2} +\exnn \, , \label{rh2sret+}\\
 \jrcFq{ret}{S}{+} &=&  \conm{S}{16}{2}
  \Big\{
 \Ff{e,1}{+}\, \Big( \vci{Q}{1}\, \ps{\vc{k}}{\vci{q}{2}}  +
  \vci{q}{2}\, (\vci{Q}{1} + 3\vci{Q}{2}) \cdot \vci{q}{2}\, \Big)
  \nonumber\\
&&   + i G_{M,1}^{+}\, \pvso{\vc{k}}\, \ps{\vc{k}}{\vci{q}{2}}\,
  \Big\} \, \propp{S}{2} +\exnn \, . \label{j2sret+}
\eea
Similarly, one obtains for the the ``$-$'' part from
ATA-(4.3b) one nonrelativistic current
\beq
 \jnrFq{mes}{S}  =
 \Ff{e,1}{-} \con{S} \, (\vci{q}{1} -  \vci{q}{2} )\,
 \fq \, . \label{j2smes1-}
\eeq
The relativistic ``pro'' and ``mes'' contributions follow from
ATA-(4.4b, 4.7d, 4.10a)
\bea
 \jrcFq{pro}{S}{-} &=&
 \Ff{e,1}{-} \mpw{4}{2} \vtos{S} ( \vc{k} + i \pvso{\vci{Q}{1}} ) +\exnn \, ,
\label{j2spro-}\\
 \rhFqQ{mes}{S}{-} &=&
  \Ff{e,1}{-} \con{S} \, (q_{10} - q_{20}) \, \fq \, , \label{rh2smes-}\\
 \jrcFq{mes}{S}{-} &=&
  - \Ff{e,1}{-}\, \conm{S}{8}{2} \fq \, (\vci{q}{1} - \vci{q}{2}\, )
 \Big(
 \vcsi{Q}{1} +  \vcsi{Q}{2}
\nonumber\\
&&
+  i \psso{\vci{q}{1}}{\vci{Q}{1}} + i \psst{\vci{q}{2}}{\vci{Q}{2}}
-8m^2\frac{q_{10}^2 - q_{20}^2}{\vcsi{q}{1} - \vcsi{q}{2}} \Big) \,
 \, , \label{j2smes-}
\eea
while again the ``ret'' currents come from
(\ref{rhret})-(\ref{jretSV}) for $\tilde \nu = 0$
\bea
 \rhFqQ{ret}{S}{-} &=&
 - \Ff{e,1}{-} \conm{S}{8}{ }\,   (\vci{Q}{1} + 3\vci{Q}{2}) \cdot \vci{q}{2}
 \ \propp{S}{2}  + \exnn
\, . \label{rh2sret-}\\
 \jrcFq{ret}{S}{-}  &=& - \conm{S}{16}{2} \,
 \Bigg\{ \Ff{e,1}{-}\, \Big(  \vci{q}{2}\, \ps{\vc{k}}{\vci{q}{2}} +
   \vci{Q}{1}\, (\vci{Q}{1} + 3\vci{Q}{2}) \cdot \vci{q}{2}
 \, \Big) \nonumber\\
&&
   + iG_{M,1}^{-}\,  \pvso{\vc{k}} \
 (\vci{Q}{1} + 3\vci{Q}{2}) \cdot \vci{q}{2}
 + \Ff{e,1}{-}\,16m^2
 \frac{\vci{q}{1} - \vci{q}{2}}{(\vcsi{q}{1} - \vcsi{q}{2})} \,
 q_{20}^2 \,  \Bigg\} \propp{S}{2}\nonumber\\
&& + \exnn \, . \label{j2sret-}
\eea

\subsubsection*{Vector meson exchange}

Large parts of the MEC for vector meson exchange can
be obtained from the scalar case by the replacements 
$m_S \rightarrow m_V$, $g_S^2 \rightarrow - g_V^2$.
Those parts will not be listed again, rather we shall refer to the
corresponding expressions for scalar exchange keeping in mind the
aforementioned replacements. Only the additional currents will now be listed
explicitly.

We start again with the ``$+$'' part for which we have only
one additional current from ATA-(4.4c)
\beq
\jrcFq{pro}{V}{+} =
  - \Ff{e,1}{+} \mpw{4}{2} \vtos{V} \Big( \vci{Q}{2} +
 i(1+\kappa_V )\, \pvsp{\vci{q}{2}} \Big) +\exnn  \, ,
\label{j2vpro+}
\eeq
where $\kappa_V$\ is the usual ratio of the normal (vector) to the
anomalous (tensor) $VNN$\ coupling constants. The retardation currents
follow from (\ref{rh2sret+}) and  (\ref{j2sret+}).

For the ``$-$'' part, we have first the retardation terms from (\ref{rh2sret-})
and (\ref{j2sret-}). For the additional currents we collect
from the expressions in
ATA-(4.4c, 4.9a)
\bea
 \jrcFq{pro}{V}{-} &=&
   \Ff{e,1}{-} \mpw{4}{2} \vtos{V}
 \Big( (1+\kappa_V) \vci{q}{2} - \kappa_V \vci{q}{1}
 - (1+\kappa_V)^2 \, \vs{1} \times (\vs{2} \times \vci{q}{2} )
 \nonumber\\
 &&
 - i \kappa_V \pvso{\vci{Q}{1}} + i (1+\kappa_V) \pvso{\vci{Q}{2}}
 \Big) +\exnn \, .
\label{j2vpro-}
\eea
For the ``mes'' currents, one has two contributions from ATA-(4.7c, 4.8).
First from (\ref{rh2smes-}) the charge density $ \rhFqQ{mes}{V}{-}$, 
and from ATA-(4.10b) plus the last term of (\ref{j2smes-}) the current 
density 
\bea
 \jrcFq{mes}{V}{-} &=&
  \Ff{e,1}{-}\, \conm{V}{4}{2} \fq \, (\vci{q}{1} - \vci{q}{2}\, )
\nonumber\\
&&
 \Big[ (\frac{1}{2} + \kappa_V) ( \vcsi{q}{1} +  \vcsi{q}{2}) +
 \ps{\vci{Q}{1}}{\vci{Q}{2}}
 - (1+\kappa_V)^2
  \pf{\vs{1}}{\vci{q}{1}}{\vs{2}}{\vci{q}{2}}
 \nonumber\\
&& + i(\frac{1}{2} + \kappa_V) \Big(
 \psso{\vci{Q}{1}}{\vci{q}{1}} + \psst{\vci{Q}{2}}{\vci{q}{2}}  \Big)
\nonumber\\
&&  - i(1+\kappa_V) \Big(
 \psso{\vci{Q}{2}}{\vci{q}{1}} + \psst{\vci{Q}{1}}{\vci{q}{2}}  \Big)
-4m^2\frac{q_{10}^2 - q_{20}^2}{\vcsi{q}{1} - \vcsi{q}{2}}
 \Big]  \, . \label{j2vmes-}
\eea
Second, from the transverse part of the $\gamma VV$\ vertex as given in
ATA-(4.7e, 4.10c)
\bea
 \rhFqQ{mes-tr}{V}{-} &=&
  \Ff{e,1}{-}\, \conm{V}{2}{} \fq \nonumber\\
&& \Big( \vc{k} \cdot (\vci{Q}{1}-\vci{Q}{2}) +
 i(1+\kappa_V) \pssp{\vci{q}{1}}{\vci{q}{2}} \Big) \, ,
\label{rh2vmestr-} \\
 \jrcFq{mes-tr}{V}{-} &=&
  \Ff{e,1}{-}\, \conm{V}{4}{2} \fq \nonumber\\
&& \Big\{ 2m k_0 \Big[ \vci{Q}{1} - \vci{Q}{2} +
 i(1+\kappa_V) (\pvso{\vci{q}{1}} - \pvst{\vci{q}{2}} ) \Big]
 \nonumber\\
&&
+ \vc{k} \times \Big[ (1+\kappa_V)^2 \, (\vs{1} \times \vci{q}{1})
 \times  (\vs{2} \times \vci{q}{2}) - \vci{Q}{1} \times \vci{Q}{2}
 \nonumber\\
&&   + i (1+\kappa_V) \Big(
 (\vs{2} \times \vci{q}{2} ) \times \vci{Q}{1} -
 (\vs{1} \times \vci{q}{1} ) \times \vci{Q}{2} \Big) \Big] \Big\}
\, .
\label{j2vmestr-}
\eea
The usual minimal form of this part
of the vertex is adopted here, in accordance with ATA. The discussion
and implications of a nonminimal form can be found in the appendix
B of ATA and in \cite{GrR}.

\subsubsection*{Pseudoscalar meson exchange}

In the case of pseudoscalar meson exchange, a unitary freedom 
shows up in the potential and in the interaction currents which 
is not present for scalar and vector exchange up to the order 
considered here. Let us recall that in order to derive the 
nuclear potential and e.m.\ current operators from a
meson-nucleon Lagrangian, one has to fix the energy transfer
at the meson-nucleon vertices. This is done in different ways
within various techniques \cite{Fr80}. In particular, in \cite{ATA}
the vertex energy transfer was replaced by the difference of the
on-mass-shell nucleon energies. For the off-energy-shell
parameter $\beta$\ introduced in \cite{AGA} this means $\beta = 0$. Most of the
other techniques, e.g.\ those of \cite{GAr,Fr77}, correspond to a symmetric
off-energy-shell continuation with $\beta = 1/2$.
Whatever choice is adopted, the final results
are unitarily equivalent \cite{Fr80}, the unitary parameter \mut\ being
given by
\beq
 \mut = 4 \beta \, \Big( \frac{1-c}{2} (\mu -1 ) + 1 \Big) - 1 \, ,
\label{mut}
\eeq
where $\mu = 0\, (1) $\ for $PS (PV) \ \pi NN $\-coupling, and
$c$ is the so-called Barnhill parameter (for a more detailed discussion
see \cite{Fr80,GAr,Fr77,AGA}).

This unitary freedom does not affect the nonrelativistic parts of the
potential and the MEC but the leading order relativistic contributions. It
means that $\tilde v^{(3)}$ and $j_\lambda^{(3)}(2;k)$ for an arbitrary
choice of the parameter $\mut$ are obtained from the operators for
$\mut=0$ by the approximate unitary transformation $(1-i\mut U_{PS})$ where
\bea
 \langle \vcpi{p}{1} \vcpi{p}{2} \, |\, U_{PS} \, |
 \vci{p}{1} \vci{p}{2} \, \rangle & = &  \tilde{U}_{PS} (\vc{q},
\vc{K}, \vc{Q}) \ps{\vci{\tau}{1}}{\vci{\tau}{2}}\, \delta(\vci{q}{1} +
\vci{q}{2}\, ) \, ,
\label{vosu1}\\
  i \tilde{U}_{PS} (\vc{q},\vc{K}, \vc{Q}) &=&  \conm{PS}{32}{3}
\Big[ \frac{1}{2}\Sigma^{(-)}(\vc{q},\, \vec K) - \Sigma^{(+)}(\vc{q},\, \vec Q)
\Big] \prop{PS}{}
\label{vosu2}\,,
\eea
where we have introduced as abbreviation for two vectors $\vec a$ and $\vec b$
\beq
\Sigma^{(\pm)}(\vec a,\,\vec b) = \ps{\vec\sigma_1}{\vec a}
\ps{\vec\sigma_2}{\vec b} \pm
\ps{\vec\sigma_1}{\vec b}\ps{\vec\sigma_2}{\vec a}\,. \label{Bab}
\eeq
Note that here $\vec P^{\,\prime}=\vec P$ and 
$\vec K=2\vec P=2(\vec p_1+\vec p_2)$. In detail one finds
\beq
\vtt{PS (\mut)} = \vtt{PS} + \vtt{\mut}\,,
\eeq
where
\bea
\vtt{\mut} &=& i\mut\langle \vcpi{p}{1} \vcpi{p}{2} \, | [t^{(1)},\, U_{PS}]
| \vci{p}{1} \vci{p}{2} \, \rangle
\nonumber\\
&=&- \mut  \conm{PS}{32}{4} \ps{\vc{q}}{\vc{Q}} \Sigma^{(+)}(\vc{q},\, \vec Q)
\prop{PS}{}\, ,
\label{v3psmut}
\eea
and
\beq
  j_{\lambda,FW} (\mbox{2};\mut;\vc{k}\, )_{PS} =
  j_{\lambda,FW} (\mbox{2};\vc{k}\, )_{PS}+
  j_{\lambda,FW\,\mut} (\mbox{2};\vc{k}\, )_{PS}\,,
\label{voscom}
\eeq
with
\beq
j_{\lambda,FW\,\mut} (\mbox{2};\vc{k}\, )_{PS} =
 i\mut\Big[ j_{\lambda,FW} (1;\vc{k}\, ) ,\,  U_{PS} \Big] \, .
\label{voscom1}
\eeq
Evaluating (\ref{voscom1}) with (\ref{vosu2}),
one finds for the additional $\mut$-dependent FW currents
\bea
 \rho^{(2)}_{FW\,\mut}(2;\vec k\,)_{PS} &=& - \mut \conm{PS}{32}{3} \, \Big\{ \,
 \Ff{e,1}{+}\, \ps{\vs{1}}{\vc{k}} \ps{\vs{2}}{\vci{q}{2}}  \nonumber\\
&&
 + \Ff{e,1}{-}\, \Big[ \ps{\vs{1}}{\vci{q}{2}} \ps{\vs{2}}{\vci{Q}{2}}
 - \ps{\vs{1}}{\vci{Q}{1}}  \ps{\vs{2}}{\vci{q}{2}} \Big]\, \Big\}
  \prop{PS}{2}
+\exnn \, ,  \label{rh2psvos}\\
 \vec \jmath^{\,\,(3)}_{FW\,\mut} (2;\vec k\,)_{PS}
   &=& - \mut \conm{PS}{64}{4} \nonumber\\
&&
   \Big\{\, \Ff{e,1}{+}\, \Big[
 \vci{Q}{1} \ps{\vs{1}}{\vc{k}} \ps{\vs{2}}{\vci{q}{2}} - \vci{q}{2}
 \Big(\ps{\vs{1}}{\vci{q}{2}} \ps{\vs{2}}{\vci{Q}{2}}
 - \ps{\vs{1}}{\vci{Q}{1}}  \ps{\vs{2}}{\vci{q}{2}} \Big) \Big]
  \nonumber\\
&&
 + \Gm{+}\, \vc{k} \times \Big[
 \vci{q}{2} \times \vs{1} \ps{\vs{2}}{\vci{Q}{2}}
 - \vci{Q}{1} \times \vs{1} \ps{\vs{2}}{\vci{q}{2}}\, \Big] \nonumber\\
&&
 + \Ff{e,1}{-}\, \Big[
 \vci{Q}{1} \Big\{\ps{\vs{1}}{\vci{q}{2}} \ps{\vs{2}}{\vci{Q}{2}}
 - \ps{\vs{1}}{\vci{Q}{1}}  \ps{\vs{2}}{\vci{q}{2}} \Big\}
  - \vci{q}{2}\, \ps{\vs{1}}{\vc{k}} \ps{\vs{2}}{\vci{q}{2}}
 \Big] \nonumber\\
&&
 + \Gm{-}\, \vc{k} \times \Big[
 \vc{k} \times \vs{1} \ps{\vs{2}}{\vci{q}{2}} +
 i \vci{Q}{1} \ps{\vs{2}}{\vci{q}{2}} -
 i \vci{q}{2} \ps{\vs{2}}{\vci{Q}{2}} \, \Big] \Big\} \prop{PS}{2}
\nonumber\\
&&+\exnn \, . \label{j2psvos}
\eea

After these remarks concerning the unitary freedom, we will now proceed to
give the explicit expressions for the FW currents. We will start with those
given in \cite{ATA} which correspond to the unitary
representation  $\mut = -1$ \cite{Fr80,AGA}. However, this unitary freedom
affects the ``pro''-currents only for which we will characterize the
expressions from \cite{ATA} by an additional subscript ``$ATA$''.
The FW currents for the
unitary representation  $\mut = 0$ are then obtained by adding
(\ref{rh2psvos}) or (\ref{j2psvos}), respectively, for $\mut =1$
to the corresponding expressions, i.e.,
\bea
\rhFqQ{pro}{PS}{\pm} &=& \rhFqQ{pro}{PS,\,ATA}{\pm} + \rho^{(2)}_{FW\,\mut=1}
(2;\vec k\,)_{PS}^{\pm}\,,\label{rprounit}
\\
\jrcFq{pro}{PS}{\pm} &=& \jrcFq{pro}{PS,\,ATA}{\pm} + \vec \jmath^{\,\,(3)}
_{FW\,\mut=1} (2;\vec k\,)_{PS}^{\pm}\,. \label{jprounit}
\eea
All other FW currents are \mut -independent.

Considering first the ``$+$'' terms,
their ``pro'' part follows from ATA-(4.4d) and (4.19a,b). Explicitly, one finds
\bea
 \rhFqQ{pro}{PS,\,ATA}{+} &=& \Ff{e,1}{+}\, \conm{PS}{8}{3} \,
 \ps{\vs{1}}{\vc{k}} \ps{\vs{2}}{\vci{q}{2}}\,
 \prop{PS}{2} +\exnn \, , \label{rh2pspro+x}\\
 \jrcFq{pro}{PS,\,ATA}{+} &=&  \conm{PS}{16}{4}
 \Big\{ \Ff{e,1}{+}\, \Big[ \vci{Q}{1} \ps{\vs{1}}{\vci{q}{2}} +
 \vs{1} \ps{\vc{k}}{\vci{Q}{1}} + \vs{1} \ps{\vci{q}{2}}{\vci{Q}{2}}
- i \pv{\vc{k}}{\vci{q}{2}} \, \Big]  \nonumber\\
&&
 + \Ff{\kappa,1}{+}\, \vc{k} \times (\vs{1} \times \vci{Q}{1} ) \Big\}
 \ps{\vs{2}}{\vci{q}{2}}\, \prop{PS}{2} + \exnn \, .
\label{j2pspro+x}
\eea
The transformations (\ref{rprounit}) and (\ref{jprounit}) give then the
FW-``pro''-current for $\mut = 0$
\bea
 \rhFqQ{pro}{PS}{+} &=& \Ff{e,1}{+}\, \conm{PS}{32}{3} \,
 3\, \ps{\vs{1}}{\vc{k}} \ps{\vs{2}}{\vci{q}{2}}\,
 \prop{PS}{2} +\exnn \, , \label{rh2pspro+}\\
 \jrcFq{pro}{PS}{+} &=&  \conm{PS}{16}{4}
 \Big\{ \Ff{e,1}{+}\,\Big( \frac{1}{4}\vec q_2\,\Big[
 \ps{\vs{1}}{\vci{q}{2}} \ps{\vs{2}}{\vci{Q}{2}}
- \ps{\vs{1}}{\vci{Q}{1}}\ps{\vs{2}}{\vci{q}{2}}\Big]\nonumber\\
&&
+ \Big[ \vci{Q}{1} \ps{\vs{1}}{(\vci{q}{2}+
\frac{3}{4}\vc{k}\,)}  + \vs{1} \ps{\vci{q}{2}}{\vci{Q}{2}}
- i \pv{\vc{k}}{\vci{q}{2}}\Big)\Big]
 \ps{\vs{2}}{\vci{q}{2}} \,\Big)  \nonumber\\
&&
 + \frac{1}{4}\Gm{+}\, \vc{k} \times 
\Big[3\,\vs{1} \times \vci{Q}{1}\,\ps{\vs{2}}{\vci{q}{2}}
- \vs{1} \times \vci{q}{2}\,\ps{\vs{2}}{\vci{Q}{2}} \Big] \Big\}
 \prop{PS}{2} \nonumber\\
&&
 + \exnn \, .
\label{j2pspro+}
\eea
The ``ret'' part we take from (\ref{rhret}) through (\ref{jret}) of Appendix
D.
\bea
 \rhFqQ{ret}{PS}{+} &=& \Ff{e,1}{+}\, \conm{PS}{32}{3} \,
 \ps{\vc{k}}{\vci{q}{2}} \ps{\vs{1}}{\vci{q}{2}} \ps{\vs{2}}{\vci{q}{2}}\,
 \propp{PS}{2} +\exnn \, , \label{rh2psret+}\\
 \jrcFq{ret}{PS}{+} &=&  \conm{PS}{64}{4}
 \Big\{ \Ff{e,1}{+}\, \ps{\vs{1}}{\vci{q}{2}}
 \Big[ \vci{Q}{1}\, \ps{\vc{k}}{\vci{q}{2}} +
 \vci{q}{2}\, (\vci{Q}{1} + 3 \vci{Q}{2} ) \cdot \vci{q}{2}
  \, \Big]   \nonumber\\
&&
 + \Gm{+}\, \vc{k} \times \,  \Big[
 \vs{1} \times \vci{q}{2}\,
 (\vci{Q}{1} + 3 \vci{Q}{2} ) \cdot \vci{q}{2}
 -i\vci{q}{2}\, \ps{\vc{k}}{\vci{q}{2}} \Big] \Big\}
 \ps{\vs{2}}{\vci{q}{2}}\, \propp{PS}{2}
\nonumber\\
&&+ \exnn \, .
\label{j2psret+}
\eea

With respect to the ``$-$'' part, there are first of all two well-known
\mut -independent nonrelativistic currents from ATA-(4.3d,e), namely
\bea
 \jnrFq{pro}{PS}  &=&
 \Ff{e,1}{-} \conm{PS}{4}{2} \, \vs{1} \ps{\vs{2}}{\vci{q}{2}} \prop{PS}{2}
 + \exnn \, , \label{j2pspro1-}\\
 \jnrFq{mes}{PS}  &=&
 - \Ff{e,1}{-} \conm{PS}{4}{2} \, (\vci{q}{1} -  \vci{q}{2} )\,
  \ps{\vs{1}}{\vci{q}{1}}   \ps{\vs{2}}{\vci{q}{2}} \fq \, .
\label{j2psmes1-}
\eea
For the relativistic  MECs the ``pro'' parts are taken from
ATA-(4.4d, 4.7b, 4.9b, 4.19).
Notice that that the first ``$-$'' term in ATA-(4.4d) should read $\vs{1}
\vcsi{q}{2}$ and
the sign in front of the $G^{-}$-terms in ATA-(4.19b) should be changed.
Explicitly, one finds that $ \rhFqQ{pro}{PS}{-} $ vanishes since ATA-(4.7b)
and  ATA-(4.19a) cancel exactly. For the current one has
\bea
 \jrcFq{pro}{PS,\,ATA}{-} &=& - \conm{PS}{32}{4} \Big\{
 \Ff{e,1}{-} \Big[ \Big( \vs{1} ( 3 \vcsi{q}{2} +
 \vcsi{Q}{1} + \vcsi{Q}{2}) + 2\vc{k} \ps{\vs{1}}{\vci{q}{2}}
 + \vci{q}{1} \ps{\vs{1}}{\vci{q}{1}}
\nonumber\\
&&
+  \vci{Q}{1} \ps{\vs{1}}{\vci{Q}{1}}
+  i (\vc{k} + \vci{q}{2}) \times \vci{Q}{1} \Big)
 \ps{\vs{2}}{\vci{q}{2}}
 + \vs{1} \ps{\vs{2}}{\vci{Q}{2}} \ps{\vci{q}{2}}{\vci{Q}{2}} \Big]
\nonumber\\
&&
 - 2 \Gm{-}\, \vc{k} \times \Big( i \vci{Q}{1} + \vc{k} \times \vs{1} \Big)
  \ps{\vs{2}}{\vci{q}{2}} \Big\} \prop{PS}{2}
 +\exnn \, .\label{j2pspro3-x}
\eea
Again the transformations (\ref{rprounit}) and (\ref{jprounit}) give then the
FW-``pro''-current for $\mut = 0$
\bea
 \rhFqQ{pro}{PS}{-} &=&- \Ff{e,1}{-}\,\conm{PS}{32}{3} \,
 \Big[ \ps{\vs{1}}{\vci{q}{2}} \ps{\vs{2}}{\vci{Q}{2}}
 - \ps{\vs{1}}{\vci{Q}{1}}  \ps{\vs{2}}{\vci{q}{2}} \Big]
  \prop{PS}{2}
\nonumber\\
&&
+\exnn \, ,  \label{rh2pspro-}\\
 \jrcFq{pro}{PS}{-} &=& - \conm{PS}{32}{4} \Big\{
 \Ff{e,1}{-} \Big[ \Big( \vs{1} ( 3 \vcsi{q}{2} +
 \vcsi{Q}{1} + \vcsi{Q}{2}) + 2\vc{k} \ps{\vs{1}}{\vci{q}{2}}
 + \vci{q}{1} \ps{\vs{1}}{\vci{q}{1}}
\nonumber\\
&&
 -\frac{1}{2}\vci{q}{2} \ps{\vs{1}}{\vc{k}}
+  i (\vc{k} + \vci{q}{2}) \times \vci{Q}{1} \Big)
 \ps{\vs{2}}{\vci{q}{2}}
 + \vs{1} \ps{\vci{q}{2}}{\vci{Q}{2}}\ps{\vs{2}}{\vci{Q}{2}} \Big]
\nonumber\\
&&
 +\frac{1}{2}\vci{Q}{1}\Big(
 \ps{\vs{1}}{\vci{Q}{1}}\ps{\vs{2}}{\vci{q}{2}}
 +\ps{\vs{1}}{\vci{q}{2}}\ps{\vs{2}}{\vci{Q}{2}}\Big)
\nonumber\\
&&
 - \frac{1}{2}\Gm{-}\, \vc{k} \times \Big[
 3\Big( i \vci{Q}{1} + \vc{k} \times \vs{1} \Big)
  \ps{\vs{2}}{\vci{q}{2}} +i \vci{q}{2}\ps{\vs{2}}{\vci{Q}{2}}
 \Big\} \prop{PS}{2}
\nonumber\\
&&
 +\exnn \, .\label{j2pspro3-}
\eea
The ``mes'' currents follow from ATA-(4.7f, 4.10d,e)
\bea
 \rhFqQ{mes}{PS}{-} &=&
 - \Ff{e,1}{-} \conm{PS}{4}{2} \, (q_{10} - q_{20}) \,
  \ps{\vs{1}}{\vci{q}{1}}  \ps{\vs{2}}{\vci{q}{2}} \fq \, , \label{rh2psmes-}\\
 \jrcFq{mes}{PS}{-} &=&  \Ff{e,1}{-}\, \conm{PS}{32}{4}\,
 \fq\, (\vci{q}{1} - \vci{q}{2})  \bigg\{
\ps{\vs{1}}{\vci{q}{1}}  \ps{\vs{2}}{\vci{Q}{2}} \ps{\vci{q}{2}}{\vci{Q}{2}}
\nonumber\\
&&
+ \ps{\vs{1}}{\vci{q}{1}} \ps{\vs{2}}{\vci{q}{2}} \Big[ \vcsi{q}{1}
 + \vcsi{Q}{1}   -4m^2\frac{q_{10}^2 - q_{20}^2}{\vcsi{q}{1} - \vcsi{q}{2}}
 \Big]
+\exnn \bigg\} \, , \label{j2psmes3-}
\eea
and finally the corrected retardation operators from (\ref{rhret})-(\ref{jret})
for $\tilde \nu = 0$
\bea
 \rhFqQ{ret}{PS}{-} &=&
 - \Ff{e,1}{-} \conm{PS}{32}{3}\, (\vci{Q}{1} + 3\vci{Q}{2}) \cdot \vci{q}{2}
\, \ps{\vs{1}}{\vci{q}{2}} \ps{\vs{2}}{\vci{q}{2}}
 \ \propp{PS}{2}
\nonumber\\
&& + \exnn \, , \label{rh2psret-}\\
  \jrcFq{ret}{PS}{-} &=& - \conm{PS}{64}{4} \bigg\{ \Ff{e,1}{-}\,
 \Big[ 16 m^2\, q_{20}^2\,
 \Big( \vs{1} -
 \frac{\vci{q}{1} - \vci{q}{2}}{\vcsi{q}{1} - \vcsi{q}{2}}
    \,  \ps{\vs{1}}{\vci{q}{1}}\, \Big)
\nonumber\\
&&
  + \ps{\vs{1}}{\vci{q}{2}}
 \Big(  \vci{q}{2}\, \ps{\vc{k}}{\vci{q}{2}} +
   \vci{Q}{1}\, (\vci{Q}{1} + 3\vci{Q}{2}) \cdot \vci{q}{2} \, \Big)
  \,  \Big] \nonumber\\
&&
   - G_{M,1}^{-}\, i \vc{k} \times \Big( \vci{q}{2}\,
 (\vci{Q}{1} + 3\vci{Q}{2}) \cdot \vci{q}{2} +
 i \vs{1} \times \vci{q}{2}\,  \ps{\vc{k}}{\vci{q}{2}}\,
\Big) \bigg\} \ps{\vs{2}}{\vci{q}{2}}  \propp{PS}{2}
\nonumber\\
&&+ \exnn \, .
\label{j2psret-}
\eea

\renewcommand{\theequation}{B.\arabic{equation}}
\setcounter{equation}{0}
\section*{Appendix B: Boost and separation contributions for scalar and
pseudoscalar meson exchange currents}

In this appendix, we collect the separation and boost current
contributions to the interaction currents for scalar and pseudoscalar
exchange only, since for vector exchange they are formally equal to the
scalar ones. The separation and kinematic boost contributions arise from
the nonrelativistic meson exchange current whereas a potential dependent
boost appears from the one-body current for pseudoscalar exchange only.

\subsection*{Scalar meson exchange}

Taking the only nonrelativistic current contribution from (\ref{j2smes1-}),
we find by separating into the ``pro'', ``mes'' and ``ret'' parts
\bea
  \jrcq{\chis}{mes}{S}{-} &=&
 i  \Ff{e,1}{-} \, \conm{S}{8}{2} \, \vc{q} \, \fq \,
 \pssm{\vc{Q}}{\vc{k}}  \, , \label{jschs}\\
 \jrcq{\chir}{pro}{S}{-}  &=&
  -\Ff{e,1}{-} \, \conm{S}{32}{2} \, \vc{k} \, \prop{S}{2} +\exnn
 \, , \label{jschrpro}\\
 \jrcq{\chir}{ret}{S}{-}  &=&
 - \Ff{e,1}{-} \, \conm{S}{16}{2} \, \vc{q} \,
 \ps{\vc{k}}{\vci{q}{2}} \, \propp{S}{2} +\exnn
 \, , \label{jschrret}\\
  \jrcq{sep}{ret}{S}{-} &=&
 \Ff{e,1}{-} \, \conm{S}{32}{2} \,\vc{q}\,\,\vcs{k} \,
\frac{\ps{\vc{k}}{\vci{q}{2}}}
{\ps{\vc{k}}{\vc{q}\, }} \propp{S}{2} +\exnn \, . \label{jssep}
\eea

\subsection*{Pseudoscalar meson exchange}

We will list
first the contributions from the kinematic boost and start with
the \chis -currents with separation according to the propagator structure
\bea
 \jrcq{\chis}{pro}{PS}{-} &=&  \Ff{e,1}{-}\, \conm{PS}{64}{4}\,
 \nonumber\\
 &&
 \Big\{ \Big[ \vec k\,\ps{\vs{1}}{\vc{q}\,} -
 i \pv{\vc{k}}{\vc{Q}} \Big] \ps{\vs{2}}{\vci{q}{2}}
 -\vec q\,\ps{\vs{1}}{\vci{q}{1}}\ps{\vs{2}}{\vc{k}}
\nonumber\\
&& \hspace*{-0.1truecm} 
+ \vs{1}\, \Big[ \pf{\vs{2}}{\vci{q}{2}}{\vc{k}}{\vc{q}\, } +
 i \pv{\vc{k}}{\vc{q}\, } \cdot \vc{Q} \Big] \Big\} \,
 \prop{PS}{2}  + \exnn \, ,
\label{jpschispro} \\
 \jrcq{\chis}{mes}{PS}{-} &=& \Ff{e,1}{-}\, \conm{PS}{32}{4}
 \vc{q}\, \fq \Big\{
 \vcs{q}\ps{\vs{1}}{\vc{k}}\ps{\vs{2}}{\vc{k}}
 + \vcs{k}\ps{\vs{1}}{\vc{q}\,}\ps{\vs{2}}{\vc{q}\, }
\nonumber\\
&&
 - \pv{\vc{k}}{\vc{q}\, } \cdot
 \Big[ i \vc{Q}\, ( \vs{1} \cdot \vci{q}{1} +  \vs{2} \cdot \vci{q}{2} )
 - \pv{\vs{1}}{\vs{2}} \ps{\vci{q}{1}}{\vci{q}{2}}\Big] \, .
\label{jpschismes}
\eea
Analogously, we get from  (\ref{chir}) and (\ref{sepj3}) for the
\chir - and separation currents
\bea
 \jrcq{\chir}{pro}{PS}{-} &=&  \Ff{e,1}{-}\, \conm{PS}{128}{4}
 \Big\{\vs{1}\Big(\vcs{k}\ps{\vs{2}}{\vci{q}{2}}
- \ps{\vc{k}}{\vc{q}\,} \ps{\vs{2}}{\vc{k}}\Big)
\nonumber\\
&&
 -\vc{q}\,\Big( \ps{\vs{1}}{\vc{k}\, } \ps{\vs{2}}{\vci{q}{2}} -
  \ps{\vs{1}}{\vci{q}{1}} \ps{\vs{2}}{\vc{k}\, } \Big)
\nonumber\\
&&
- \vc{k}\, \ps{\vs{1}}{\vci{q}{1}} \ps{\vs{2}}{\vci{q}{2}}
 \Big\} \prop{PS}{2} + \exnn  \, ,
\label{jpschirrpro} \\
 \jrcq{\chir}{ret}{PS}{-} &=& -\Ff{e,1}{-}\, \conm{PS}{64}{4}
 \Big[ \vs{1} - \vc{q} \,
   \frac{\ps{\vs{1}}{\vci{q}{1}}}{\ps{\vc{k}}{\vc{q}\, }} \Big]\,
\nonumber\\
&&   \ps{\vs{2}}{\vci{q}{2}}\ps{\vc{k}}{\vc{q}\,} \ps{\vc{k}}{\vci{q}{2}}
   \propp{PS}{2} + \exnn \, ,
\label{jpsrsrret}\\
 \jrcq{sep}{pro}{PS}{-} &=& \Ff{e,1}{-}\, \conm{PS}{128}{4} \vs{1}\,
 \vcs{k}\ps{\vs{2}}{(\vc{k}-\vc{q}\,)}\prop{PS}{2}+ \exnn \, ,
\label{jpsrsrpro1}\\
 \jrcq{sep}{mes}{PS}{-} &=&  - \Ff{e,1}{-}\, \conm{PS}{128}{4}\,
 \vc{q}\,\,\vcs{k}\fq \nonumber\\
&& \Big[  \ps{\vs{1}}{\vc{k}\, } \ps{\vs{2}}{\vci{q}{2}} +
  \ps{\vs{1}}{\vci{q}{1}} \ps{\vs{2}}{\vc{k}\, } \Big]  \, ,
\label{jpsrsrmes}\\
 \jrcq{sep}{ret}{PS}{-} &=&  \Ff{e,1}{-}\, \conm{PS}{128}{4}
   \Big[ \vs{1} - \vc{q} \,
   \frac{\ps{\vs{1}}{\vci{q}{1}}}{\ps{\vc{k}}{\vc{q}\, }} \Big]\,
\nonumber\\
&&   \ps{\vs{2}}{\vci{q}{2}}\,\vcs{k}\, \ps{\vc{k}}{\vci{q}{2}}
   \propp{PS}{2} + \exnn \, ,
\label{jpsrsrret1}
\eea
where we have again separated the contributions with respect to their
propagator structure.

Finally, we list the interaction dependent boost contribution 
of the one-body current
\beq
  j_{\lambda,\chi_V} (2;\vc{k}\, )_{PS} =
-i \Big[ j_{\lambda,FW} (1;\vc{k}\, )  , \,  (\mut-1)\chi_V + \delta
\chi_V\Big] \, ,
\label{chivcom}
\eeq
where $\chi_V$ is given by
\bea
 \langle \vcpi{p}{1} \vcpi{p}{2} \, |\, \chi_V \, |
 \vci{p}{1} \vci{p}{2} \, \rangle & = & \tilde{\chi}_V (\vc{q},
\vc{K},  \vc{Q}) \ps{\vci{\tau}{1}}{\vci{\tau}{2}}\, \delta(\vci{q}{1} +
\vci{q}{2}\, ) \, ,
\label{chiv1}\\
  i \tilde{\chi}_V (\vc{q},\vc{K}, \vc{Q}) & = &
  \frac{g^2_{PS}}{(2\pi )^3 64 m^3}
  \Sigma^{(-)}(\vec q,\,\vec K)
\prop{PS}{}
\, ,\label{chiv2}
\eea
and $\delta \chi_V $ is the model-independent
interaction boost which vanishes for a two-particle system with equal
masses \cite{Fr75,Fr77}.

Evaluation of (\ref{chivcom}) gives for the boost contributions
\bea
 \rho_{\chi_V}^{(2)} (2;\vc{k}, \vec q,\vec K, \vec Q\, )_{PS}
 &=& (\mut-1) \conm{PS}{64}{3} \, \Big\{ \,
 \Ff{e,1}{+}\, \pf{\vc{k}}{\vci{q}{2}}{\vs{1}}{\vs{2}}  \nonumber\\
&&
  - \Ff{e,1}{-}\,
 \pf{\vc{K}}{\vci{q}{2}}{\vs{1}}{\vs{2}} \Big\} \,
  \prop{PS}{2} +\exnn \, ,  \label{rh2pschiv}\\
 \vec{\jmath}_{\chi_V}^{\,\,(3)} (2;\vc{k}, \vec q,\vec K, \vec Q\, )_{PS}
 &=& (\mut - 1) \conm{PS}{128}{4} \nonumber\\
&&
   \Big\{\, \Ff{e,1}{+}\, \Big[
 \vci{Q}{1}\,  \pf{\vc{k}}{\vci{q}{2}}{\vs{1}}{\vs{2}}
 + \vci{q}{2}\,  \pf{\vc{K}}{\vci{q}{2}}{\vs{1}}{\vs{2}} \Big]
  \nonumber\\
&&
 + \Gm{+}\, \vc{k} \times \Big[
 \vci{q}{2} \times \vs{1} \ps{\vs{2}}{\vc{K}}
 - \vc{K} \times \vs{1} \ps{\vs{2}}{\vci{q}{2}}
 + i \vci{q}{2} \ps{\vs{2}}{\vc{k}} \, \Big] \nonumber\\
&&
   - \Ff{e,1}{-}\, \Big[
 \vci{Q}{1}\,  \pf{\vc{K}}{\vci{q}{2}}{\vs{1}}{\vs{2}}
 + \vci{q}{2}\,  \pf{\vc{k}}{\vci{q}{2}}{\vs{1}}{\vs{2}} \Big]
 \nonumber\\
&&
 + \Gm{-}\, \vc{k} \times \Big[
 \vc{k} \times \vs{1} \ps{\vs{2}}{\vci{q}{2}} -
 \vci{q}{2} \times \vs{1} \ps{\vs{2}}{\vc{k}}
 \nonumber\\
&&+ i \vc{K} \ps{\vs{2}}{\vci{q}{2}} -
 i \vci{q}{2} \ps{\vs{2}}{\vc{K}} \, \Big] \Big\} \prop{PS}{2}
+\exnn \, . \label{j2pschiv}
\eea
In the Breit frame, these expressions yield then
\bea
\rhqQa{\chi_V}{pro}{PS}{+}
 &=& (\mut-1)  \Ff{e,1}{+}\,\conm{PS}{64}{3} \,
\pf{\vc{k}}{\vci{q}{2}}{\vs{1}}{\vs{2}}\prop{PS}{2}  \nonumber\\
&& +\exnn \label{rh2pschiv+}\, ,\\
 \jrcq{\chi_V}{pro}{PS}{+} &=&  (\mut -1) \conm{PS}{128}{4}\, \Big\{
\Ff{e,1}{+}\, \vc{Q}\,  \pf{\vc{k}}{\vci{q}{2}}{\vs{1}}{\vs{2}}
\nonumber\\
&& + i\Gm{+}\, \vc{k} \times
\vci{q}{2}\, \ps{\vs{2}}{\vc{k}} \, \Big\}\prop{PS}{2} +\exnn \,,
\label{jpschiv+}\\
 \jrcq{\chi_V}{pro}{PS}{-} &=&  -(\mut-1)  \conm{PS}{128}{4}\, \Big\{
 \Ff{e,1}{-}\, \vci{q}{2}\,
 \pf{\vs{1}}{\vs{2}}{\vc{k}}{\vci{q}{2}} \nonumber\\
&&
+ \Gm{-}\, \vc{k} \times \big[
 \vci{q}{2} \times \vs{1} \, \ps{\vs{2}}{\vc{k}\, } -
 \vc{k} \times \vs{1} \, \ps{\vs{2}}{\vci{q}{2}}\, \big] \Big\}\,
 \prop{PS}{2} \nonumber\\
&&+ \exnn \, .
\label{jpschiv-}
\eea

\subsection*{Alternative representation of \chir\ and separation currents} 

We would like to remind the reader that the separation currents 
were introduced because for the one-body currents a
useful cancellation occurs between them and the \chir-ones. 
Also for the exchange currents it is convenient to consider the 
following combination of the $\chir$, $sep$ and $recoil$ currents, 
\beq
 \vc{\jmath}_{\chi s r}^{\, \, (3)} (\rm{2};\vc{k}\, )=
 \vc{\jmath}_{\chir}^{\, \, (3)} (\rm{2};\vc{k}\, )
+ \vc{\jmath}_{sep}^{\, \, (3)} (\rm{2};\vc{k}\, )
- \vc{\jmath}_{rec}^{\, \, (3)} (\rm{2};\vc{k}\, )\,,
\eeq
where the recoil current is given by
\beq
 \vc{\jmath}^{\,\,(3)}_{rec} (\vc{k}) =  - \frac{\vc{k}}{32 m^2} \, \, 
\vec k \cdot \vc{\jmath}^{\,\, (1)} (\vc{k}) \, .
\label{jrec}
\eeq
Explicitly, one finds from (\ref{sepj3}), (\ref{chir}) and (\ref{jrec}) 
the following expression for this combination 
\bea
 \vc{\jmath}_{\chi s r}^{\, \, (3)} (\rm{2};\vc{k}\, )
 & = &
\frac{\vc{k}}{32m^2}\, \vc{k} \cdot
 \vc{\jmath}^{\, \, (1)} (2;\vc{k}\, )  +
 \frac{\vcs{k}}{16m^2}\, \vc{\jmath}^{\, \, (1)} (2;\vc{k}\, )
 \nonumber\\
&&
 + \frac{1}{32m^2}
 \Big[ \ps{\vc{k}}{\vci{q}{1}} \ps{\vc{k}}{\vnab{q_1}\,} +
  \ps{\vc{k}}{\vci{q}{2}} \ps{\vc{k}}{\vnab{q_2}\,} \Big]\,
  \vc{\jmath}^{\, \, (1)} (\rm{2};\vc{k}\, )  \, .
\label{jrsr}
\eea
For a ``mes''-type current, having the generic form
\beq
  \vc{\jmath}^{\, \, (1)} (\rm{2;mes};\vc{k}\, ) =
 (\vci{q}{1} - \vci{q}{2} )\, \fq \, g(\vci{q}{1}, \vci{q}{2}) \, ,
\label{jmesgen}
\eeq
the expression (\ref{jrsr}) becomes
\bea
 \vc{\jmath}_{\chi s r}^{\, \, (3)}
 (\rm{2};\vc{k}\, )
 & = &  \frac{\vc{k}}{16m^2}\,  \vc{k} \cdot
 \vc{\jmath}^{\, \, (1)} (\rm{2;mes};\vc{k}\, ) \nonumber\\
 && 
 + \frac{1}{16m^2}\, g(\vci{q}{1}, \vci{q}{2})\,
 \frac{\vc{q}}{\ps{\vc{k}}{\vc{q}\, }}\, \ps{\vc{k}}{\vci{q}{2}}^2\,
 \propp{B}{2} \nonumber\\
 && + 2\, \vc{q}\, \fq \,
 \Big[ \ps{\vc{k}}{\vci{q}{1}} \ps{\vc{k}}{\vnab{q_1}\,} +
  \ps{\vc{k}}{\vci{q}{2}} \ps{\vc{k}}{\vnab{q_2}\,} \Big]\,
   g(\vci{q}{1}, \vci{q}{2}) \, .
\label{jrsrmes}
\eea
For scalar and vector 
exchanges there is only a ``mes''-type nonrelativistic exchange current 
given in (\ref{j2smes1-}) where
$g(\vci{q}{1}, \vci{q}{2})$\ of (\ref{jmesgen}) is just a constant. One finds
without separating into ``pro'', ``mes'' and ``ret'' parts
\bea
 \vc{\jmath}_{\chi s r}^{\, \, (3)}
 (\rm{2};\vc{k},\,\vc{q},\,\vc{Q} )_{B}^{-} &=&  
  \frac{\vc{k}}{16m^2}\, \vc{k} \cdot \jnrq{mes}{B} \nonumber\\
 && \pm \Ff{e,1}{-} \, \conm{B}{16}{2} \,\vc{q}\,\,
\frac{\ps{\vc{k}}{\vci{q}{2}}^2}
{\ps{\vc{k}}{\vc{q}\, }} \propp{B}{2} +\exnn \, ,
\label{jsrsr}
\eea
where $B=S$ or $V$ and the minus sign applies for vector exchange. 

For pseudoscalar exchange one has to apply (\ref{jrsr}) and
(\ref{jrsrmes}) to the nonrelativistic currents (\ref{j2pspro1-}) and
(\ref{j2psmes1-}), respectively. Collecting all terms one finds 
\bea
 \vc{\jmath}_{\chi s r}^{\, \, (3)}
 (\rm{2};\vc{k},\,\vc{q},\,\vc{Q} )_{PS}^{-} &=&  
\jrcq{sep}{mes}{PS}{-} \nonumber\\
&&
+ \Ff{e,1}{-} \, \conm{PS}{64}{4} \Big[ \vs{1} -\ps{\vs{1}}{\vci{q}{1} } 
\frac{\vc{q}}{\ps{\vc{k}}{\vc{q}\, }}\Big] \ps{\vs{2}}{\vci{q}{2} } 
\ps{\vc{k}}{\vci{q}{2}}^2  \, \propp{PS}{2}\nonumber \\
&&
+ \Ff{e,1}{-} \, \conm{PS}{128}{4} \, \vc{D}  \prop{PS}{2} +\exnn \, ,
\label{jpsrsrpro}
\eea
where 
$\jrcq{sep}{mes}{PS}{-}$ is given in (\ref{jpsrsrmes}) and 
\bea
 \vc{D} &=& \vs{1} \Big[ - 2 \vcs{k} \ps{\vs{2}}{\vc{q}\, } + 
 \Big(\frac{3}{2} \vcs{k} - \ps{\vc{k}}{\vc{q}\, }\Big) 
\ps{\vs{2}}{\vc{k}} \Big]
 - 2 \vc{k} \, \ps{\vs{1}}{\vc{q}\, }  \ps{\vs{2}}{\vci{q}{2}} \nonumber\\
&&  - \vc{q}\, \Big[  \ps{\vs{1}}{\vc{k}\, }  \ps{\vs{2}}{\vci{q}{2}} -
  \ps{\vs{1}}{\vci{q}{1}}  \ps{\vs{2}}{\vc{k}\, } \Big] \label{D1}
\\
&=&  2 \vc{k}\, \ps{\vs{1}}{\vci{q}{2}}  \ps{\vs{2}}{\vci{q}{2}}
 + \vci{q}{2}\, \Sigma^{(+)}(\vc{k},\,\vci{q}{2}) + 
 \frac{1}{2} \vc{k} \, \Big[ \ps{\vs{1}}{\vc{k}\, }  \ps{\vs{2}}{\vci{q}{2}} 
 +  \ps{\vs{1}}{\vci{q}{1}}  \ps{\vs{2}}{\vc{k}\, } \Big] \nonumber\\
&& - \vc{k} \times 
 \Big[\,  2\, \vc{k} \times \vs{1} \, \ps{\vs{2}}{\vci{q}{2}} 
  +  \vci{q}{2} \times \vs{1} \, \ps{\vs{2}}{\vc{k}} \Big] \, ,\label{D2}
\eea
The first form of $\vec{D}$\ in (\ref{D1}) is more convenient for 
comparison with the results given earlier, whereas the second one in (\ref{D2})
is used in the appendix E in order to obtain the ``pro-IV'' current.

\renewcommand{\theequation}{C.\arabic{equation}}
\setcounter{equation}{0}
\section*{Appendix C: \mut-dependent currents}

Collecting the \mut-dependent terms from (\ref{rh2psvos}), (\ref{j2psvos}),
and (\ref{rh2pschiv+})-(\ref{jpschiv-}), we get for the total \mut-dependent
current
\bea
 \rho_{\mut } (2;\vc{k},\vc{q},\vc{Q}\, )_{PS}
 &=& \mut \conm{PS}{32}{3} \, \Big[ \,
 - \frac{1}{2} \Ff{e,1}{+}\, \Sigma^{(+)}(\vci{q}{2},\, \vec k)
+ \Ff{e,1}{-}\, \Sigma^{(+)}(\vci{q}{2},\, \vec Q) \Big]
 \,  \prop{PS}{2} \nonumber\\
&&+\exnn \, ,  \label{rh2psmut}\\
 \vec{\jmath}_{\mut } (2;\vc{k},\vc{q},\vc{K},\vc{Q}\,)_{PS}
 &=& \frac{\vci{Q}{1}}{2m} \rho_{\mut } (2;\vc{k},\vc{q},\vc{Q}\, )_{PS}
\nonumber\\
&&+\mut \conm{PS}{64}{4}    \Big\{\, \vci{q}{2}\,\Big(
\frac{1}{2} \Ff{e,1}{-}\,\Sigma^{(+)}(\vci{q}{2},\, \vec k)
- \Ff{e,1}{+} \Sigma^{(+)}(\vci{q}{2},\, \vec Q) \Big)
  \nonumber\\
&&
 + \Gm{+}\, \vc{k} \times \Big[
 \vci{q}{2} \times \vs{1} \ps{\vs{2}}{\vc{Q}}
 + \vc{Q} \times \vs{1} \ps{\vs{2}}{\vci{q}{2}}
 + \frac{i}{2} \vci{q}{2} \ps{\vs{2}}{\vc{k}} \, \Big] \nonumber\\
&&
 - \Gm{-}\, \vc{k} \times \Big[
 \frac{1}{2} \vci{q}{2} \times \vs{1} \ps{\vs{2}}{\vc{k}} +
 \frac{1}{2} \vc{k} \times \vs{1} \ps{\vs{2}}{\vci{q}{2}} +
  i \vci{q}{2} \ps{\vs{2}}{\vc{Q}}  \nonumber\\
&&+ i \vc{Q} \ps{\vs{2}}{\vci{q}{2}}
 \, \Big] \Big\} \prop{PS}{2} +\exnn \, . \label{j2psmut}
\eea

The charge density $\rho_{\mut } (2;\vc{k}\, )_{PS}$ is frame-independent
and adds to the intrinsic charge density operator.
The current $\vec{\jmath}_{\mut } (2;\vc{k},\vc{q},\vc{K},\vc{Q})_{PS}$\
depends on the reference frame, i.e. on \vc{K}, only through
$\vci{Q}{1} = \vc{Q} + \vc{K}/2$ in the first term of (\ref{j2psmut}).
Obviously,
\beq
 \vec{\jmath}_{\mut } (2;\vc{k},\vc{q},\vc{K},\vc{Q})_{PS} -
 \vec{\jmath}_{\mut } (2;\vc{k},\vc{q},\vc{0},\vc{Q})_{PS} =
 \frac{\vc{K}}{4m} \, \rho_{\mut } (2;\vc{k},\vc{q},\vc{Q}\, )_{PS} \, ,
\eeq
which is consistent with the general expression (\ref{fj3I}). Furthermore,
for the divergence of the current one easily finds from
(\ref{rh2psmut}) and (\ref{j2psmut})
\bea
 \vc{k} \cdot \vec{\jmath}_{\mut } (2;\vc{k},\vc{0})_{PS} &=&
 \frac{\ps{\vc{q}}{\vc{Q}}}{m} \rho_{\mut } (2;\vc{k}\, )_{PS} \nonumber\\
&&+ \mut \conm{PS}{32}{4}
\Big\{ - \frac{1}{2} \Ff{e,1}{+}\, \Big[
\ps{\vc{k}}{\vci{q}{2}} \Sigma^{(+)}(\vci{q}{2},\, \vec Q)
+ \ps{\vc{Q}}{\vci{q}{2}} \Sigma^{(+)}(\vci{q}{2},\, \vec k) \Big]
  \nonumber\\
&&
 + \Ff{e,1}{-}\, \Big[ \ps{\vc{Q}}{\vci{q}{2}} \Sigma^{(+)}(\vci{q}{2},\, \vec 
Q)
 + \frac{1}{4} \ps{\vc{k}}{\vci{q}{2}} \Sigma^{(+)}(\vci{q}{2},\, \vec k) 
\Big]\,
 \Big\}  \prop{PS}{2} +\exnn \nonumber\\
 &=& \langle \vcp{p}\, |\Big[t^{(1)}\,,\,\rho_{\mut } (2;\vc{k}\, )_{PS}
\Big]+ \Big[ v^{(3)}_{\mut } \, , \,
 \rho^{(0)} (1;\vc{k}\, ) \Big] \, | \vc{p} \, \rangle \, ,
\eea
where the \mut -dependent intrinsic potential $\tilde{v}^{(3)}_{\mut } (\vc
{q},\vc{Q})$ is given in (\ref{v3psmut}).

According to (\ref{voscom1}) and (\ref{chivcom}), the
\mut-dependent operators of (\ref{v3psmut}), (\ref{rh2psmut}) and
(\ref{j2psmut}) as well as the foregoing continuity equation
can be obtained from the nonrelativistic one-nucleon current and
the kinetic energy with the help of another (approximate)
unitary transformation $(1-i\mut U_{\mut })$ \cite{AGA}, where
\beq
U_{\mut }= U_{PS} -\chi_V\,.
\eeq
In detail, this gives for $U_{\mut }$
\bea
 \langle \vcpi{p}{1} \vcpi{p}{2} \, |\, U_{\mut} \, |
 \vci{p}{1} \vci{p}{2} \, \rangle & = &  \tilde{U}_{\mut} (\vc{q},
\vc{Q}) \ps{\vci{\tau}{1}}{\vci{\tau}{2}}\, \delta(\vci{q}{1} +
\vci{q}{2}\, ) \, ,\\
 i\tilde{U}_{\mut } (\vc{q},\vc{Q}) &=& - \conm{PS}{32}{3}\,
\Sigma^{(+)}(\vec q,\, \vec Q) \prop{PS}{}\,,
\eea
where we already have used the fact that there is no 
$\vec K$-dependence.
Also for more than two nucleons the difference 
$U_{\mut }= U_{PS} -\chi_V\ $\
gives the operator depending only on the relative coordinates
$\vci{p}{a} - \frac{m}{M} \vc{P}$. 

One can use any unitary transformation acting in the intrinsic
space only 
with the generator $U = {\cal O}(1/m^2)$ in two ways. First, one
can apply it only to the intrinsic Hamiltonian and currents. The transformed
operators still obey the intrinsic continuity equation. 
Then, inserting these into the full operators
(\ref{frh0I})-(\ref{fj3I})  and inspecting the $\vc{K}$-dependent 
terms, one finds that only the convection current
$(\vc{K}/2M)\,\rho^{(2)}(\vc{k}\, )$\ is affected by the unitary
transformation resulting in an additional contribution 
$(\vc{K}/2M)\,\rho^{(2)}_U (2;\vc{k}\, )$ with
\beq
 \rho^{(2)}_U (2;\vc{k}\, ) = i\Big[\, \rho^{(0)} (1;\vc{k}\, )\, , \, U\,
\Big] \, .
\label{intunit}
\eeq
Since the general parametrization
(\ref{frh0I})-(\ref{fj3I}) is preserved, the full current is still 
conserved and 
transformes properly under the Poincar\'{e} transformation.
Alternatively, one may let act the transformation in the full
Hilbert space including the c.m.\ motion part. That is, one transforms
also the nonrelativistic convection current 
$(\vc{K}/2M)\,\rho^{(0)}(\vc{k}\, )$\  and gets the correction to 
its relativistic part as given by (\ref{intunit}). Both views are
equivalent and consistent, and it implies that it is sufficient to 
transform the conserved intrinsic current only. 
Another example of such a unitary transformation is encountered 
in the next appendix.

\renewcommand{\theequation}{D.\arabic{equation}}
\setcounter{equation}{0}
\section*{Appendix D: Retardation currents for $\nu \neq 1/2 $ }

In the framework of the $1/m$-expansion techniques \cite{Fr80}, the effects
of retardation of the exchanged mesons are included via the Taylor
expansion of the meson propagators (see also (\ref{propp}) and 
(\ref{ffret}))
\beq
 \tilde{\Delta}_B (\vcs{q} - q_0^2 ) \simeq
 \prop{B}{} - q_0^2\, \propp{B}{} \, ,
\eeq
and expressing the meson energy in terms of the energies of the
on-mass-shell nucleons. This procedure is unambiguous for the nonrelativistic
reduction of the Feynman amplitudes corresponding to the MEC operators,
but in order to define the general nuclear potential one has to allow for
an
off-energy-shell continuation of the corresponding amplitude, and then
$q_0$  is no longer fixed by energy conservation at the vertex. It has been
argued
in \cite{ATA} that up to the order considered the most general expression
symmetric with respect to nucleon interchange reads
\beq
 q_0^2  \rightarrow  - q_{10}\,  q_{20} + \frac{1-\nu}{2}
 (q_{10} + q_{20} )^2 =
\mpw{4}{2} \Big[ \ps{\vc{q}}{\vc{P}}^2 - 2 \nut \ps{\vc{q}}{\vc{Q}}^2
 \Big] +\ord{-4} \, ,
\eeq
where $\nu$\ is an arbitrary parameter and $\tilde \nu = \nu - 1/2$.
In \cite{ATA} it is described in detail
how this freedom translates into the $\nu$-dependent retardation
contribution to the MECs from the positive-energy nuclear pole Born diagram.

It is convenient to introduce the following notation
\bea
 \wt{S}{} &=& - \frac{g_S^2 }{(2\pi )^3} \, \propp{S} \, , \\
 \wt{V}{} &=&  \frac{g_V^2 }{(2\pi )^3} \, \propp{V} \, , \\
 \wt{PS}{} &=& - \conm{PS}{4}{2} \ps{\vs{1}}{\vc{q}\, }
 \ps{\vs{2}}{\vc{q}\, } \, \propp{PS} \, .
\eea
The retardation contribution to the potential in an arbitrary frame
is then given by
\beq
 \tilde{v}_{B,ret}^{\,(3)} (\vc{q},\vc{Q},\vc{P}) = -
 \mpw{4}{2} \wt{B}{} \ \Big[ \ps{\vc{q}}{\vc{P}}^2
- 2 \nut \ps{\vc{q}}{\vc{Q}}^2 \Big] \, ,\label{potretnu}
\eeq
where the first \vc{P}-dependent term is required by the Foldy condition
and the second one is the retardation contribution to the intrinsic potential
\beq
 \tilde{v}_{B,\nut}^{(3)} (\vc{q},\vc{Q}) =
  \frac{\nut}{2m^2} \wt{B}{} \  \ps{\vc{q}}{\vc{Q}}^2 \, .
\label{vnut}
\eeq
Obviously, (\ref{vnut}) vanishes for $\nut = 0$, i.e. $\nu = 1/2$, which is
the choice for the intrinsic potentials and the retardation e.m.\ operators
adopted in the main part of the paper. It is also common for
the construction of realistic $NN$-potentials which usually do not
include the retardation terms in the c.m.\ frame.
In this appendix we present for completeness the currents for arbitrary
\nut.

Since the expressions for the retardation operators ATA-(4.21-4.24)
contain a number of misprints and do not contain the strong form factors,
we find it useful to present the correct form here
\bea
 \rhFqQ{ret}{B}{} &=& - \mpw{4}{}\, \wt{B}{2}\,
 \Big( \Ff{e,1}{+} R_k -  \Ff{e,1}{-} R_Q \Big) + \exnn \, ,
\label{rhret}\\
 \jrcFq{ret}{B}{} &=& - \mpw{8}{2}\, \wt{B}{2}\,
 \Big( \Ff{e,1}{+} [R_k \vci{Q}{1} + R_Q \vci{q}{2} ]  \nonumber\\
&&   -  \Ff{e,1}{-} [R_Q \vci{Q}{1} + R_k \vci{q}{2} ] \Big)
 + \exnn +  \jrcFq{ret-tr}{B}{}\nonumber\\
&=& 
\jrcFq{ret-tr}{B}{} + 
\frac{\vci{Q}{1}}{2m}\rhFqQ{ret}{B}{} 
\nonumber\\
&&
- \mpw{8}{2}\, \wt{B}{2}\, \vec q_2\,
\Big( \Ff{e,1}{+} R_Q -  \Ff{e,1}{-} R_k \Big) + \exnn 
\, ,
\label{jretB}\\
 \jrcFq{ret-tr}{S,V}{} &=&  - \mpw{8}{2}\, \wt{S,V}{2}\,
 \Big( \Gm{+} R_k - \Gm{-} R_Q \Big) i \pv{\vs{1}}{\vc{k}} + \exnn \, ,
\label{jretSV}\\
  \jrcFq{ret-tr}{PS}{} &=& - \conm{PS}{32}{4}\, i \vc{k} \times
 \Big\{ \Gm{+} [ R_k \vci{q}{2} + R_Q i \vs{1} \times \vci{q}{2} ]
  \nonumber\\
&&  - \Gm{-} [ R_Q \vci{q}{2} + R_k i \vs{1} \times \vci{q}{2} ]
 \Big\}\, \ps{\vs{2}}{\vci{q}{2}} \propp{PS}{2} + \exnn \, ,
\eea
with
\bea
 R_k &=& (\frac{1}{2} - \nut) \ps{\vc{k}}{\vci{q}{2}} \, , \\
 R_Q &=& \frac{1}{2} (\vci{Q}{1} + 3 \vci{Q}{2}) \cdot \vci{q}{2} -
         \nut \, (\vci{Q}{1} - \vci{Q}{2}) \cdot \vci{q}{2} \, .
\label{jret}
\eea
The \nut-independent part of these operators is included in the retardation
currents in Appendix A and then added to the intrinsic currents in the
main part of the paper. Separating the \nut-dependent part we get
\bea
 \rho_{\nut }^{(2)} (\mbox{2;ret};\vc{k}\, )_{B}
 &=& \frac{\nut}{4m} \wt{B}{2} \, \Big[
    \Ff{e,1}{+}\, \ps{\vc{k}}{\vci{q}{2}}
 - 2 \Ff{e,1}{-} \ps{\vc{Q}}{\vci{q}{2}} \Big]
   +\exnn \, ,  \label{rh2nut}\\
 \vec{\jmath}_{\nut }^{\,\,(3)} (\mbox{2;ret};\vc{k},\vc{K})_{B}
 &=& 
 \vec{\jmath}_{\nut } (\mbox{2;ret-tr};\vc{k}\, )_{B} +
\frac{\vci{Q}{1}}{2m}\rho_{\nut }^{(2)} (\mbox{2;ret};\vc{k}\, )_{B}
\nonumber\\
&&
+\frac{\nut}{8m^2} \wt{B}{2} \,\vec q_2\, \Big(
 2 \Ff{e,1}{+}\,\ps{\vc{Q}}{\vci{q}{2}} 
 - \Ff{e,1}{-}\,\ps{\vc{k}}{\vci{q}{2}} \Big)
+ \exnn \, ,
 \label{j2nut}\\
  \vec{\jmath}_{\nut }^{\,\,(3)} (\mbox{2;ret-tr};\vc{k}\, )_{S,V} &=&
  \frac{\nut}{8m^2} \wt{S,V}{2} \, \Big[ \Gm{+}\,  \ps{\vc{k}}{\vci{q}{2}} -
  2\Gm{-} \ps{\vc{Q}}{\vci{q}{2}} \Big] \, i \vs{1} \times \vc{k}
\nonumber\\
&&
+ \exnn \, ,  \label{j2btrnut}\\
  \vec{\jmath}_{\nut }^{\,\,(3)} (\mbox{2;ret-tr};\vc{k}\, )_{PS} &=&
 \nut\,  \conm{PS}{32}{4}\, i \vc{k} \times
 \Big\{ \Gm{+} \Big[ \vci{q}{2}\,  \ps{\vc{k}}{\vci{q}{2}} +
  2 i \vs{1} \times \vci{q}{2} \ps{\vc{Q}}{\vci{q}{2}}  \Big]
  \nonumber\\
&&
  - \Gm{-} \Big[ 2\vci{q}{2}\, \ps{\vc{Q}}{\vci{q}{2}}  +
  i \vs{1} \times \vci{q}{2}\,  \ps{\vc{k}}{\vci{q}{2}} \Big]
 \Big\}\, \ps{\vs{2}}{\vci{q}{2}} \propp{PS}{2} \nonumber\\
&&+ \exnn \, .
\label{j2pstrnut}
\eea
It is easy to show that
\beq
 \vc{k} \cdot  \vec{\jmath}_{\nut }^{\,\,(3)} (\mbox{2;ret};\vc{k},\vc{0})_{B}
 = \langle \vcp{p}\, |\Big[ t^{(1)},\, \rho_{\nut }^{(2)}
(\mbox{2;ret};\vc{k}\,)_{B} \Big]+ \Big[ v^{(3)}_{B,\nut } \, , \,
 \rho^{(0)} (1;\vc{k}\, ) \Big]| \vc{p} \, \rangle \, ,
\eeq
because one finds for the commutator
\bea
\langle \vcp{p}\, |\Big[ \tilde v^{(3)}_{B,\nut } \, , \,
 \rho^{(0)} (1;\vc{k}\, ) \Big] | \vc{p} \, \rangle
&=&  \frac{\nut}{2m^2} \wt{B}{2} \Big( \Ff{e,1}{+}
 \ps{\vc{k}}{\vci{q}{2}} \ps{\vc{Q}}{\vci{q}{2}} - \Ff{e,1}{-}
( \ps{\vc{Q}}{\vci{q}{2}}^2 +  \frac{1}{4}  \ps{\vc{k}}{\vci{q}{2}}^2) \Big)
\nonumber\\
&& + \exnn
\eea
One can also check that, similar to the \mut-dependent ones discussed at
the end of the previous appendix,
 all \nut -dependent operators can be generated with
the help of a unitary transformation $(1-i  U_{B,\nut})$ from the
nonrelativistic one-nucleon currents and the kinetic energy
\bea
  j_{\lambda,\nut}^{(3)} (2;\vc{k}\, )_{B} &=&
i \Big[ j_{\lambda,FW}^{(1)} (1;\vc{k}\, )  \, , \,  U_{B,\nut} \Big] \, , \\
  \tilde{v}_{B,\nut}^{(3)} &=&   i\Big[ \frac{\vcs{p}}{m} \, , \,  U_{B,\nut}
\Big]
\, ,
\eea
where
\bea
 i \tilde{U}_{B,\nut} (\vc{q},\vc{Q}) &=& \tilde \nu
  \frac{\ps{\vc{q}}{\vc{Q}}}{2m} \, \wt{B}{} \nonumber\\
&=& \frac{\tilde \nu}{2}
\langle \vcp{p}\, |\Big[ t^{(1)},\, \wt{B}{}
\Big]| \vc{p} \, \rangle \, .
\eea

\renewcommand{\theequation}{E.\arabic{equation}}
\setcounter{equation}{0}
\section*{Appendix E: Continuity equation for the \Fmn\ currents
for pseudoscalar exchange}

In this appendix we want to identify the various pieces of the ``pro'' 
and ``mes'' currents corresponding to the various commutators in 
(\ref{cont23}). To this end 
we split the total ``pro'' and ``mes'' currents into several
parts (labelled by I, II,..) which will be specified in the following
\bea
 \jrcq{F}{pro}{PS}{-} &=& \sum_{i= I}^{V} \jrcq{F}{pro-i}{PS}{-} \, , 
\\
 \jrcq{F}{mes}{PS}{-} &=& \sum_{i= I}^{IV} \jrcq{F}{mes-i}{PS}{-} \, .
\eea
As first current, we single out the purely transverse $G_M^-$ term
since it drops out from the continuity equation 
\bea
  \jrcq{F}{pro-V}{PS}{-} &=&   \Gm{-} \,  \conm{PS}{128}{4}\,
  \vc{k} \times \Big[
 \Big( 6i\, \vc{Q} + 5\, \vc{k} \times \vs{1} \Big) \ps{\vs{2}}{\vci{q}{2}} 
\nonumber\\
&& + \vci{q}{2} \times \vs{1}\, \ps{\vs{2}}{\vc{k}\, } 
 - 2i \vci{q}{2}\,  \ps{\vs{2}}{\vc{Q}} \, \Big]
 \Big\}\prop{PS}{2} + \exnn \, .
\label{intjpsproIV-}  
\eea
Now we start with 
the first commutator in (\ref{cont23}) which has two contributions 
from the ``pro'' and ``mes'' parts of $\rho_F^{(2)}(2;\vec k\,)$, 
namely
\bea
 \langle \vcp{p} \, | \Big[ t^{(1)}, \, \rho^{(2)}_{F} (2;\mbox{pro};\vc{k}\, )
\Big]^- | \vc{p} \, \rangle&=&\Ff{e,1}{-}\conm{PS}{32}{4}\ps{\vec q\,}{\vec Q}
\,\Sigma^{(+)}(\vec q_2, \,\vec Q)
\prop{PS}{2}
\nonumber\\ 
&&
 +\exnn \label{cpst1pro}
\eea
and
\beq 
 \langle \vcp{p} \, | \Big[ t^{(1)}, \, \rho^{(2)}_{F} (2;\mbox{mes};\vc{k}\, )
 \Big]^- | \vc{p} \, \rangle = -\Ff{e,1}{-}\conm{PS}{8}{4}
 \ps{\vec k}{\vec Q}\ps{\vec q}{\vec Q}
 \ps{\vs{1}}{\vci{q}{1}}  \ps{\vs{2}}{\vci{q}{2}} \fq \label{cpst1mes}
\,.
\eeq
Their current counterparts are easy to identify, namely 
\bea
 \jrcq{F}{pro-I}{PS}{-} &=&  
  - \Ff{e,1}{-}\, \conm{PS}{64}{4}\, \Big\{ \vc{Q}\, 
   \Sigma^{(-)} ( \vc{Q}, \vci{q}{2} ) \, +
   2\, \vs{1}\, \ps{\vc{Q}}{\vci{q}{2}} \ps{\vs{2}}{\vc{Q}} \, \Big\}\,   
 \nonumber\\ 
 &&  \prop{PS}{2} + \exnn \, ,  
\label{intjpsproII-}\\  
\jrcq{F}{mes-I}{PS}{-} &=&  \Ff{e,1}{-}\, \conm{PS}{16}{4}\, \vc{q}\, \fq 
 \nonumber\\
 &&
\Big\{ \ps{\vci{q}{2}}{\vc{Q}} \ps{\vs{1}}{\vci{q}{1}}  \ps{\vs{2}}{\vc{Q}}
 + \ps{\vci{q}{1}}{\vc{Q}} \ps{\vs{1}}{\vc{Q}}  \ps{\vs{2}}{\vci{q}{2}} 
\Big\} \, ,
\label{intjpsmesII-} 
\eea 
with the divergence
\beq
\vc{k} \cdot\Big( \jrcq{F}{pro-I}{PS}{-}+ \jrcq{F}{mes-I}{PS}{-}\Big)=
 \langle \vcp{p} \, | \Big[ t^{(1)}, \, \rho^{(2)}_{F} (2;\mbox{pro};\vc{k}\, )
 \Big]^- | \vc{p} \, \rangle\,,
\eeq
and
\beq
\jrcq{F}{mes-II}{PS}{-} =  - \Ff{e,1}{-}\, \conm{PS}{8}{4}\,  \fq\
   \frac{\vc{q}}{\ps{\vc{k}}{\vc{q}\, }}\, \, 
  \ps{\vc{k}}{\vc{Q}} \ps{\vc{q}}{\vc{Q}} 
   \ps{\vs{1}}{\vci{q}{1}} \ps{\vs{2}}{\vci{q}{2}} \, ,
\label{intjpsmesIII-} 
\eeq
with
\beq
 \vc{k} \cdot \jrcq{F}{mes-II}{PS}{-} =   
 \langle \vcp{p} \, | \Big[ t^{(1)}, \, \rho^{(2)}_{F} (2;\mbox{mes};\vc{k}\, )
 \Big]^- | \vc{p} \, \rangle\,. 
\label{conpsmesIII} 
\eeq
Furthermore, the divergence of the ``recoil'' current
\beq
  \jrcq{F}{pro-II}{PS}{-} =
   - \Ff{e,1}{-}\, \conm{PS}{128}{4}\, \vc{k}\, \, 
   \ps{\vs{1}}{\vci{q}{2}} \ps{\vs{2}}{\vci{q}{2}} \,
   \prop{PS}{2} + \exnn \, , 
\label{intjpsproIII-} \\    
\eeq
yields the recoil contribution to (\ref{cont23}), i.e.\ 
\beq
\vc{k} \cdot \jrcq{F}{pro-II}{PS}{-} = 
 -  \frac{\vcs{k}}{32m^2}\,  
\langle \vcp{p} \, | \Big[ v^{(1)}_{PS}
  , \, \rho^{(0)}_{F} (1;\vc{k}\, )\Big]^- | \vc{p} \,  \rangle \, .
\label{conpsproIII} 
\eeq
Of the two remaining commutators of (\ref{cont23}) we will first 
consider the commutator of the nonrelativistic potential with the 
relativistic charge density $\rho^{(2)}_{F,e} (1;\vc{k}\, )$ 
\bea
 \langle \vcp{p} \, | \Big[ v^{(1)}_{PS} \, , \, \rho^{(2)}_{F,e} (1;\vc{k}\, )
\Big]^- | \vc{p} \, \rangle &=&
 - \Ff{e,1}{-} \conm{PS}{64}{4}\, \Big\{ i \pth{\vc{k}}{\vci{q}{2}}{\vc{Q}}
 (\vs{1} + \vs{2}) \cdot \vci{q}{2}  
+ \vcsi{q}{2} \Sigma^{(+)}(\vc{k},\vci{q}{2})
\nonumber\\
&&
  + 2  \ps{\vc{k}}{\vci{q}{1}}  \ps{\vs{1}}{\vci{q}{2}}  \ps{\vs{2}}{\vci{q}{2}}
 \Big\} \prop{PS}{2} +\exnn \, . \label{conpsIa}
\eea
Recall that the current satisfying the continuity equation with the
similar commutator of the $\kappa$-part 
$[ v^{(1)}_{PS} \, , \, \rho^{(2)}_{F,\kappa} (1;\vc{k}\, ) ] $\ has been
absorbed into the single-nucleon current, as discussed in the Sect.\ IIIA.
Explicitly, it is given in (ATA-4.17b). Note that this current does 
not change in the transition from the FW currents to the intrinsic ones, since 
$ \rho^{(2)}_{FW,\kappa} (1;\vc{k}\, ) = \rho^{(2)}_{F,\kappa} (1;\vc{k}\, )$. 
We can get the corresponding FW current proportional to \eh\ 
replacing $ \Ff{\kappa,1}{-} \rightarrow  \Ff{e,1}{-}/2$\ in  (ATA-4.17b) 
giving only a ``pro'' contribution
\bea
 \jrcq{FW,e}{pro}{PS}{-} &=&  -  \Ff{e,1}{-}\, \conm{PS}{32}{4}\, 
\Big[ \vci{q}{1}\, \ps{\vs{1}}{\vci{q}{2}}  \nonumber\\
&& 
+ \vs{1} \, \vcsi{q}{2} + i \vci{q}{2} \times \vc{Q} \Big] \,
 \ps{\vs{2}}{\vci{q}{2}}\, \prop{PS}{2} + \exnn \, .
\label{jpsexte}  
\eea
But since $\rho^{(2)}_{F,e} (1;\vc{k}\, ) = \rho^{(2)}_{FW,e} (1;\vc{k}\, )
+ \rho^{(2)}_{\chis} (1;\vc{k}\, )$\ one has to add to (\ref{jpsexte}) 
in this case the \chis\ currents from (\ref{jpschispro}) and 
(\ref{jpschismes}) in order to obtain 
the intrinsic ``pro'' and ``mes'' currents saturating (\ref{conpsIa})
\bea
 \jrcq{F}{pro-III}{PS}{-} &=& \jrcq{FW,e}{pro}{PS}{-}  +
                          \jrcq{\chis}{pro}{PS}{-} \, , \nonumber\\        
 &=& - \Ff{e,1}{-} \, 
 \conm{PS}{64}{4} \, 
\Big\{\vs{1} \Big[ (2 \vcsi{q}{2} - \ps{\vc{k}}{\vci{q}{2}} )
   \ps{\vs{2}}{\vci{q}{2}} + \vcsi{q}{2} \ps{\vs{2}}{\vc{k}\, } \nonumber\\
&&
- i \vc{k} \times \vc{q} \cdot \vc{Q}\, \Big] 
 + \vc{q}\, \Big[ \, 2 \ps{\vs{1}}{\vci{q}{2}} \ps{\vs{2}}{\vci{q}{2}} +
  \ps{\vs{1}}{\vci{q}{1}} \ps{\vs{2}}{\vc{k}\, } \Big] \nonumber\\
 &&
  + \Big[ \vc{k}\, \vs{1} \cdot ( 2 \vci{q}{2} - \frac{1}{2} \vc{k}\, )   
 + 2i (\vc{k} - \vc{q}\, ) \times \vc{Q} \, \Big] \ps{\vs{2}}{\vci{q}{2}} 
\Big\}
\, \prop{PS}{2} \nonumber\\
&&+\exnn \label{jpsproIc} \\
&=& - \Ff{e,1}{-} \, 
 \conm{PS}{64}{4} \, 
\Big\{\vs{1} \Big[ 2 \ps{\vc{q}}{\vci{q}{2}} \ps{\vs{2}}{\vc{q}\, }
 + ( \vcsi{q}{2}  - \ps{\vc{q}}{\vci{q}{2}} ) \ps{\vs{2}}{\vc{k}\, }
\nonumber\\
&&  - i \vc{k} \times \vc{q} \cdot \vc{Q}\,  \Big]
+ 2 \vci{q}{1} \, \ps{\vs{1}}{\vci{q}{2}} \ps{\vs{2}}{\vci{q}{2}} +
  \vc{q}\, \ps{\vs{1}}{\vci{q}{1}} \ps{\vs{2}}{\vc{k}\, } \nonumber\\
 && - \vc{k}\, \ps{\vs{1}}{\vc{q}\, } \ps{\vs{2}}{\vci{q}{2}}
 + 2i (\vc{k} - \vc{q}\, ) \times \vc{Q} \, \Big] \ps{\vs{2}}{\vci{q}{2}} 
\Big\}
\, \prop{PS}{2}\nonumber\\
&&
 +\exnn \, ,
\label{jpsproIa} 
\\
\jrcq{F}{mes-III}{PS}{-} &=&  \Ff{e,1}{-}\, \conm{PS}{32}{4}\, \vc{q}\, \fq\
 \nonumber\\
 && 
\Big\{ \vcs{q}\, \ps{\vs{1}}{\vc{k}\, } \ps{\vs{2}}{\vc{k}\, } +
 \vcs{k}\, \ps{\vs{1}}{\vc{q}\, } \ps{\vs{2}}{\vc{q}\, } \nonumber\\
&&
+ \vc{k} \times \vc{q} \cdot 
 \Big[ (\vs{1} \times \vs{2}\, ) \ps{\vci{q}{1}}{\vci{q}{2}} 
 - i \vc{Q}\, (\ps{\vs{1}}{\vci{q}{1}} + \ps{\vs{2}}{\vci{q}{2}} )
 \Big] \Big\} \, . 
\label{jpsmesIa}  
\eea 
The first form (\ref{jpsproIc}) is convenient for reconstructing 
the total ``pro'' current, the second one (\ref{jpsproIa}) for 
verifying the continuity equation.

Finally, the currents belonging to the remaining commutator
of (\ref{cont23}) 
\bea
 \langle \vcp{p} \, | \Big[ v^{(3)}_{PS} , \, 
 \rho^{(0)}_{F} (1;\vc{k}\, )\Big]^- | \vc{p} \, \rangle &=&
 - \Ff{e,1}{-} \conm{PS}{32}{4}\,
\Big[ 2 \vcs{Q} + 2 \vcsi{q}{2} + \frac{1}{2} \vcs{k} \Big]
\nonumber\\
&&
 \ps{\vs{1}}{\vci{q}{2}}  \ps{\vs{2}}{\vci{q}{2}} \prop{PS}{2}
 +\exnn \, , \label{conpsIb}
\eea
are made up from the remaining pieces of the total ``pro'' and 
``mes'' currents that have not yet been singled out. 
Let us write them in the form
\bea
\jrcq{F}{pro-IV}{PS}{-} &=& - \Ff{e,1}{-} \, 
 \conm{PS}{32}{4} \, \Big\{ \Big[\,  2 \, \vs{1}\, ( \vcsi{q}{2} + \vcs{Q} ) 
 + \frac{1}{2} \vc{k} \, \ps{\vs{1}}{\vci{q}{2}} 
 + 2 \vc{q}\, \ps{\vs{1}}{\vci{q}{1}} \nonumber\\
 &&
+ i \vc{k} \times \vc{Q} \Big]
  \ps{\vs{2}}{\vci{q}{2}} 
  - \frac{1}{8} \vc{k}\, 
  \Big[ \ps{\vs{1}}{\vc{k}\, } \ps{\vs{2}}{\vci{q}{2}}\, +
   \ps{\vs{1}}{\vci{q}{1}} \ps{\vs{2}}{\vc{k}\, }\, \Big] \nonumber\\
 && 
  + \frac{1}{2} \vc{k} \times \Big[ \,  
   \vc{k} \times \vs{1} \ps{\vs{2}}{\vci{q}{2}}\,
   + \frac{1}{2} \vci{q}{2} \times \vs{1} \ps{\vs{2}}{\vc{k}\, }\, \Big]
   \Big\} \nonumber\\
 &&
 \prop{PS}{2}  +\exnn \label{jpsproId}\\
&=& - \Ff{e,1}{-} \, 
 \conm{PS}{32}{4} \, \Big\{ \, \vs{1}\, 
 \Big[\,  ( 2 \vcsi{q}{2} + 2 \vcs{Q} - \frac{1}{2} \vcs{k} ) 
 \ps{\vs{2}}{\vci{q}{2}}  \nonumber\\
&&  
  - \frac{1}{4} \ps{\vc{k}}{\vci{q}{2}} \ps{\vs{2}}{\vc{k}\, } \, \Big]
    + i \vc{k} \times \vc{Q} \ps{\vs{2}}{\vci{q}{2}} 
 \nonumber\\
 &&
  + \vc{q}\, \Big[\, 2 \ps{\vs{1}}{\vci{q}{1}} \ps{\vs{2}}{\vci{q}{2}}
  -  \frac{1}{4} \ps{\vs{1}}{\vc{k}\, } \ps{\vs{2}}{\vc{k}\, }\, \Big]  
    \nonumber\\
 && 
 +  \vc{k} \, \Big[\,    
 \frac{1}{2} \ps{\vs{1}}{\vci{q}{2}} \ps{\vs{2}}{\vci{q}{2}}\, +
  \frac{3}{8} \ps{\vs{1}}{\vc{k}\, } \ps{\vs{2}}{\vci{q}{2}}
  \nonumber\\
  &&
   + \frac{1}{8} \ps{\vs{1}}{\vci{q}{2}} \ps{\vs{2}}{\vc{k}\, }\, 
   \Big]\, \Big\} \prop{PS}{2}  +\exnn \, ,   
\label{jpsproIb}\\
\jrcq{F}{mes-IV}{PS}{-} &=& \Ff{e,1}{-} \, 
 \conm{PS}{32}{4} \, \vc{q}\, \fq 
 \Big\{ \Big[ 4 \vcs{q} + 4 \vcs{Q} + \vcs{k}\, \Big]
  \ps{\vs{1}}{\vci{q}{1}}  \ps{\vs{2}}{\vci{q}{2}} \nonumber\\
&&  - \frac{\vcs{k}}{4} \Big[  
  \ps{\vs{1}}{\vc{k}\, } \ps{\vs{2}}{\vci{q}{2}} +
  \ps{\vs{1}}{\vci{q}{1}} \ps{\vs{2}}{\vc{k}\, } \Big] \Big\} \, . 
\label{jpsmesIb}
\eea 
Again, the first form (\ref{jpsproId}) is more convenient for 
checking the continuity equation, the second one (\ref{jpsproIb}) 
for constructing the the total current of (\ref{intjpspro-}).

\begin{table}
\caption{Survey on the contributons to the two-body charge density
$\rho (2;\vec k\,)$ with a listing of the corresponding equations.
}
\begin{center}
\begin{tabular}{cccccc}
meson & type & {$\rho^{(2)\,+}_{FW}$} & {$\rho^{(2)\,+}_{\chi_V}$} &
{$\rho^{(2)\,+}_{F}$} &
{$\rho^{(2)\,-}_{F}$} \\
\hline
S  & mes &  &  &  &
(\ref{intrhsmes-}) \\
S  & ret & (\ref{rh2sret+}) &  & (\ref{intrhsret+}) &
(\ref{intrhsret-}) \\
\hline
V  & mes &  &  &  &
(\ref{intrhsmes-}) \\
V  & ret & (\ref{rh2sret+}) &  & (\ref{intrhsret+}) &
(\ref{intrhsret-}) \\
V  & mes-tr &  &  &  & (\ref{intrhvmestr-}) \\
\hline
PS & pro & (\ref{rh2pspro+}) & (\ref{rh2pschiv+}) & (\ref{intrhpspro+}) &
(\ref{intrhpspro-}) \\
PS & mes &  &  &  &
(\ref{intrhpsmes-}) \\
PS & ret & (\ref{rh2psret+}) &  & (\ref{intrhpsret+}) &
(\ref{intrhpsret-})\\
\end{tabular}
\end{center}
\label{tab1}
\end{table}

\begin{table}
\caption{Survey on the contributions to the
two-body current density $\vec\jmath\,(2;\vec k\,)$ with a
listing of the corresponding equations.
}
\begin{center}
\begin{tabular}{cccccccccccc}
 meson & type &
{$\vec \jmath^{\,\,(1)\,-}_{F}$} &
{$\vec \jmath^{\,\,(3)\,+}_{FW}$} &
{$\vec \jmath^{\,\,(3)\,+}_{\chi_V}$} &
{$\vec \jmath^{\,\,(3)\,+}_{F}$} &
{$\vec \jmath^{\,\,(3)\,-}_{FW}$} &
{$\vec \jmath^{\,\,(3)\,-}_{sep}$} &
{$\vec \jmath^{\,\,(3)\,-}_{\chi_r}$} &
{$\vec \jmath^{\,\,(3)\,-}_{\chi_\sigma}$} &
{$\vec \jmath^{\,\,(3)\,-}_{\chi_V}$} &
{$\vec \jmath^{\,\,(3)\,-}_{F}$}
\\
\hline
S  & pro &  & (\ref{j2spro+}) &  & (\ref{intjspro+}) & (\ref{j2spro-}) &  &
 (\ref{jschrpro}) &
 &  & (\ref{intjspro-}) \\
S  & mes & (\ref{j2smes1-}) &  &  &  & (\ref{j2smes-}) &  &  & (\ref{jschs}) &
         & (\ref{intjsmes3-}) \\
S  & ret &  & (\ref{j2sret+}) &  & (\ref{intjsret+}) & (\ref{j2sret-}) &
(\ref{jssep}) & (\ref{jschrret}) &  &  & (\ref{intjsret-}) \\
\hline
V  & pro &  & (\ref{j2vpro+}) &  & (\ref{intjvpro+}) & (\ref{j2vpro-}) &  &
 (\ref{jschrpro}) &
 &  & (\ref{intjvpro-}) \\
V  & mes & (\ref{j2smes1-}) &  &  &  & (\ref{j2vmes-}) &  &  & (\ref{jschs}) &
         & (\ref{intjvmes-}) \\
V  & ret &  & (\ref{j2sret+}) &  & (\ref{intjsret+}) & (\ref{j2sret-}) &
(\ref{jssep}) & (\ref{jschrret}) &  &  & (\ref{intjsret-}) \\
V  & mes-tr &  &  &  &  &  (\ref{j2vmestr-}) &  &  &  &  &
(\ref{intjvmestr-}) \\
\hline
PS & pro & (\ref{j2pspro1-}) & (\ref{j2pspro+}) & (\ref{jpschiv+}) &
(\ref{intjpspro+}) &
(\ref{j2pspro3-}) & (\ref{jpsrsrpro1}) & (\ref{jpschirrpro}) &
(\ref{jpschispro}) & (\ref{jpschiv-}) &
(\ref{intjpspro-}) \\
PS & mes & (\ref{j2psmes1-}) &  &  &  & (\ref{j2psmes3-}) & (\ref{jpsrsrmes}) &
  & (\ref{jpschismes}) &  & (\ref{intjpsmes-}) \\
PS & ret &  & (\ref{j2psret+}) &  & (\ref{intjpsret+}) & (\ref{j2psret-}) &
(\ref{jpsrsrret1}) & (\ref{jpsrsrret}) &  &  & (\ref{intjpsret-}) \\
\end{tabular}
\end{center}
\label{tab2}
\end{table}

\end{document}